\documentclass[11pt]{article}   

\usepackage[left=2.54cm, right=2.54cm, top=2.54cm]{geometry}
\usepackage{setspace} 
\usepackage{appendix}

\usepackage{hyperref}
\usepackage{nameref}
\hypersetup{colorlinks,
	linkcolor=blue,
	citecolor=blue,
	urlcolor=magenta,
	linktocpage,
	plainpages=false}

\usepackage{natbib}
\usepackage{graphicx}
\graphicspath{ {figures/} } 
\usepackage{amsmath,amsfonts,bm} 
\usepackage{algorithm}
\usepackage[noend]{algpseudocode} 
\usepackage{multirow} 
\usepackage{adjustbox} 
\usepackage[makeroom]{cancel} 

\hyphenchar\font=-1
\providecommand{\keywords}[1]
{
	\small	
	\textbf{\textit{Keywords:}} #1
}

\usepackage[T1]{fontenc}
\usepackage[utf8]{inputenc}
\usepackage[affil-it]{authblk} 
\usepackage{etoolbox}
\usepackage{lmodern}

\makeatletter
\patchcmd{\@maketitle}{\LARGE \@title}{\fontsize{16}{19.2}\selectfont\@title}{}{}
\makeatother

\title{Estimating Dynamic Conditional Spread Densities to Optimise Daily Storage Trading of Electricity}
\author[$\dagger$]{Ekaterina Abramova}
\author[$\ddagger$]{Derek Bunn}
\affil[$\dagger$]{Department of Finance, London Business School, London, UK}
\affil[$\ddagger$]{Department of Management Science and Operations, London Business School, London, UK}
\affil[ ]{\textit {\{eabramova,dbunn\}@london.edu}}
\date{01 March 2019}

\begin{document}
	\maketitle
	\bigskip
	
	\begin{abstract}
		This paper formulates dynamic density functions, based upon skewed-t and similar representations, to model and forecast electricity price spreads between different hours of the day. 
		This supports an optimal day ahead storage and discharge schedule, and thereby facilitates a bidding strategy for a merchant arbitrage facility into the day-ahead auctions for wholesale electricity. 
		The four latent moments of the density functions are dynamic and conditional upon exogenous drivers, thereby permitting the mean, variance, skewness and kurtosis of the densities to respond hourly to such factors as weather and demand forecasts. 
		The best specification for each spread is selected based on the Pinball Loss function, following the closed form analytical solutions of the cumulative density functions. 
		Those analytical properties also allow the calculation of risk associated with the spread arbitrages. 
		From these spread densities, the optimal daily operation of a battery storage facility is determined.
	\end{abstract}
	
	\keywords{Electricity, Spreads, Forecasting, GAMLSS}
	
	\newpage
	\section{Introduction} \label{sec_Intro}
	Whilst day ahead electricity price forecasting has been a topic of substantial and wide ranging research in terms of methods, the focus has mostly been upon price levels for the delivery periods (usually hourly) in the following day.  
	More recently there has been an interest in density forecasts for the hourly prices, motivated by considerations of risk management, see \citep{weron2014electricity, nowotarski2017recent} for extensive reviews. 
	In this paper we provide a new formulation with a focus upon price spreads, and specifically we forecast the density functions for the intraday spreads in the day-ahead prices.
	The optimal operation of storage facilities, e.g. batteries, or load shifting programmes, e.g. demand-side management, over daily cycles depends upon these spreads if they are operated as merchants, arbitraging buying and selling from the wholesale market. 
	Furthermore if risk is a consideration, analysis of the mean-differences in price levels would be inadequate, and we therefore directly estimate the density functions of all hourly spreads in prices at the day-ahead stage. 
	Our specification, estimation and forecasting of these arbitrage spreads is new and computationally-intensive. 
	We then show how these spread densities can support the optimal daily operation of a risk-constrained merchant battery facility.
	
	In contrast to the body of work on gas and other storable commodities, e.g. \citep{boogert2011gas, secomandi2018improved}, 
	because of the daily periodicity in electricity prices and the predominance of the day-ahead auctions, which typically set all hourly prices for the following day simultaneously, the operational horizon for storage and load shifting is usually episodic on a daily basis. 
	So the continuous time, dynamic optimisation formulations used for gas and other commodity storage operations are less appropriate for electricity, and furthermore the stochastic simplifications generally required in the analysis of these other models would not meet the requirements of adequate fit to the more complex power price dynamics. 
	Thus, based upon day-ahead forecasts for the drivers of electricity prices, such as demand, wind and solar production, gas and coal prices, forecasts for electricity price levels have been proposed from various methods, e.g. \citep{nowotarski2015computing, garcia2012forecasting, karakatsani2010fundamental} but, apparently, no methods have been developed specifically for forecasting intraday spread densities. 
	Until recently, storage assets, such as pumped hydro storage would regularly store energy at night and discharge at the daily peak demand periods, which were quite predictable. 
	But with the penetration of wind and especially solar generating facilities, the peaks and troughs in prices move around the day and in sunny locations with substantial solar energy, e.g. California, the lowest prices may often be in the middle of the day \citep{denholm2015overgeneration}.
	Thus, the expected daily spreads in prices, and their consequent arbitrage opportunities for storage or load shifting, will depend upon wind and solar forecasts, as well as demand and supply considerations. 
	Furthermore the price density functions are non-normal with skewness switching between positive and negative depending upon the dynamics of production of renewable energy \citep{gianfreda2017stochastic}.
	
	We apply our spread densities formulation to the German market. 
	This is the largest and the main daily reference market for wholesale power in Europe. 
	It is also strongly influenced by wind and solar production, as well as providing a context where batteries and demand-side management are active innovations.  
	The day-ahead auction has been actively researched and closes at noon each day, with the vector of 24 hourly prices for the next day being released an hour later. 
	In this context, a reasonable question might be why is there a need in the morning to forecast these prices, if they are available by 13:00, and apply to the following day? For the operation of spread-based arbitrage facilities in particular, it is clear that the traders need to have a forward plan in order to decide whether they will be entering the auction to buy or to sell at particular hours and thereby formulate their bids and offers into the auction accordingly. 
	
	For modelling the spread densities, we adapt the Generalised Additive Model for Location, Scale and Shape (GAMLSS) parametric regression model \citep{stasinopoulos2007generalized}, which has already been used effectively to form day-ahead densities of price levels in the German context \citep{gianfreda2017stochastic}. 
	Within this framework, the hourly electricity price spreads form a response variable, whose distribution function varies according to multiple exogenous factors.
	The GAMLSS framework allows choice from a wide range of distributions, whose moments change according to the exogenous variables specified using (non)linear relationships. 
	The dynamic location, scale and shape parameters ("latent moments" related to the mean, volatility, skewness and kurtosis of price spreads) are therefore explicitly incorporated into the forecasting model.
	
	The paper therefore proceeds by first describing the data and the density estimation process. 
	In section 3 we use the Pinball Loss function to select the best fitting density model with four latent moments. 
	Then in section 4 we undertake a rolling window forecasting evaluation and demonstrate the value of the dynamic, conditional latent moment estimates. 
	Section 5 uses these density functions to devise the optimal daily scheduling of battery storage and demonstrates through backtesting the superior profitability compared to using normal densities. 
	Section 6 concludes.

	\section{Data and Methodology}
	\subsection{Data Processing}
	The German hourly day-ahead electricity price, wind forecast, solar forecast and actual total load data were downloaded from the Open Power System Data website \url{https://data.open-power-system-data.org} for the period of 01.01.2012 - 31.03.2017 (resulting in $t=1, ..., 1917$ time steps).
	The data is comprised of four German control areas in MW: 50 Hertz, Amprion, TenneT and TransnetBW.
	The summer time hour change was accounted for by creating hour 02 with interpolation between hours 01 and 03.
	For the clock change back the (later) 02 hour was deleted.
	
	The German day-ahead total load data is calculated as an average of 4 x 15 min segments following the beginning of each hour.  
	The "actual total load" for each control area (obtained at the end of the 15 min real bidding time in the balancing markets) is averaged and results across four regions are summed up to give the total actual load. 

	We use the steam coal ARA 1 month forward benchmark index for steam coal (one price per month) and the Germany Gaspool (GPL) natural gas day-ahead forward (one price per day) for gas. 
	The weekly seasonality and holidays are included into a single dummy variable which takes on value 1 for Saturday/Sunday and the following German state holidays: New Year's Day, Good Friday, Easter Monday, Labour Day, Ascension Day, Whit Monday, German Unity day, Christmas Day, Boxing Day and New Years Eve (see Appendix \ref{app:dataCollection} for further details).

	\subsection{Creating Spreads}
	The intraday hourly electricity spot price data displays high volatility, fast changing dynamics and highly skewed distributions (see Appendix Figures \ref{fig:App1_histHr00_spotPrice} - \ref{fig:App1_histHr20_spotPrice}). 
	The variance-covariance matrix of the hourly prices displays strong correlations, as demonstrated  by highly-positive off-diagonal entries (see Appendix Figure \ref{fig:App1_prices_VarCov}). 
	Thus, there is no independence between the hourly prices, and so the spread densities need to be modelled directly.
	
	Because the hourly spread data possesses high degrees of skewness (see Appendix Figure \ref{fig:App1_spreads_skewness}) and kurtosis (see Appendix Figure \ref{fig:App1_spreads_kurtosis}), we model the full four parameter distribution of electricity price spreads at each time step (day), $t$.      
	An intraday spread, $Y^{(s)}_{t}$, between two hours, denoted as spread number $s$, of the day-ahead electricity spot prices, is calculated by taking away the later hour price information from the earlier one, resulting in a positive spread if later hour is less expensive and negative otherwise. 
	The full spread data set is comprised of a total possible $\frac{n(n+1)}{2}= \frac{24\times25}{2}=276$ non-duplicate entries, and is denoted by $\mathbf{Y} \in \mathbb{R}^{1917 \times 276}$.
	Each electricity spread time series data is tested for stationarity using the Augmented Dickey Fuller test (see Appendix Section \ref{App:subsec_ADF_spreads}) and is confirmed to be stationary at the 1\% significance level.
	We plot example histograms for intraday spreads obtained between hours $00-08, 08-12, 12-16, 16-20$ (see Appendix Figures \ref{fig:App1_histHr00-08} - \ref{fig:App1_histHr16-20}) which depict high skewness and kurtosis of the spreads.
	
	The spreads between exogenous variables are likewise formed by taking away the later hour values from the earlier ones in all cases except for the gas forward prices, coal forward prices and the dummy variable, for which only the daily, rather than hourly, values are available. 
	Hence the following exogenous variables are considered for modelling $Y^{(s)}_{t}$:
	(1) spread of the lagged intraday electricity price,
	(2) gas Gaspool forward daily price, 
	(3) coal ARA forward daily price,
	(4) spread of wind day-ahead forecast,
	(5) spread of solar day-ahead forecast,
	(6) dummy variable taking value of 1 for weekends/holidays,
	(7) spread of the day-ahead total load forecast, and
	(8) an interaction load variable, calculated as $Load_{spread} * \frac{1}{2}(Load_{earlierHr} + Load_{laterHr}) =  \frac{1}{2}*(Load_{earlierHr}^2 - Load_{laterHr}^2)$ i.e. average of the load for the two hours from which the spread is calculated, weighted by the load spread obtained for those hours.
	This variable provides interaction of $\Delta Load * {avLoad}$ in order to account for the rate of change in load.
	The full exogenous variables spread data is stored in design matrix $\mathbf{X} \in \mathbb{R}^{1917 \times 9 \times 276}$, where the first column for each spread number $s=1,...,276$ is a vector of 1s needed for the calculation of an intercept.

	\subsection{GAMLSS Framework}
	\subsubsection{Mathematical Formulation}
	The Generalised Additive Model for Location, Scale and Shape (GAMLSS) framework allows modelling each parameter (${\mu}, {\sigma}, {\nu}, {\tau}$) of the response variable's distribution as a function of explanatory variables.
	The range of possible distributions includes both exponential family (as per Generalised Linear Model (GLM)) and general family distributions, which allow for both discrete/continuous and highly skewed/kurtotic distributions (unlike GLM). 
	The probability density function of response variable for $T$ observations is given by $f_{Y}(y^{(s)}_{t} | \boldsymbol{\theta}^{(s)}_{t}) \sim D(\boldsymbol{\theta}^{(s)}_{t})$, where $s=1,...,276$ is the spread number, $\boldsymbol{\theta}^{(s)}_{t} = [\theta^{(s)}_{t,1}, \theta^{(s)}_{t,2}, \theta^{(s)}_{t,3}, \theta^{(s)}_{t,4}]^T = [\mu^{(s)}_t, \sigma^{(s)}_t, \nu^{(s)}_t, \tau^{(s)}_t]^T$ is a vector of distribution parameters, and $D(\boldsymbol{\theta}^{(s)}_t)$ represents the distribution of spread $s$ on day $t$.
	We use \textit{parametric linear GAMLSS} framework which relates distribution parameters to explanatory variables by
	\begin{equation} \label{eq:general_linearGamlss}
	g_{k}(\boldsymbol{\theta}^{(s)}_k) = \boldsymbol{\eta}^{(s)}_k = \mathbf{X}^{(s)}_k \boldsymbol{\beta}^{(s)}_k
	\end{equation}
	where 
	$s = 1, ..., 276$ indicates the number given to each unique spread between two intra-day hours;
	$k=1,...,4$ specifies the distribution parameter corresponding to ${\mu}, {\sigma}, {\nu}, {\tau}$ respectively;
	$\boldsymbol{\theta}^{(s)}_k \in \mathbb{R}^T$ is a vector comprised of values for distribution parameter $k$ over $t=1,...,T$ time steps (i.e. $\boldsymbol{\theta}^{(s)}_1 = \boldsymbol{\mu}^{(s)}, \boldsymbol{\theta}^{(s)}_2 = \boldsymbol{\sigma}^{(s)}$, $\boldsymbol{\theta}^{(s)}_3 = \boldsymbol{\nu}^{(s)}, \boldsymbol{\theta}^{(s)}_4 = \boldsymbol{\tau}^{(s)}$); 
	$T$ is the number of observations;
	$g_{k}(\cdot)$ is the monotonic link function of distribution parameter $k$;
	$\boldsymbol{\eta}^{(s)}_k \in \mathbb{R}^T$ is the linear predictor vector for distribution parameter $k$;
	$ \boldsymbol{\beta}^{(s)}_k =[\beta^{(s)}_{0,k},\beta^{(s)}_{1,k}, ..., \beta^{(s)}_{J_k,k}]^T \in \mathbb{R}^{J_k + 1}$ vector of coefficients learnt for parameter $k$;
	$J_k$ is the number of significant exogenous variables for parameter $k$ obtained at 5\% significance level following the estimation and specification steps; and
	$\mathbf{X}^{(s)}_k \in \mathbb{R}^{T \times J_k+1}$ is the design matrix with each column containing spread data for significant independent variables.\\
	Equation \ref{eq:general_linearGamlss} can be re-written for each distribution parameter
	\begin{align}
	g_{1}(\boldsymbol{\mu}^{(s)})      &=  \boldsymbol{\eta}^{(s)}_1  = \mathbf{X}^{(s)}_1 \boldsymbol{\beta}^{(s)}_1 \\
	g_{2}(\boldsymbol{\sigma}^{(s)}) &= \boldsymbol{\eta}^{(s)}_2  = \mathbf{X}^{(s)}_2 \boldsymbol{\beta}^{(s)}_2 \\
	g_{3}(\boldsymbol{\nu}^{(s)})       &= \boldsymbol{\eta}^{(s)}_3 = \mathbf{X}^{(s)}_3 \boldsymbol{\beta}^{(s)}_3 \\
	g_{4}(\boldsymbol{\tau}^{(s)})     &= \boldsymbol{\eta}^{(s)}_4  = \mathbf{X}^{(s)}_4 \boldsymbol{\beta}^{(s)}_4 
	\end{align}
	The choice of the link function influences how the linear predictor (i.e. systematic component $\mathbf{X}^{(s)}_k \boldsymbol{\beta}^{(s)}_k$) relates to each parameter.
	For example: a log link function for the standard deviation $g_{2}(\boldsymbol{\sigma}^{(s)}) = log(\boldsymbol{\sigma}^{(s)})$ results in the relationship $log(\boldsymbol{\sigma}^{(s)})= \boldsymbol{\eta}^{(s)}_2 = \mathbf{X}^{(s)}_2 \boldsymbol{\beta}^{(s)}_2$, hence the distribution parameter itself is obtained through a transformation $\boldsymbol{\sigma}^{(s)} = \exp(\mathbf{X}^{(s)}_2 \boldsymbol{\beta}^{(s)}_2)$.
	
	Parameters $\boldsymbol{\theta}^{(s)}_k$ are calculated by maximizing the penalized likelihood using an algorithm which does not require calculation of the likelihood function's cross derivatives, but is a generalisation of the MADAM algorithm used for fitting mean and dispersion additive models, as in \citep{rigby1996mean}.

	\subsubsection{Suitable Four Parameter Distributions} 
	The GAMLSS family provides a number of four parameter continuous distributions with pre-determined link functions.
	In order to model electricity spreads, the parameters of the distribution of choice must meet the following criteria: the standard deviation and kurtosis should take on positive values only, while the mean and skewness should be able to take on negative values since the spread data shows that both parameters can be positive or negative. 
	Therefore distributions with the following link functions for each parameter are considered (see Table \ref{tab:possibleDists}): 
	mean: \texttt{identity}, standard deviation: \texttt{log} or \texttt{logit}, skewness: \texttt{identity}; kurtosis: \texttt{log} or \texttt{logit}. 
	
	There are seven continuous distributions within the GAMLSS framework which correspond to the necessary link function specifications:
	Johnson's SU ($\mu$ the mean),
	Johnson's original SU (JSU),
	Skew power exponential type 1 (SEP1),
	Skew power exponential type 2 (SEP2),
	Skew t type 1 (ST1),
	Skew t type 2 (ST2), and
	Skew t type 5 (ST5).
	The Box-Cox power exponential and Box-Cox t distributions are not suitable due to only being defined on the positive interval $Y_t \in [0,+\infty)$, while the electricity spread data is highly skewed and kurtotic (see Appendix Figures \ref{fig:App1_spreads_skewness} and \ref{fig:App1_spreads_kurtosis}). 
	The Skew t type 3 and 4 are also unsuitable since they are only defined for positive skewness yet the spread data is often negatively skewed.
	\begin{table}[h!]
		\centering 
		\begin{tabular}{ |l|l|l|l|l| }  
			\hline
			\textbf{Continuous Distribution} & $\boldsymbol{\mu}$ & $\boldsymbol{\sigma}$ & $\boldsymbol{\nu}$ & $\boldsymbol{\tau}$ \\ \hline
			\textbf{Johnson's SU ($\mu$ the mean) (JSU)} & \texttt{identity}  & \texttt{log}       & \texttt{identity} & \texttt{log} \\
			\textbf{Johnson's original SU (JSUo)}                  & \texttt{identity}  & \texttt{log}       & \texttt{identity}  & \texttt{log} \\
			\textbf{Skew power exponential type 1 (SEP1)}  & \texttt{identity}  & \texttt{log}       & \texttt{identity}  & \texttt{log} \\
			\textbf{Skew power exponential type 2 (SEP2)} & \texttt{identity}  & \texttt{log}       & \texttt{identity}  & \texttt{log} \\
			\textbf{Skew t type 1 (ST1)}                                    & \texttt{identity}  & \texttt{log}       & \texttt{identity}  & \texttt{log} \\
			\textbf{Skew t type 2 (ST2)}                                   & \texttt{identity}  & \texttt{log}       & \texttt{identity}  & \texttt{log} \\
			\textbf{Skew t type 5 (ST5)}                                   & \texttt{identity}  & \texttt{log}       & \texttt{identity}  & \texttt{log} \\
			\hline
		\end{tabular}
		\caption{\textbf{Continuous Four Parameter Distributions} - Suitable Distributions in GAMLSS.} \label{tab:possibleDists}
	\end{table}

	
	The expected value of a random variable $Y_t$ for each distribution in Table \ref{tab:possibleDists} is given by $E(Y_t) = \mu_t + \sigma_t E(Z_t)$, where $z_t = \frac{y_t - \mu_t}{\sigma_t}$ is a normalised value of $y_t$, $Z_t$ is specified for each distribution separately, and $\mu_t,\sigma_t,\nu_t,\tau_t$ are the mean, standard deviation, skewness and kurtosis of the given distribution at time step $t$. 
	See Appendices \ref{app:SEP1} - \ref{app:ST5} for details on expected value calculations for each distribution. 
	
	\section{Model Selection}
	In order to build accurate models of the spreads, we begin by analysing which of the seven possible distributions in Table \ref{tab:possibleDists} fits each price spread data the best.
	We also analyse whether using a single distribution for modelling all of the spread data is a viable possibility and provide reasoning for when this may be beneficial.
	The analysis is divided into two main steps: 
	(a) \textit{simple distribution fit}, where each distribution of Table \ref{tab:possibleDists} is fitted to the spreads in the training data set and the Akaiki Information Criterion (AIC) is used to assess the goodness-of-fit; and 
	(b) \textit{factor-based distribution fit}, where for each spread hour, the candidate distributions resulting from the analysis of step (a) are utilised to build a model using exogenous factors. 
	The fit of these models is assessed using a validation data set and a number of goodness-of-fit measures.
	
	\subsection{Simple Distribution Fit} \label{sec:simple_dist_fit}
	The simple distribution fit is performed using GAMLSS function \texttt{gamlss <- $\mathbf{y} \sim \mathbf{1}$}, which results in a model under the specified distribution and time series $\mathbf{y}$.
	The analysis is performed using \textit{training data}, comprised of the first 60\% of the spread time series, $\mathbf{Y}_{train} \in R^{1150 \times 276}$.
	Therefore, for each spread number $s=1,...,276$ and each distribution $i=1,...,7$ of Table \ref{tab:possibleDists}, we build a model $\widehat{M}^{(s,i )} \leftarrow \mathbf{y}^{(s)}_{train} \sim \mathbf{1}$, resulting in $\widehat{ \mathbf{M} } \in \mathbb{R}^{276 \times 7}$ models.
	The AIC criteria of models $\widehat{ \mathbf{M} } ^{(s)} \in \mathbb{R}^{7}$, obtained for spread number $s$, are ranked in ascending order and the distribution corresponding to the model with the lowest AIC criterion is selected as the 'distribution of best fit'. 
	
	The results show that \textit{Skew t type 5} distribution is selected most often as the distribution of best fit based on this simple selection criterion (see Table \ref{tab:bestTrainDists_SimpleFit}).
	It is closely followed by the same family \textit{Skew t type 1} distribution.
	Overall, all of the distributions except JSUo are indicated as potentially of the best fit for some spread hours.
	A more detailed breakdown of the spreads for which each distribution was selected is displayed in Appendix Figure \ref{fig:App1_spreads_bestDistSimple}.
	We use the result to further analyse the six possible distributions with a factor-based distribution fit method.
	\begin{table}[h!]
		\centering 
		\begin{tabular}{ |l|l|l|l|l|l|l| }  
			\hline
			\textbf{JSU} & \textbf{JSUo} & \textbf{SEP1} & \textbf{SEP2} & \textbf{ST1} & \textbf{ST2} & \textbf{ST5} \\ 
			\hline
			$49$ & $0$ & $38$ & $14$ & $68$ & $33$ & $74$ \\
			\hline
		\end{tabular}
		\caption{\textbf{Simple Distribution Fit (Training Data)} - Best Distribution Selection Based on AIC.} \label{tab:bestTrainDists_SimpleFit}
	\end{table}

	\subsection{Factor-based Distribution Fit}
	The results of simple distribution fit indicated that six out of seven continuous four parameter distributions could be used for modelling electricity spread data (see Table \ref{tab:bestTrainDists_SimpleFit}), which are $D^{(1)}$ - JSU, $D^{(2)}$ - SEP 1, $D^{(3)}$ - SEP 2, $D^{(4)}$ - ST1,  $D^{(5)}$ - ST2, and $D^{(6)}$ - ST5.
	Hence we proceed with a factor-based analysis by fitting a regression to each parameter of candidate distributions using exogenous variables within the GAMLSS framework.
	The training data set is comprised of the dependent $\mathbf{Y}_{train} \in R^{1150 \times 276}$ and independent $\mathbf{X}_{train} \in R^{1150 \times 9 \times 276}$ variables, where for each spread number $s$, under distribution $D^{(i)}, i=1,...,6$, initial equations for the central moments are
	\begin{align}
	\widehat{\mu}_t  &= \widehat{\beta}_{1,0} + \widehat{\beta}_{1,1} x_{t}  + ... + \widehat{\beta}_{1,8} x_{8,t} 
	= \mathbf{x}_t^T \bm{\widehat{\beta}}_1                
	\label{eq:mu} \\
	log(\widehat{\sigma}_t) &= \widehat{\beta}_{2,0} + \widehat{\beta}_{2,1} x_{t}  + ... + \widehat{\beta}_{2,8} x_{8,t} 
	= \mathbf{x}_t^T  \bm{\widehat{\beta}}_2                
	\label{eq:sigma} \\
	\widehat{\nu}_t &= \widehat{\beta}_{3,0} + \widehat{\beta}_{3,1} x_{t}  + ... + \widehat{\beta}_{3,8} x_{8,t} 
	= \mathbf{x}_t^T  \bm{\widehat{\beta}}_3              
	\label{eq:nu} \\
	log(\widehat{\tau}_t) &= \widehat{\beta}_{4,0} + \widehat{\beta}_{4,1} x_{t}  + ... + \widehat{\beta}_{4,8} x_{8,t} 
	= \mathbf{x}_t^T \bm{\widehat{\beta}}_4               
	\label{eq:tau}
	\end{align} 
	where 
	$\bm{\widehat{\beta}}_k   =[\widehat{\beta}_{k,0},\widehat{\beta}_{k,1}, ..., \widehat{\beta}_{k,8}]^T \in \mathbb{R}^{9}$ is initial vector of coefficients for distribution parameter $k$,
	$\mathbf{x}_t =[1,x_{1,t},...,x_{8,t}]^T \in \mathbb{R}^{9}$ is the initial vector of independent variables where
	$x_1$ is the spread of lagged day-ahead electricity price,
	$x_2$ is the gas Gaspool forward daily price, 
	$x_3$ is the coal ARA forward daily price, 
	$x_4$ is the spread of wind day-ahead forecast,
	$x_5$ is the spread of solar day-ahead forecast, 
	$x_6$ is the dummy variable taking value 1 for weekends/holidays, 
	$x_7$ is the spread of the day-ahead total load forecast,
	$x_8$ is the interaction load variable.
	
	The models are specified using iterative updating of the equations for each moment, where the refinement is performed by deleting the most insignificant variable one-by-one and re-estimating the model, until all variables are significant at 5\% (see Algorithm \ref{algo:learningModels}).	
	This results in $\mathbf{\widehat{M}} \in \mathbb{R}^{276 \times 6}$ models containing the estimated coefficients.
	Note: the process revealed that some distributions were not suitable for modelling certain spreads within the GAMLSS framework.
	Convergence was not achieved for some distribution parameters, typically $\tau$ and occasionally $\mu$ (see Appendix Section \ref{app:DistnFit_SpreadsAdv}) and when this happened the distribution was omitted from candidacy for that spread.	
	
	The models estimated using the factor-based distribution fit are analysed in four ways: (1) producing the expected value fit over the training data, (2) producing the expected value fit over \textit{validation data} (comprised of the next 20\% of unseen time series, at data points $t=1151,...,1534$), (3) analysing the goodness-of-fit over the validation data using Root Mean Squared Error, and (4) analysing the goodness-of-fit over validation data using Pinball Loss function measure.
	
	\begin{algorithm}[h!] 
		\caption{Factor-based Distribution Fit - Model Specification and Estimation}
		\label{algo:learningModels}
		\begin{algorithmic}[1]
			\For{each spread number $s=1,...,276$}
			\State Extract full training data design matrix $\mathbf{X}^{(s)}_{train} \in \mathbb{R}^{1150 \times 9}$ i.e. time steps $t=1,...,1150$
			\For{each distribution $D^{(i)}, i=1,...,6$}
			\State Initialise iteration number $j \gets 0$
			\State Initialise model $\widehat{M}^{(s,i)}_j$ $\gets \{ \widehat{\boldsymbol{\beta}}^{(s,i)}_{k,j} \} _{k=1}^4 \in \mathbb{R}^{J_k+1 \times 4}$ using RS algorithm
			\While{any $\boldsymbol{\widehat{\beta}}^{(s,i)}_{k,j}$ insignificant at 5\%} 
			\State $j \gets j + 1$
			\State Find most insignificant coeff. ${\widehat{\beta}^\dagger}$ (except intercept) across all $k$ 
			\State Remove exogenous variable associated with ${\widehat{\beta}^\dagger}$ from Eq. for $k$ and from $\mathbf{X}^{(s)}_{train}$
			\State Re-estimate model $\widehat{M}^{(s,i)}_j \gets \{ \boldsymbol{\widehat{\beta}}^{(s,i)}_{k,j} \} _{k=1}^4$ using RS algorithm
			\EndWhile
			\EndFor
			\EndFor
		\end{algorithmic}
	\end{algorithm}

	\subsubsection{Expected Value Fit Over Training Data}
	The fitted distribution parameters $\widehat{\boldsymbol{\theta}}^{(i)}_{train} \in \mathbb{R}^{1150 \times 4 \times 276}$ for each distribution $i=1,...,6$ are used to find the fitted expected value, $E(\widehat{\mathbf{Y}}_{train})$, of the training price spread data for $s=1,...,276$ over training data points $t=1,...,1150$ using Equations \ref{eq:SEP1_E(Y)} - \ref{eq:ST5_E(Y)}.
	Illustrative examples of the fitted expected values for spreads between hours 00-08 and 08-12 across the six possible distributions are depicted in Appendix Figures \ref{fig:app1_muTrain_00_08} and \ref{fig:app1_muTrain_08_12} respectively.
	The true price spread values, $E(\mathbf{Y}_{train})$, are given by blue lines and the fitted, $E(\widehat{\mathbf{Y}}_{train})$, by red lines. 
	The plots show a good fit to the true data across all distributions with slight underestimation in the spread price. 
	In the case of the two example spreads, the spikes in the data seem to be fitted best by different distributions.
	The spread hour 00-08 spike at time step $t=360$ is only fitted using the SEP2 distribution, while spike of spread 08-12 at time step $t=409$ is fitted best with the ST2 distribution. 
	This supports the analysis of the six possible distributions for candidacy of best fit to individual spread data. 
	
	\subsubsection{Expected Value Fit Over Validation Data}
	The fitted models $\widehat{\mathbf{M}} \in \mathbb{R}^{276 \times 6}$, containing the estimated coefficients for each spread number $s=1,...,276$ under each distribution $D^{(i)}, i=1,...,6$, are used to forecast the distribution parameters, $\widehat{\boldsymbol{\theta}}^{(i)}_{validate} \in \mathbb{R}^{ 383 \times 4 \times 276}$ over validation data.
	We note that the same estimated model $\widehat{M}^{(s,i)}$ is used to build the forecasts over validation time series, i.e. each vector of estimated coefficients $\boldsymbol{\widehat{\beta}}^{(s,i)}_k$ for distribution parameters $k=1,...,4$ is re-used at each time step $t$ to make the predictions.
	Once the distribution parameters are forecasted the expected value of the spread is calculated using Equations \ref{eq:SEP1_E(Y)} - \ref{eq:ST5_E(Y)}.
	Illustrative examples of the forecasted expected values for spread hours 00-08 and 08-12 across the six possible distributions are depicted in Appendix Figures \ref{fig:app1_muVal_00_08} and \ref{fig:app1_muVal_08_12} respectively.
	The true spread prices, $E(\mathbf{Y}_{validate})$, are given by blue lines and the forecasted, $E(\widehat{\mathbf{Y}}_{validate})$, by red lines. 
	The results for spread 00-08 show that all distributions are able to fit the validation data well, however the downward spikes seem to be underestimated across all distributions. 
	%
	The expected value forecast for the spread hour 08-12 shows that some distributions fit the data better than others, for example ST1 does not fit the data as well as ST5.
	
	\subsubsection{Goodness of Fit Measure - Root Mean Squared Error}
	A common performance measure, Root Mean Squared Error (RMSE), is used to assess the goodness of fit for the forecasted expected value of the spreads over the validation data set. 
	The RMSE is calculated for the forecasted expected values of each spread, $s=1,...,276$, obtained under each distribution $D^{(i)},i=1,...,6$, using
	\begin{equation}
	\text{RMSE}^{(s,i)} = \sqrt{ \sum^{383}_{t=1} \bigg[ E\Big(Y^{(s,i)}_t\Big) -  E\Big(\widehat{Y}^{(s,i)}_t\Big) \bigg]^2 }
	\end{equation}
	
	Since the expected value at each time step $t$ is calculated from all of the four forecasted parameters $\big( \widehat{\mu}^{(s,i)}_t, \widehat{\sigma}^{(s,i)}_t, \widehat{\nu}^{(s,i)}_t, \widehat{\tau}^{(s,i)}_t \big)$, the RMSE measure provides a goodness of fit based on the forecast for the entire distribution specification.
	The results are summarised in Table \ref{tab:bestValidDists_AdvFit_basedonRMSE}, and show that ST5 distribution was used to form models corresponding to the lowest error across $\frac{110}{276}*100=40\%$ of the spreads. 
	This is in line with the finding of the simple distribution fit over the spread training data, which indicated that ST5 was chosen as the best distribution most often among suitable four parameter distributions (see Section \ref{sec:simple_dist_fit}).
	A detailed breakdown of best distribution assignment for each spread number and the corresponding RMSE values are shown in Appendix Figures \ref{fig:app1_muVal_RMSEbasedBestDistn} and \ref{fig:app1_muVal_RMSEbasedBestDistn-ActualRMSE} respectively.
	The results show that spreads with hours 13.00, 14.00 are the hardest to forecast since they have the highest RMSE error (dark red). 
	\begin{table}[h!]
		\centering 
		\begin{tabular}{ |l|l|l|l|l|l| }  
			\hline
			\textbf{JSU} & \textbf{SEP1} & \textbf{SEP2} & \textbf{ST1} & \textbf{ST2} & \textbf{ST5} \\ 
			\hline
			$49$ & $31$ & $21$ & $11$ & $60$ & $110$ \\
			\hline
		\end{tabular}
		\caption{\textbf{Factor-based Dist. Fit (Validation Data)} - Best Distribution Based on RMSE.}
		\label{tab:bestValidDists_AdvFit_basedonRMSE}
	\end{table}

	\subsubsection{Goodness of Fit Measure - Pinball Loss Score}
	Our model fits the entire four parameter distribution to the spread data at each time step. 
	Therefore we consider it more appropriate to evaluate the goodness-of-fit using an entire fitted distribution as represented by quantiles. 
	The Pinball Loss (PL) function is typically used as the objective function in quantile regression and can also be interpreted as the accuracy of a quantile forecasting model.
	We adopt this measure as our main performance metric to select the best distribution to each spread. 
	We follow Algorithm \ref{algo2:bestModelVal} when making an assessment of each predictive density power and outline the steps involved in calculating the performance measure below.
	\begin{itemize}
		%
		\item [\textbf{1}] \textbf{Pinball Loss Values} 
		At each time step $t$, the target quantiles $q_a, a = 1, 2, . . . , 99$ are extracted by inverting the cumulative distribution function of distribution $D(\boldsymbol{\theta}^{(s,i)}_t), i=1,...,6$ specified with forecasted distribution parameters $\big( \widehat{\mu}^{(s,i)}_t, \widehat{\sigma}^{(s,i)}_t, \widehat{\nu}^{(s,i)}_t, \widehat{\tau}^{(s,i)}_t \big)$ and compared to the true output $y_t$ with:
		\begin{eqnarray} \label{eq:pinball_loss_value}
		L^{(s,i)}_t(q_a,y_t) = \left\{
		\begin{array}{ll}
		\frac{1- a/100}{q_{a,t} - y_t} \quad \text{if} \quad y<q_a \\
		\frac{a/100}{y_t - q_{a,t}}     \quad \text{if} \quad y \geq q_a
		\end{array}
		\right.
		\end{eqnarray}
		This results in Pinball Loss Values vector, $\mathbf{L}_t^{(s,i)}(\mathbf{q}, y_t) \in \mathbb{R}^{J_a}$, where $J_a=\{99, 97, 95\}$ is the number of quantiles extracted.
		Certain quantiles failed to converge when calculating the full set of 99 values due to convergence issues around distribution tails.
		To resolve this we removed the tail quantiles one pair at a time and attempted to extract the quantiles again (i.e. resulting in 97 quantiles: $[q_2, q_{98}]$ and if still not convergent 95 quantiles $[q_3, q_{97}]$). 
		If this still did not resolve the issue, we omitted calculating the Pinball Loss value for that time step and recorded number of such occurrences, $n$.
		\textit{Note}: JSU and ST5 extracted quantiles with 100\% success, making these two distributions the most stable out of the 6 tested.
		This is in contrast to ST1 and ST2 distributions which failed to converge often (see Appendix Table \ref{app_tab:failedQuantilesValidationData} for the number of times each distribution failed to provide a quantile answer).
		\item [\textbf{2}] \textbf{Pinball Loss Scores} At each time step, $t$, the Pinball Loss Values are averaged resulting in a single Pinball Loss Score, $\bar{L}_t^{(s,i)}(q_a,y_t)$, calculated as:
		\begin{equation} \label{eq:pinball_loss_score}
		\bar{L}_t^{(s,i)}(q_a,y_t) = \sum_{a=1}^{J_a} L^{(s,i)}_t(q_a,y_t)
		\end{equation}
		Across all evaluated time steps, this results in a vector $\mathbf{\bar{L}}^{(s,i)}(\mathbf{q},\mathbf{y}) \in \mathbb{R}^{J_b}$, where  $J_b = 383-n$ ($n$ denotes the number of occurrences when $L^{(s,i)}_t(q_a,y_t)$ was not available due to convergence issues).
		\item [\textbf{3}] \textbf{ Pinball Loss Performance Measure} The final step in calculating a single value describing the goodness-of-fit measure using the Pinball Loss function is finding the average of the Pinball Loss Scores across the full validation forecast horizon at time steps $t=1151,...,1534$ (i.e. $383$ days):
		\begin{equation} \label{eq:pinball_loss_performance_measure}
		{\cal{L}}^{(s,i)} = \sum_{t=1}^{J_b} \bar{L}^{(s,i)}_t(q_a,y_t)
		\end{equation}
		The distribution $D^{(i)}$ corresponding to the model with the lowest \textit{Pinball Loss Performance Measure}, ${\cal{L}}^{(s,i)} \in \mathbb{R}$, is selected as the distribution of best fit for spread number $s$.
	\end{itemize}
	
	\begin{algorithm} 
		\caption{Model Selection using Pinball Loss Function}
		\label{algo2:bestModelVal}
		\begin{algorithmic}[1]
			\For{each spread number $s=1:276$}
			\State Extract validation data design matrix $\mathbf{X}^{(s)}_{validate} \in \mathbb{R}^{383 \times 9}$ i.e. time steps $t=1151,...,1534$ 
			\For{each distribution $D^{(i)}, i=1:6$}
			\State Forecast moments of $D^{(i)}$ resulting in $\boldsymbol{\theta}^{(s,i)} = [\boldsymbol{\mu}^{(s,i)}, \boldsymbol{\sigma}^{(s,i)}, \boldsymbol{\nu}^{(s,i)}, \boldsymbol{\tau}^{(s,i)}]^T$ each of size $\mathbb{R}^{383}$
			\For{time step $t=1,...,383$} 
			\State Obtain vector of quantiles $\mathbf{q}_t^{(s,i)} \in \mathbb{R}^{J_a}$ using $\boldsymbol{\theta}_t^{(s,i)}$
			\For { each quantile $q_{a,t}$ }
			\State Calculate Pinball Loss value $L^{(s,i)}_t(q_a,y_t)$ using Eq. \ref{eq:pinball_loss_value}
			\EndFor
			\State Calculate Pinball Loss Score $\bar{L}_t^{(s,i)}(q_a,y_t)$ using Eq. \ref{eq:pinball_loss_score}
			\EndFor
			\State Calculate Pinball Loss Performance Measure ${\cal{L}}^{(s,i)}$ using Eq. \ref{eq:pinball_loss_performance_measure}
			\EndFor
			\State Find $n=\text{argmin}_{i} {\cal{L}}^{(s)}$ and select corresponding distribution $D^{(n)}$ as best fit for $s$
			\EndFor
		\end{algorithmic}
	\end{algorithm}
	
\begin{figure}[h!]
		\centering
		\includegraphics[width=1.0\textwidth]{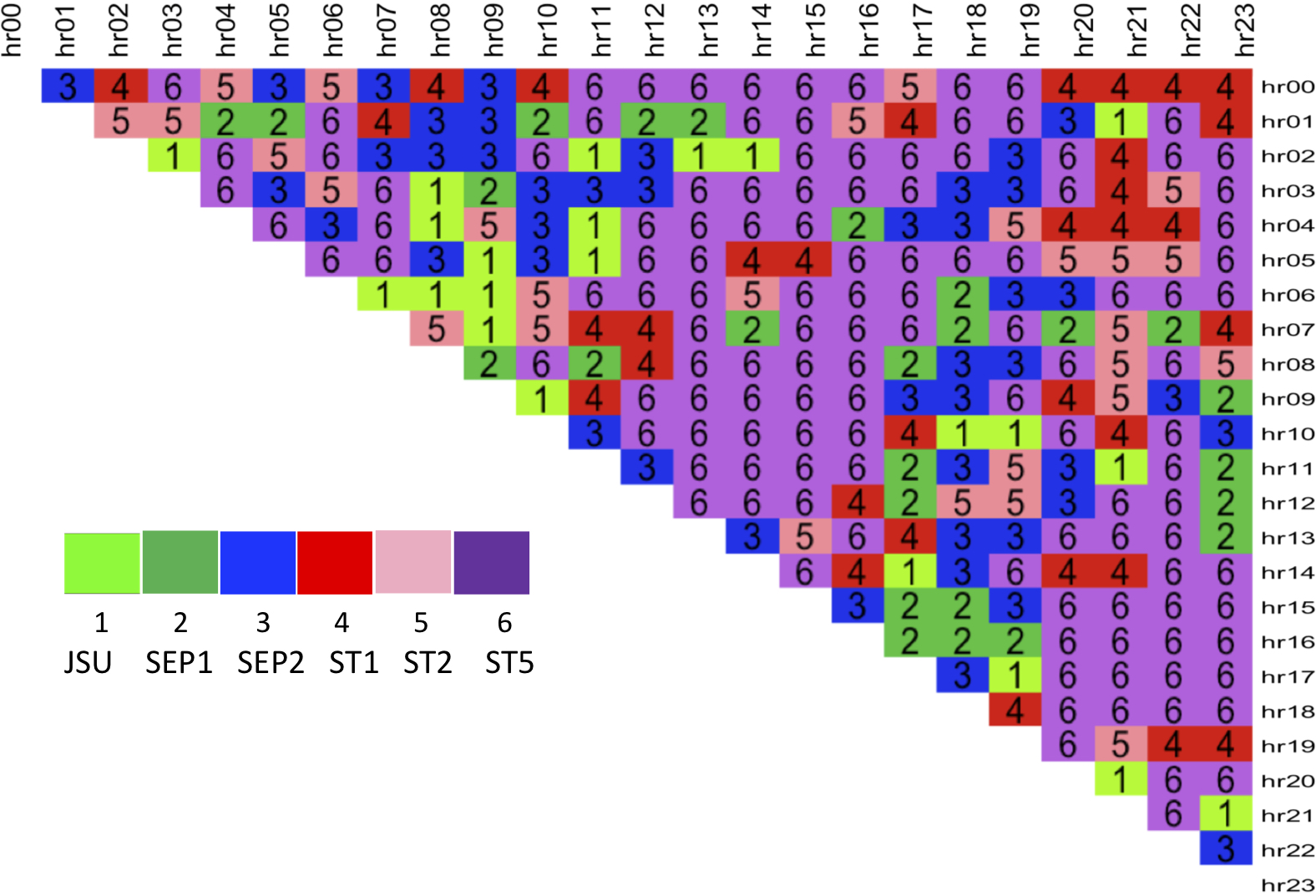}
		\caption{Best Distribution Based on Pinball Loss Function (Validation Data).}
		\label{fig:bestDistnAdvanced_PinballLossBased}
	\end{figure}

	The Pinball Loss Performance Measures are analysed for each spread $s$ and the results are presented in a $24 \times 24$ upper diagonal matrix, containing the best distribution number selected for each intraday spread between the hours indicated by row and column labels (see Figure \ref{fig:bestDistnAdvanced_PinballLossBased}). 
	The figure shows that majority of the spreads are fitted best with ST5 distribution (number 6 - purple colour). 
	This is in line with the results obtained from the factor-based distribution fit using the RMSE measure (see Appendix Figure \ref{fig:app1_muVal_RMSEbasedBestDistn}), and with the simple best distribution fit given by $\mathbf{y} \sim 1$ function (see Appendix Figure \ref{fig:App1_spreads_bestDistSimple}), which both favoured the ST5 distribution.
	
	Table \ref{tab:bestValidDists_AdvFit_basedonPINBALL} specifies the number of times each distribution was selected as best from the 276 possible spreads. 
	While ST5 is selected for almost 50\% of the spreads, the other 5 distributions were selected nearly with the same proportion for the remaining half of the spread data.
	\begin{table}[h!]
		\centering 
		\begin{tabular}{ |l|l|l|l|l|l| }  
			\hline
			\textbf{JSU} & \textbf{SEP1} & \textbf{SEP2} & \textbf{ST1} & \textbf{ST2} & \textbf{ST5} \\ 
			\hline
			$22$ & $26$ & $45$ & $33$ & $27$ & $123$ \\
			\hline
		\end{tabular}
		\caption{\textbf{Factor-based Dist. Fit (Validation Data)} - Best Distribution Based on PL.} 
		\label{tab:bestValidDists_AdvFit_basedonPINBALL}
	\end{table}
	Further to this, we calculate the \% difference of the best distribution PL performance measure value vs that of the next best distribution, $\frac{|{\cal{L}}^{(s)}_{1} - {\cal{L}}^{(s)}_{2}|}{{\cal{L}}^{(s)}_{2}} * 100\%$, for each spread number $s$.
	When ST5 was selected as the best distribution, the average difference of the performance measure was 4.84\% compared to, when other distributions were selected as best, the average difference was 1.44\% (with approx 1/3 of these cases containing ST5 is the second best).
	This makes the case for ST5 to potentially be used as a general best fit distribution across all spreads.
	
	Next, we re-estimate models based on the best distributions established for each spread number, analyse performance using Rolling Window forecast technique and evaluate results using Pinball Loss function utilised within a formal significance test framework, based on Diebold-Mariano test.

	\section{Rolling Window Forecast Analysis} 
	\begin{figure}[h!]
		\centering
		\includegraphics[width=0.6\textwidth]{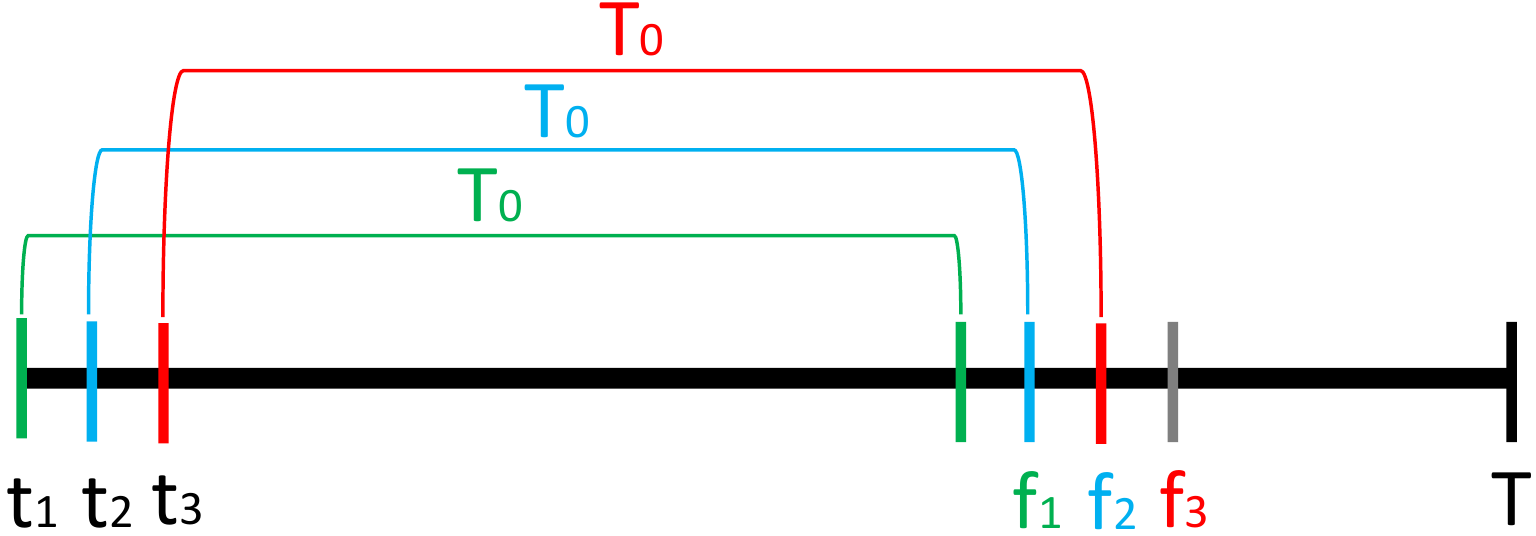}
		\caption{Rolling window procedure.}
		\label{fig:rollingWindow}
	\end{figure}
	We perform a careful forecast analysis based on the best distribution selected for each spread number (see Figure \ref{fig:bestDistnAdvanced_PinballLossBased}) using rolling window technique, and compare the results to the performance of models obtained with a Normal two-parameter benchmark using Diebold-Mariano test. 
	The estimation horizon (rolling-window size, $T_0$) comprises of 80\% of the data ($1534$ observations), which is moved up 1 time step at a time.
	For each spread number $s=1,...,276$ we \textit{re-specify} and \textit{re-estimate} the model over rolling window time frame $T_0$ using the chosen distribution $D^{(s)}$ and create day-ahead forecasts.
	
	The rolling window forecast analysis is performed over the unseen data points $t=1535,...,1917$ and depicted in Figure \ref{fig:rollingWindow}, with steps outlined in Algorithm \ref{algo:test_learningModels}.
	For each spread $s=1,...,276$ a model is specified and estimated over a fixed length horizon of $T_0=1534$ observations starting at time step $t_1$ (green coloured bracket).
	The equation specification for each distribution parameter is obtained using iterative improvements, where upon each iteration the most insignificant variables are deleted one-by-one (except intercept) until all variables in each equation are significant at 5\% level.
	Following this a 1-step ahead forecast, f$_1$ (green colour), is made using the estimated model and a full predictive density is obtained. 
	The window is moved by 1 time step to $t_2$ (blue colour) and the procedure is repeated.
	The rolling window analysis is continued until all forecasts for the current spread are created, comprising 383 data points.
	The procedure was parallelised across 8 cores on an Intel Core i7 2.9 GHz processor and took 216 hours to complete.
	
	The rolling window analysis of Algorithm \ref{algo:test_learningModels} was repeated using Normal distribution for all $D^{(s)}$ resulting in benchmark forecasts. 
	We note that the estimation of models was not always convergent within the GAMLSS framework and certain spreads failed to be fitted with our models, with Normal distribution based models or both, which was mostly due to the dataset.
	For example, the spread obtained between hours 02-03 is dominated by noise, with $\frac{1}{3}$ of the data set comprising of $0$'s, which creates a problem during weight update calculations.

	\begin{algorithm} 
		\caption{Rolling Window Analysis}
		\label{algo:test_learningModels}
		\begin{algorithmic}[1]
			\For{each spread number $s=1,...,276$}
				\State Use the best chosen distribution $D^{(s)}$ (see Figure \ref{fig:bestDistnAdvanced_PinballLossBased})
				\For {forecast number $t=1,...,383$}
					\State Extract training data design matrix $\mathbf{X}^{(s)}_{train} \in \mathbb{R}^{T_0 \times 9}$, where $T_0=[t,t+1534-1]$
					\State Initialise iteration number $j \gets 0$
					\State Initialise model $\widehat{M}^{(s,t)}_j$ $\gets \{ \widehat{\boldsymbol{\beta}}^{(s,t)}_{k,j} \} _{k=1}^4 \in \mathbb{R}^{J_k+1 \times 4}$ using RS algorithm
					\While{any $\boldsymbol{\widehat{\beta}}^{(s,t)}_{k,j}$ insignificant at 5\%} 
						\State $j \gets j + 1$
						\State Find most insignificant coeff. ${\widehat{\beta}^\dagger}$ (except intercept) across all $k$ 
						\State Remove exogenous variable associated with ${\widehat{\beta}^\dagger}$ from eq. for $k$ and from $\mathbf{X}^{(s)}_{train}$
						\State Re-estimate model $\widehat{M}^{(s,t)}_j \gets \{ \boldsymbol{\widehat{\beta}}^{(s,t)}_{k,j} \} _{k=1}^4$ using RS algorithm
					\EndWhile
					\State Using model $\widehat{M}^{(s,t)}$ create 1-step ahead forecast resulting in $\boldsymbol{\theta}_t^{(s)} = [\mu^{(s)}_t, \sigma^{(s)}_t, \nu^{(s)}_t, \tau^{(s)}_t]^T$
					\State Calculate forecasted expected value $E(\widehat{Y}_t)$ for time step $T_0+1$ using Eq. for $D^{s}$ and $\boldsymbol{\theta}_t^{(s)}$
				\EndFor
			\EndFor
		\end{algorithmic}
	\end{algorithm}

	\subsection{Analysis of Results}
	The results are analysed using two stages: (1) obtaining the difference between Pinball Loss performance measures of the forecasts made with our models and those made by the benchmark; and (2) performing an official statistical significance (Diebold-Mariano) test on whether the difference between two Pinball Loss performance measures is statistically significant.
	
	First we obtain the PL performance measures of our models and benchmark using Algorithm \ref{algo2:bestModelVal2}, where for the benchmark $D^{(s)}$ we use the Normal distribution for all spreads.
	\begin{algorithm} 
		\caption{Model Evaluation using Pinball Loss Function}
		\label{algo2:bestModelVal2}
		\begin{algorithmic}[1]
			\For{each spread number $s=1:276$}
				\State Extract parameters forecasted over $t=1535,...,1917$ using $D^{(s)}$, $\boldsymbol{\theta}^{(s)} \in \mathbb{R}^{383 \times 4}$
				\For{forecasted time step $t=1,...,383$} 
					\State Obtain vector of quantiles $\mathbf{q}_t^{(s)} \in \mathbb{R}^{J_a}$ using $\boldsymbol{\theta}_t^{(s)}$
					\For { each quantile $q_{a,t}$ }
						\State Calculate PL value $L^{(s)}_t(q_a,y_t)$ using Eq. \ref{eq:pinball_loss_value}
					\EndFor
					\State Calculate PL Score $\bar{L}_t^{(s)}(q_a,y_t)$ using Eq. \ref{eq:pinball_loss_score}
				\EndFor
				\State Calculate PL Performance Measure ${\cal{L}}^{(s)}$ using Eq.  \ref{eq:pinball_loss_performance_measure}
			\EndFor
		\end{algorithmic}
	\end{algorithm}
	
	If for a given spread number $s$ convergence issues were experienced during analysis of our models, the forecast for that time step was omitted (and accounted for in PL average calculations).
	Number of times at least 1 forecast was missing for skew type distribution was 46 times (at most 70 time steps out of 383).
	If more than 200 out of 383 forecasted time steps were missing, the result was judged as unreliable and forecasts for that spread were treated as unavailable; this happened 4 out of 276 times for spreads between hours: 02-03/12, 03-05, 13-14. 
	The convergence issue was also experienced for the Normal benchmark, and the number of times at least 1 forecast was missing was 5 times (at most 42 time steps out of 383).
	The number of times forecasts were missing for more than 200 time steps were observed more frequently, which happened 13 out of 276 times for spreads between hours: 01-12/13/14/23;  02-03/23; 05-13; 06-12/13/14/15/16; 07-08.
	When this happened we judged our model to have a superior result for that spread number $s$.
	
	\begin{figure}[h!]
		\centering
		\includegraphics[width=1.0\textwidth]{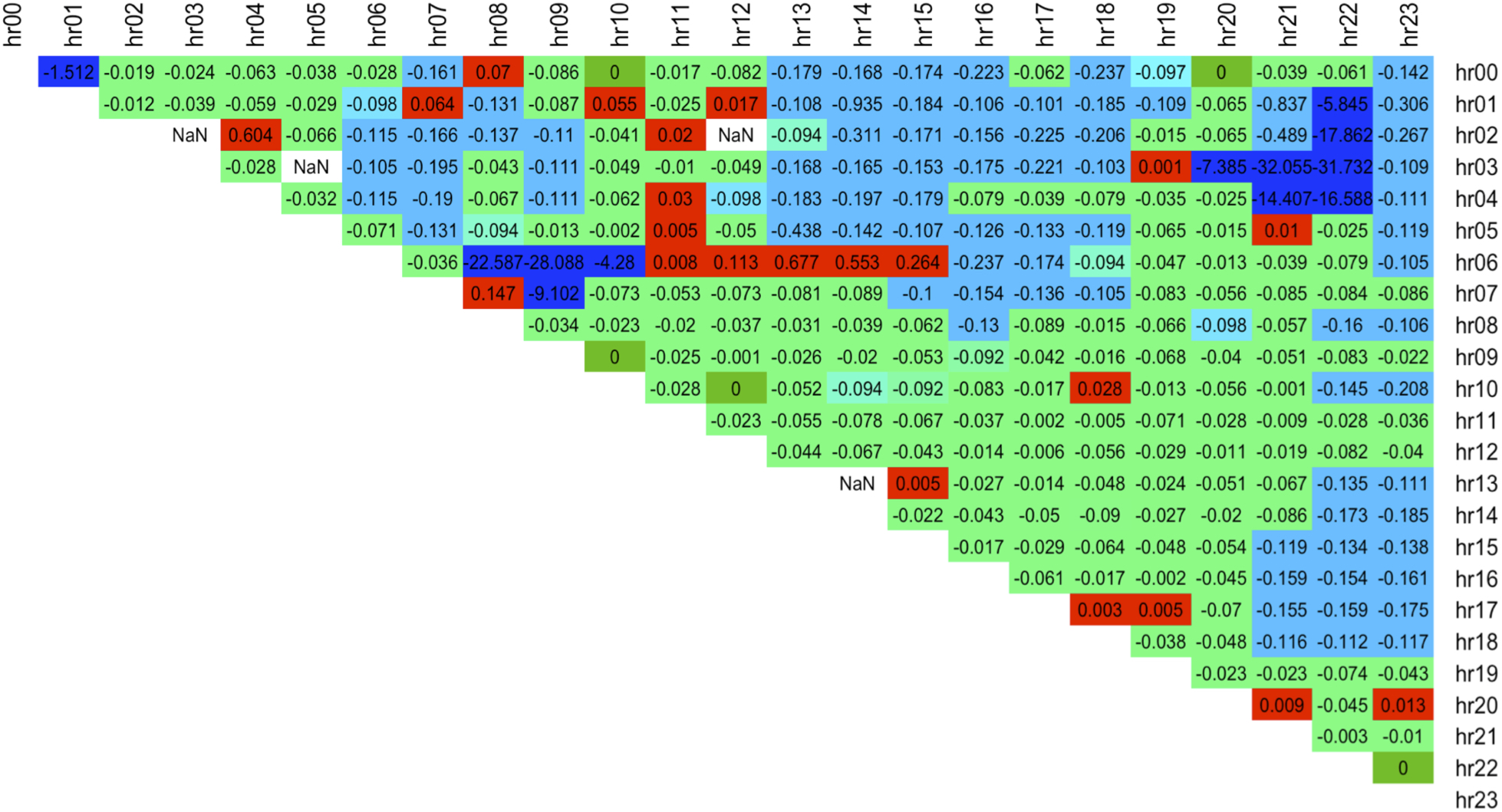}
		\caption{Forecasting power of skew vs Normal distributions - difference of PL scores.}
		\label{fig:forecastingResults_diffPL}
	\end{figure}
	
	We report the difference in PL performance measures between those obtained with our models and the benchmark for each spread number $s$ (see Figure \ref{fig:forecastingResults_diffPL}, where negative values (blue, green) indicate that our model outperformed the benchmark obtained with the Normal distribution for that spread number $s$).
	Initially results are obtained to 5 d.p. and these show that for 276 spreads our models forecasted the full density more accurately: 258 times vs 18 times for the Normal distribution.
	Reducing decimal places to 3, results in the same number of Normal distributions having smaller error values (18 times), however now results show that for 5 spreads the Normal distribution produces as good a fit as the distributions used in our models, while for 253 cases our models produce more accurate forecasts.
	The overwhelming majority preference for models obtained with skew type distributions is evident through negative values, as expected due to highly skewed and kurtotic nature of spread price data.
	Next, we seek to establish a formal significance test on the difference between the PL performance measures between our models and benchmark.

	\subsubsection{Diebold-Mariano Test}
	In order to draw statistically significant conclusions over the outperformance of the best selected distributions when forecasting unseen data points over the accuracy of forecasts made with the Normal distribution, we use the Diebold-Mariano (DB) test \citep{diebold2002comparing}.
	The test is applicable to forecast errors that do not have zero mean, that are not Gaussian, and that may be serially / contemporaneously correlated.
	
	We use a variation of the standard DB test with implementation proposed by \citep{harvey1997testing}.
	For each spread number $s$, there are $t=1,...,383$  forecasts produced by two models ($\widehat{M}^{(s,t)}_1$ - estimated with the best chosen distribution and $\widehat{M}^{(s,t)}_2$ - estimated with Normal distribution), which are tested against each other using a one-sided test at 5\% significance level.
	The null hypothesis is the that the two models have the same forecast accuracy, with a one-sided alternative hypothesis that forecasting power of the best distribution outperforms that of the Normal, $H_0 : E(\Delta_{M_1,M_2,t,s}) \leq 0$, where the Loss Differential Series $\Delta_{M_1,M_2,t,s}$ is
	\begin{equation}
	\Delta_{M_1,M_2,t,s} = |\bar{L}_t^{(s,1)}| - |\bar{L}_t^{(s,2)}|
	\end{equation}
	where
	$s$ is the spread number,
	$t$ is the forecast time step,
	$\bar{L}_t^{(s,1)}$ average quantile score of model $\widehat{M}^{(s,t)}_1$ obtained with the best chosen distribution at forecast step $t$,
	$\bar{L}_t^{(s,2)}$ average quantile score of model $\widehat{M}^{(s,t)}_2$ obtained with Normal distribution at forecast step $t$. 
	
\begin{figure}[h!]
		\centering
		\includegraphics[width=1.0\textwidth]{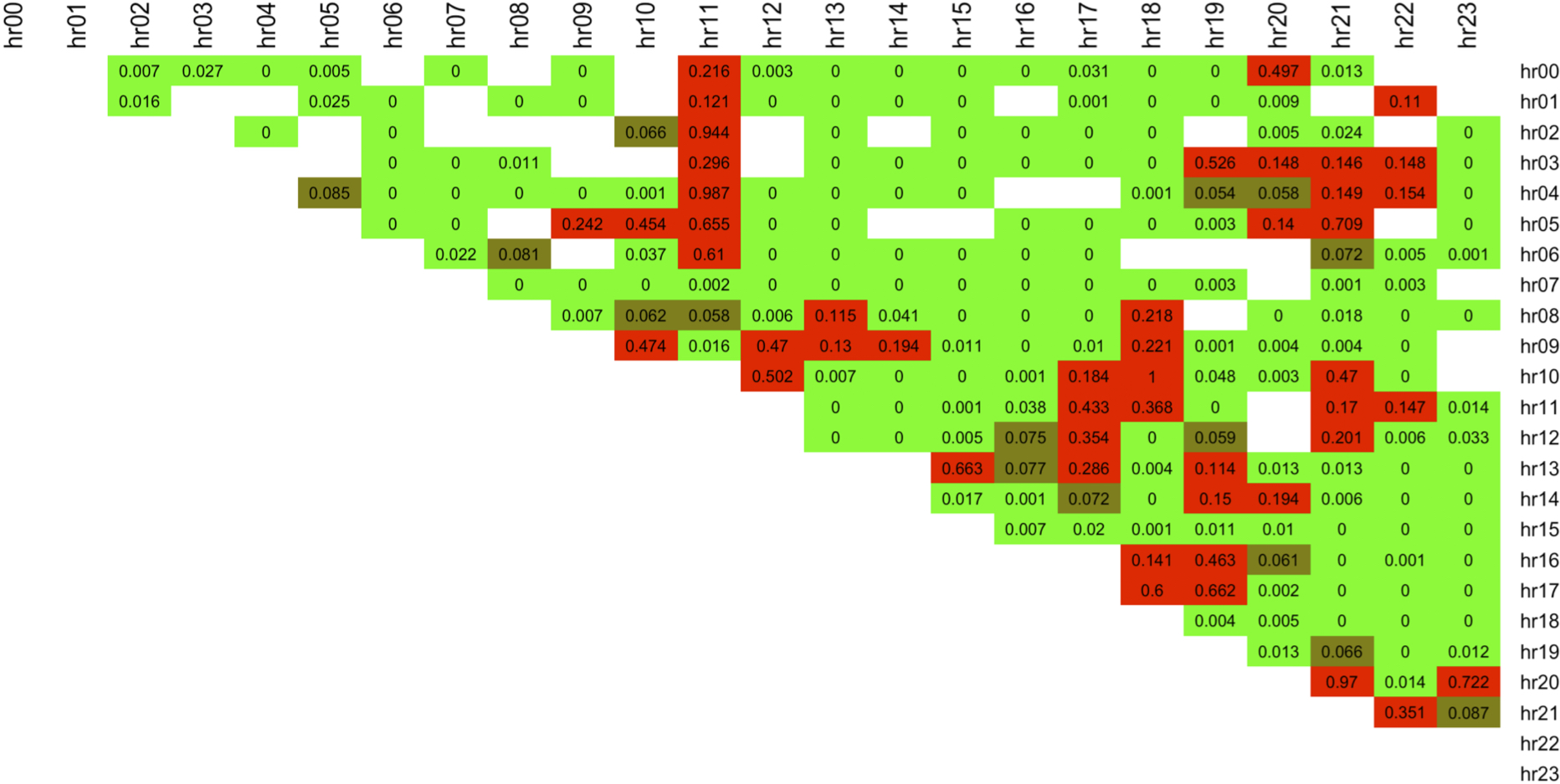}
		\caption{Diebold-Mariano test p-values (green at 5\% significance, olive at 10\% significance, red insignificant).}
		\label{fig:forecastingResults_DM}
	\end{figure}

	The p-values are displayed in Figure \ref{fig:forecastingResults_DM} and the results show that out of 276 spreads:
	\textbf{(a)} the best chosen distributions are significantly better at forecasting the spreads: 161 times at 5\% (bright green) and 15 times at 10\% (olive green).
	Note: the distributions which were used to learn these 176 models are: JSU - 9 times, SEP1 - 14 times, SEP2 - 16 times, ST1 - 13 times, ST2 - 18 times and ST5 - 106 times ($\frac{106}{164}*100 = 64.5\%$).
	\textbf{(b)} the models have the same forecasting power for 48 spreads (i.e. the null hypothesis could not be rejected at 10\%).
	Note: the distributions which were used to learn the corresponding 48 models are: JSU - 9 times, SEP1 - 4 times, SEP2 - 8 times, ST1 - 8 times, ST2 - 4 times and ST5 - 15  times ($\frac{15}{48}*100 = 31.25\%$).
	\textbf{(c)} the results could not be obtained for the models obtained with best chosen distributions 49 times (white spaces of upper triangular) due to at least 1 forecast missing due to Quantile estimate convergence issues.
	Note: the distributions which were used to learn these 49 models are: JSU - 4 times, SEP1 - 8 times, SEP2 - 21 times, ST1 - 9 times, ST2 - 5 times and ST5 - 2  times ($\frac{2}{49}*100 = 4\%$) (i.e. this typically happened for distrubitons other than ST5).
	
	We focused on selecting suitable four parameter distributions which fitted best the individual spread data and performed detailed analysis of the forecasting power of such distributions.
	As the result of the above analysis it was found that:
	\textbf{(a)} the ST5 distribution was selected most frequently as the best distribution by simple distribution fit based on $\mathbf{y} \sim \mathbf{1}$ and factor-based distribution fit based on both RMSE and PL functions;
	\textbf{(b)} the ST5 distribution was the most reliable for convergence of quantile estimates using GAMLSS function \texttt{qFUN} (e.g. \texttt{qST5}), especially for the extreme quantiles of $q_1, q_2, q_3, q_{97}, q_{98}, q_{99}$ where other distributions such as SEP1, SEP2 often failed.
	Quantile estimates are important for our research because they are utilised in statistical testing for comparing performance of two models and in the Value-At-Risk calculations used in trading strategy optimisation.
	For example, the PL calculations revealed that for 49 spreads at least one forecast step (out of 383 steps) failed to extract 95 quantiles from the estimated model.
	Upon further examination it was revealed that out of the 49 occurrences only 4\% had ST5 as the underlying distribution for which the  model was estimated.
	This supports our claim that ST5 is a reliable distribution for the quantile estimates;
	\textbf{(c)} the best distribution for each spread was chosen based on the PL performance measure calculated over the validation data.
	Each spread had 6 possible distributions from which the best distribution was chosen and the one with the lowest score was taken as the best distribution for that spread.
	Further analysis reveals that on 1/3 of occasions when other distributions than ST5 were selected as 'best', the ST5 was the second best distribution, which on average was only worse by 1.44\%.
	However for the spreads where the ST5 was the best distribution, the PL performance measure was on average better by 4.84\%, which points to the possibility that the best distribution did not have a significantly difference performance when compared to ST5.
	Therefore we conclude that if one wishes to use a single distribution across all spreads, the ST5 distribution forms a robust choice. 
	We continue our analysis using a more detailed approach, where individual spreads have established distributions of best fit assigned to them.

	\subsubsection{Illustrative Examples of Estimated Models}
\begin{figure}[h!]
		\centering
		\includegraphics[width=0.85\textwidth]{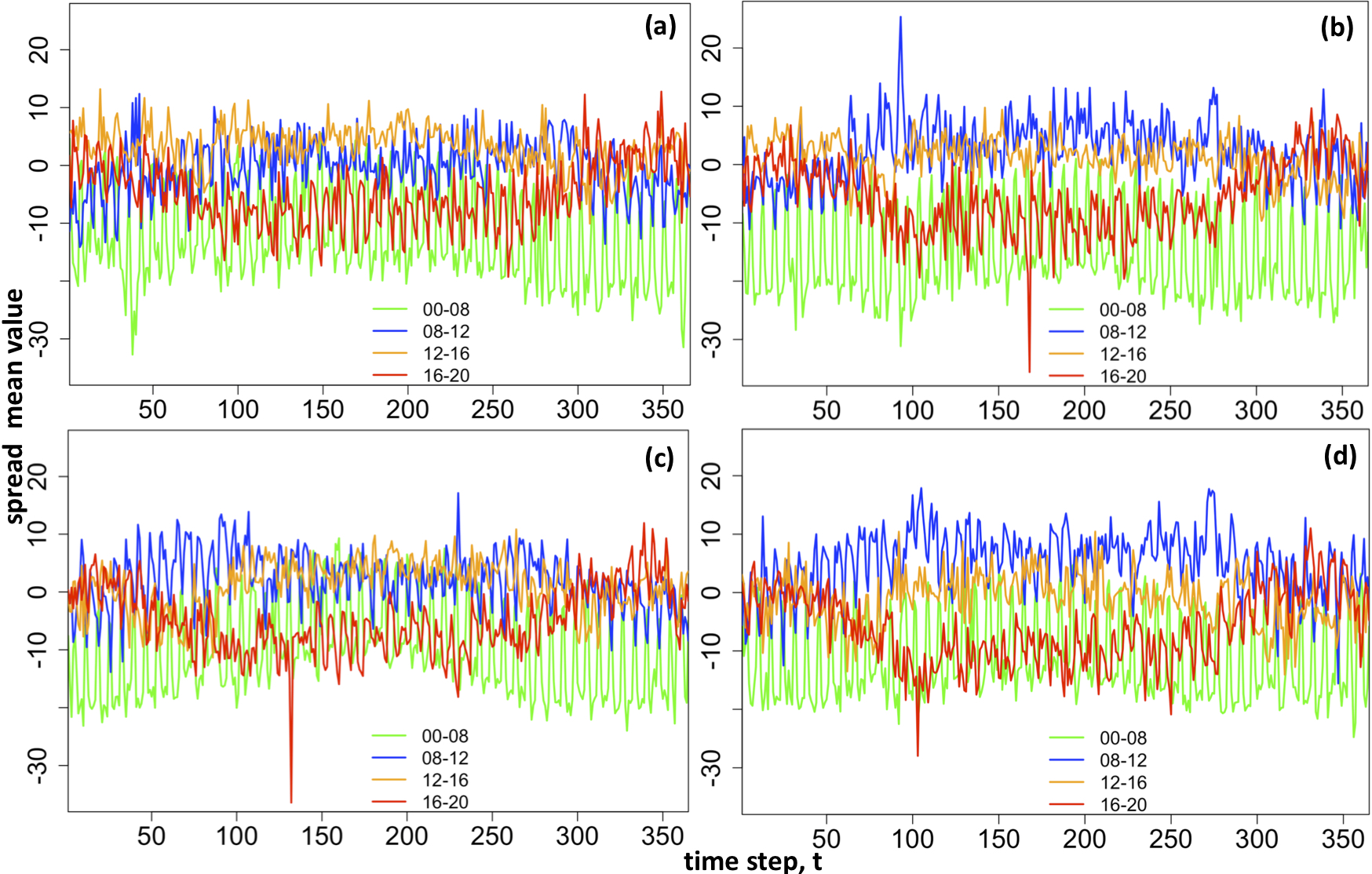}
		\caption{Evolution of fitted $\hat{\boldsymbol{\mu}}$ of the best dist, years (a) 2012, (b) 2013, (c) 2014, (d) 2015.}
		\label{fig:fittedMU}
	\end{figure}

\begin{figure}[h!]
		\centering
		\includegraphics[width=0.9\textwidth]{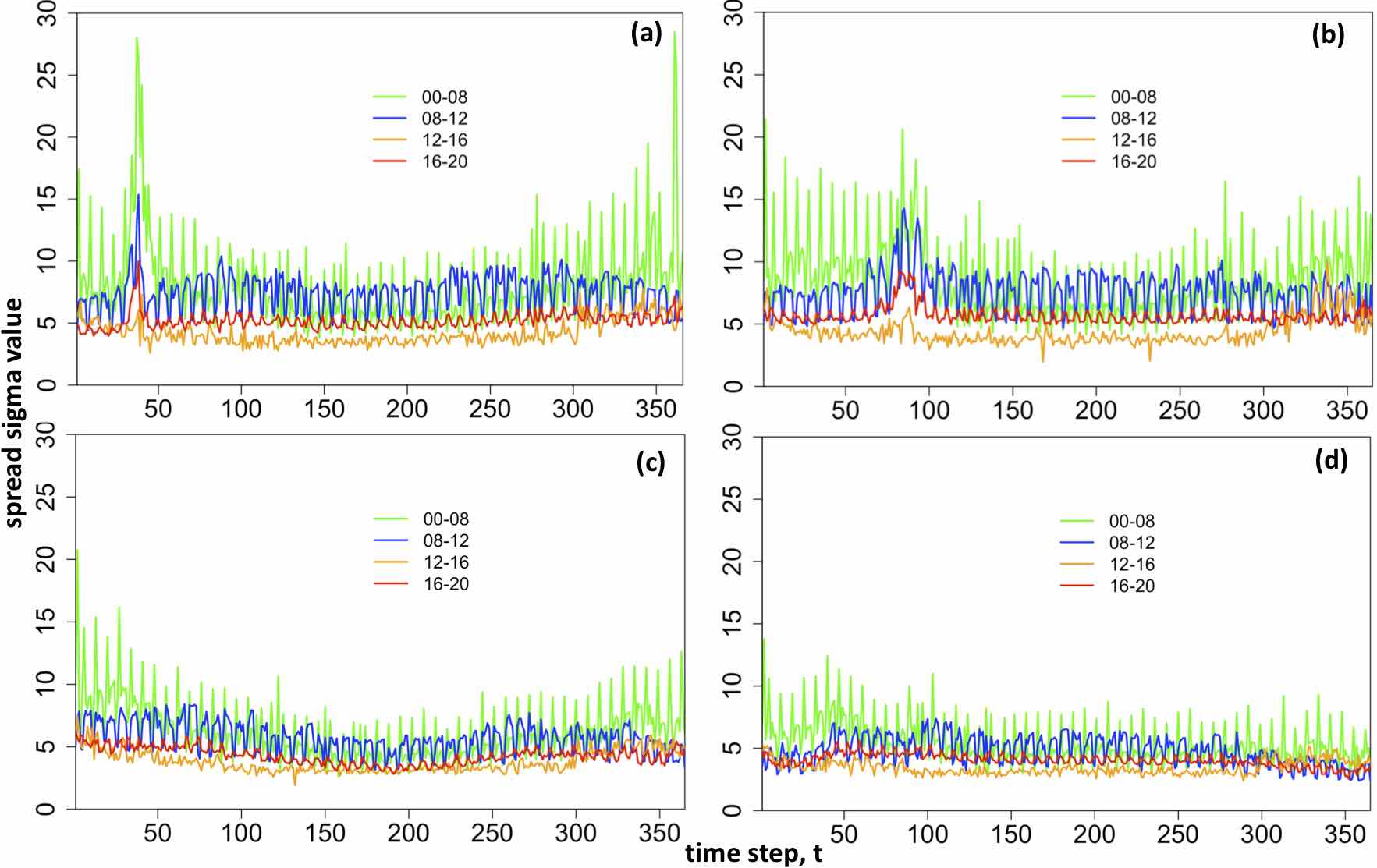}
		\caption{Evolution of fitted $\hat{\boldsymbol{\sigma}}$ of the best dist, years: (a) 2012, (b) 2013, (c) 2014, (d) 2015.}
		\label{fig:fittedSIGMA}
	\end{figure}

\begin{figure}[h!]
		\centering
		\includegraphics[width=0.9\textwidth]{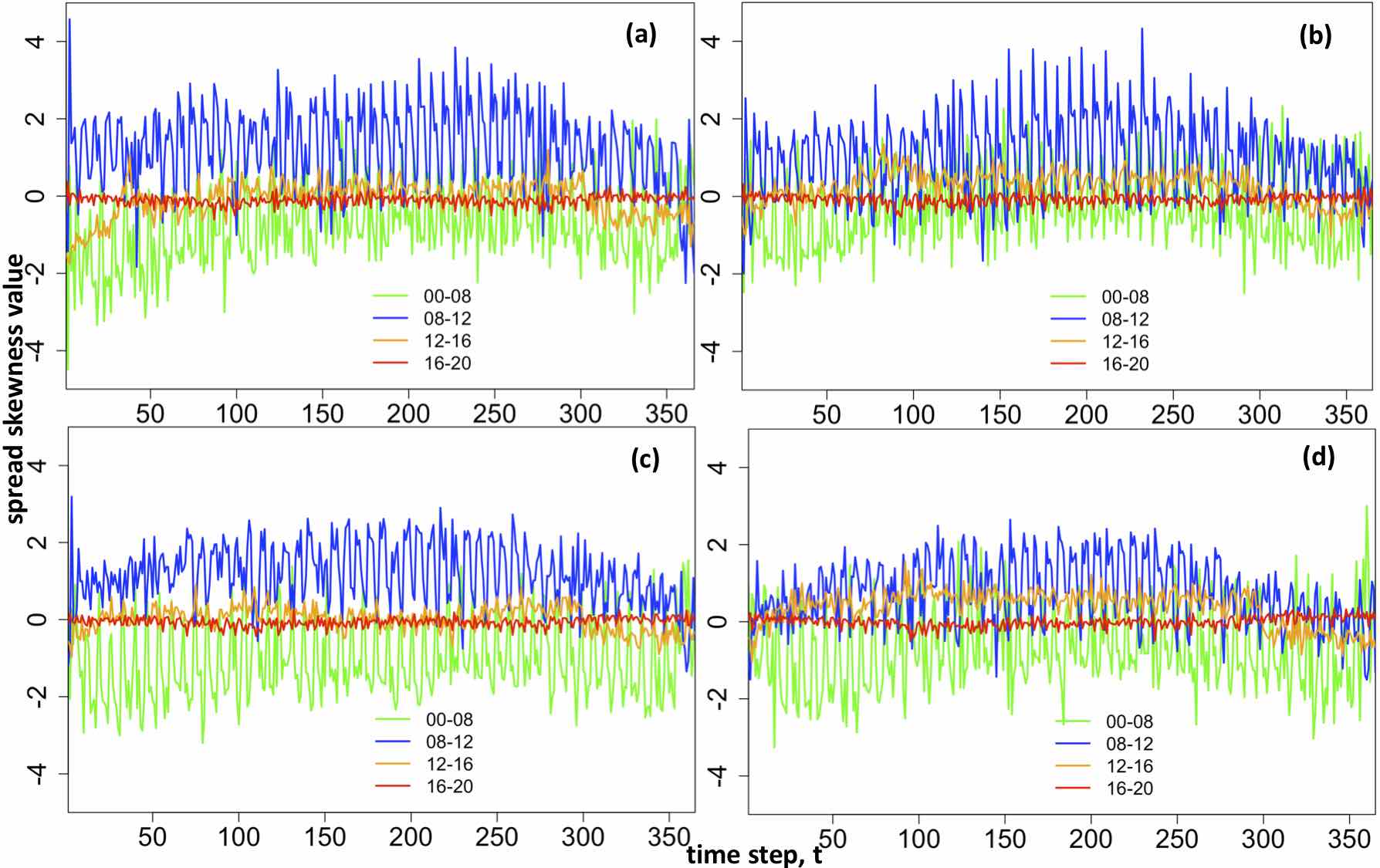}
		\caption{Evolution of fitted $\hat{\boldsymbol{\nu}}$ of the best dist, years: (a) 2012, (b) 2013, (c) 2014, (d) 2015.}
		\label{fig:fittedNU}
	\end{figure}

\begin{figure}[h!]
		\centering
		\includegraphics[width=0.9\textwidth]{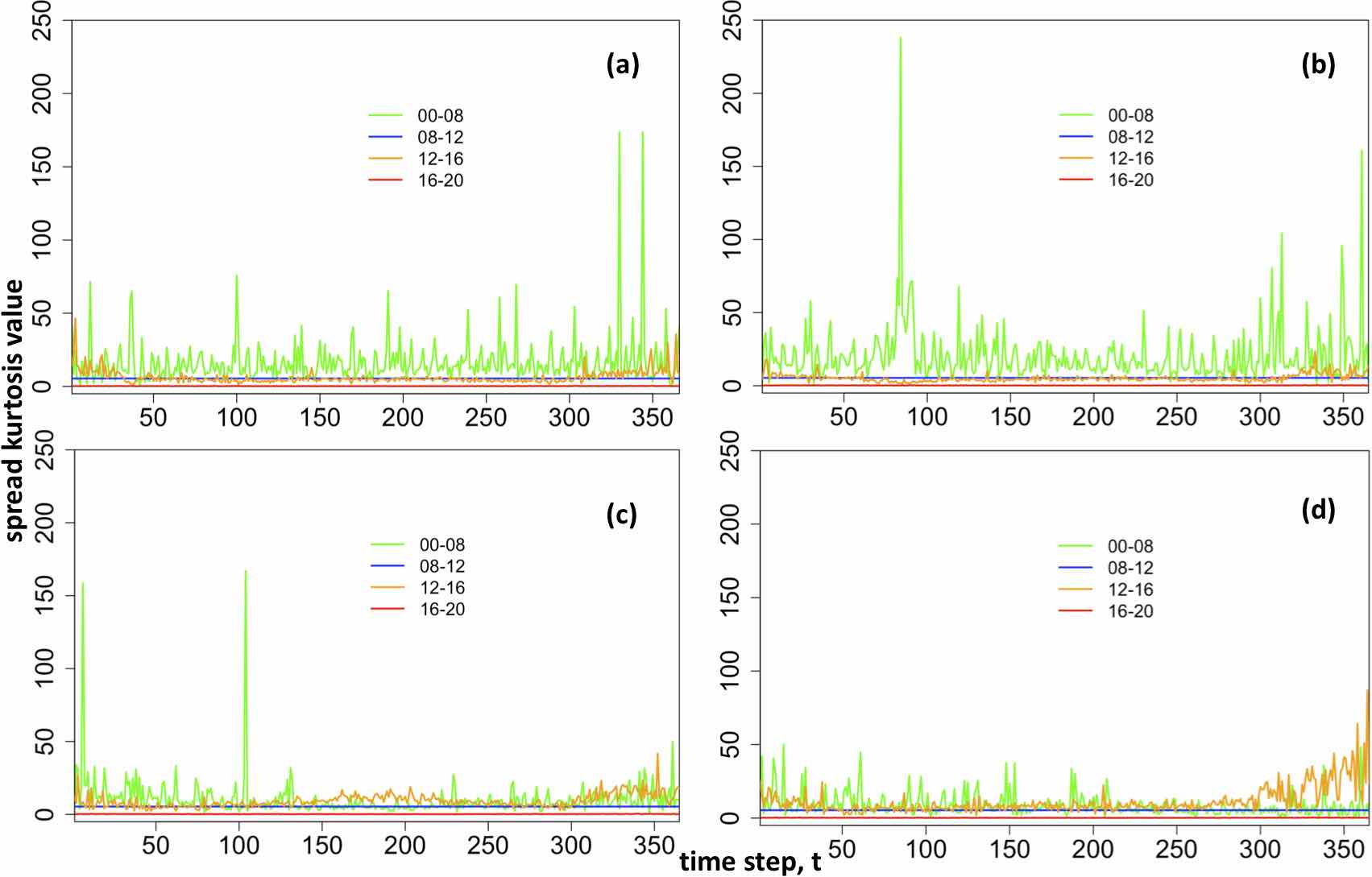}
		\caption{Evolution of fitted $\hat{\boldsymbol{\tau}}$ of the best dist, years: (a) 2012, (b) 2013, (c) 2014, (d) 2015.}
		\label{fig:fittedTAU}
	\end{figure}

	In order to validate the need for dynamic modelling, we display the evolution of \textbf{fitted} moments for four example spreads obtained for the first rolling window (i.e. first 1534 data points used for specification and estimation phase), with their associated distributions: 00-08 (ST1); 08-12 (ST1); 12-16 (ST1); 16-20 (ST5).
	\\
	\textbf{Evolution of latent 4 central \textit{moments} throughout four years} i.e. examining how selected 4 spreads behave throughout each year.
	We selected four years: 2012, 2013, 2014, 2015 in order to depict the evolution and the changing dynamics of the moments with time. 
	The evolution of each distribution parameter is plotted on separate graphs (see Figure \ref{fig:fittedMU} for the evolution of $\hat{\boldsymbol{\mu}}$, Figure \ref{fig:fittedSIGMA} for evolution of $\hat{\boldsymbol{\sigma}}$, Figure \ref{fig:fittedNU} for the evolution of $\hat{\boldsymbol{\nu}}$, Figure \ref{fig:fittedTAU} for evolution of $\hat{\boldsymbol{\tau}}$).
	The results show that the mean is fitted in line with what would be expected for the true spread price, where the spreads between 08-12 hours tend to be positive (i.e. later hour is cheaper), while the 16-20 spreads are negative, i.e. later hour is more expensive. 
	The standard deviation is highest for the spreads between night-time (less busy) and early morning / afternoon (green and blue lines).
	While the skewness tends to be positive for the 08-12 hours, and negative for the 00-08 hours.

\begin{figure}[h!]
		\centering
		\includegraphics[width=0.85\textwidth]{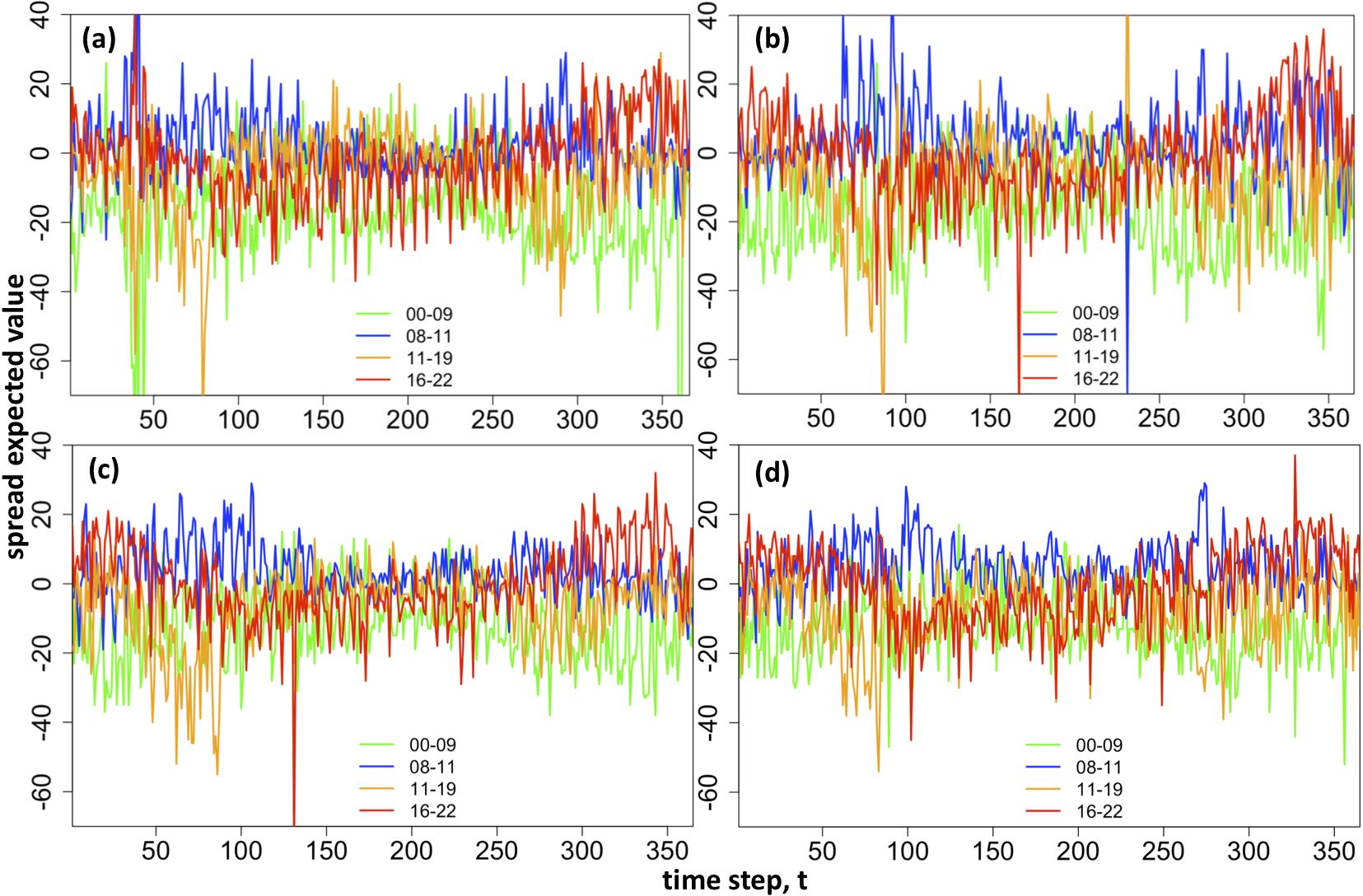}
		\caption{Evolution of \textbf{true} spread $E(\mathbf{Y})$, years: (a) 2012 (b) 2013 (c) 2014 (d) 2015.}
		\label{fig:trueEY}
	\end{figure}

\begin{figure}[h!]
		\centering
		\includegraphics[width=0.85\textwidth]{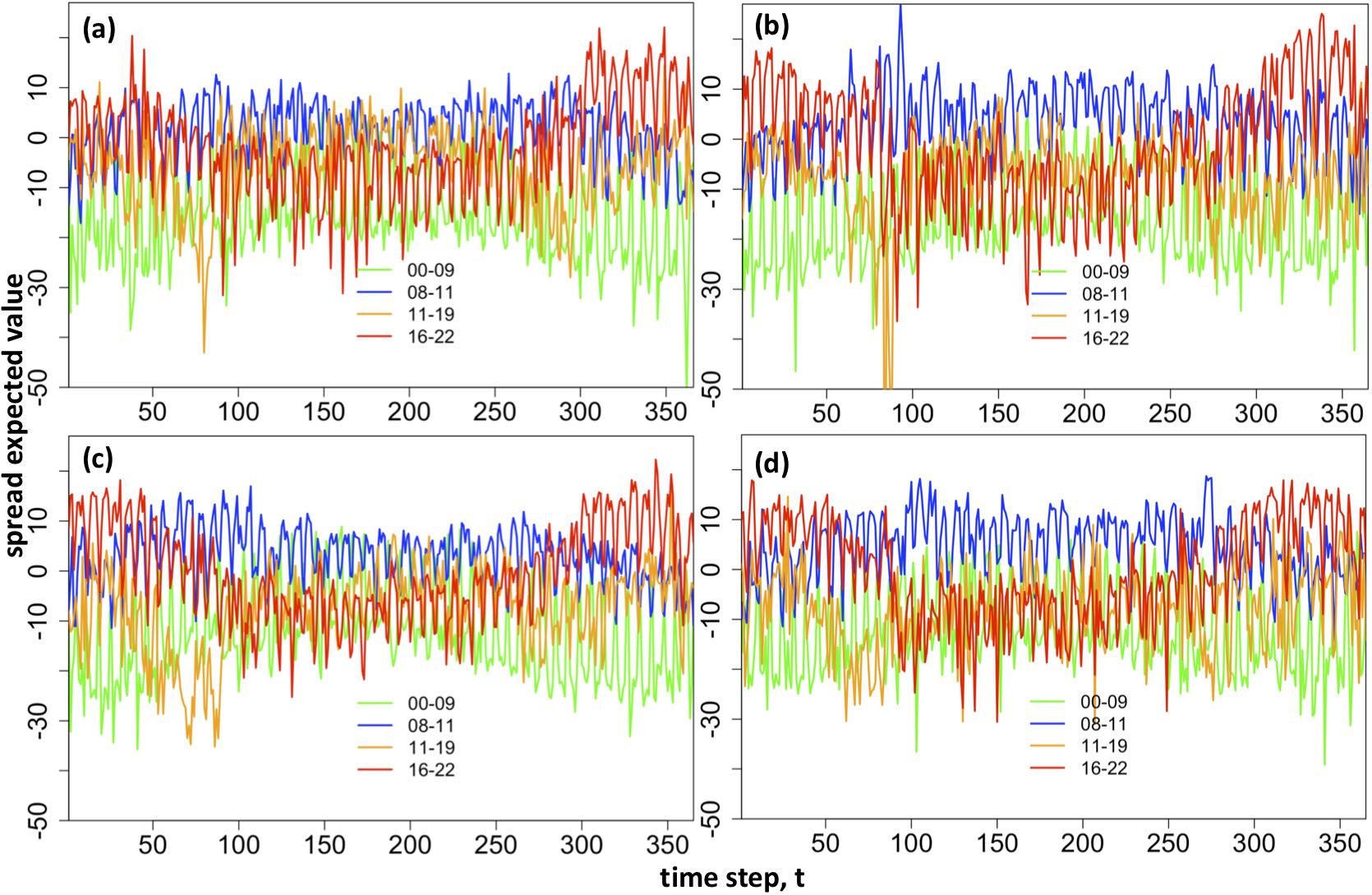}
		\caption{Evolution of \textbf{fitted} spread $E(\widehat{\mathbf{Y}})$, years: (a) 2012 (b) 2013 (c) 2014 (d) 2015.}
		\label{fig:fittedEY}
	\end{figure}

	\newpage
	Next, we plot the true $E(\mathbf{Y})$, vs fitted $E(\widehat{\mathbf{Y}})$, expected values of spreads over the four years, selecting different spread hours to the ones used above, in orer to show a variety of underlying distributions used: 00-09 (SEP2), 08-11 (SEP1), 11-19 (ST2), 16-22 (ST5). 
	Figure \ref{fig:trueEY} shows the true evolution of the spreads, compared to the expected values produced by the estimated models (see Figure \ref{fig:fittedEY}), note slightly smaller scale.
	The fitted expected values follow the true pattern throughout each year, for example the 16-22 hour spread tends to have more negative values in the summer time (i.e. electricity at earlier hour is less expensive) and positive values in the winter time (i.e. electricity at earlier hour is more expensive).
	It can be seen that the fitted spread values are slightly under-fitted as indicated by the difference in plot scale. 

\begin{figure}[h!]
		\centering
		\includegraphics[width=0.75\textwidth]{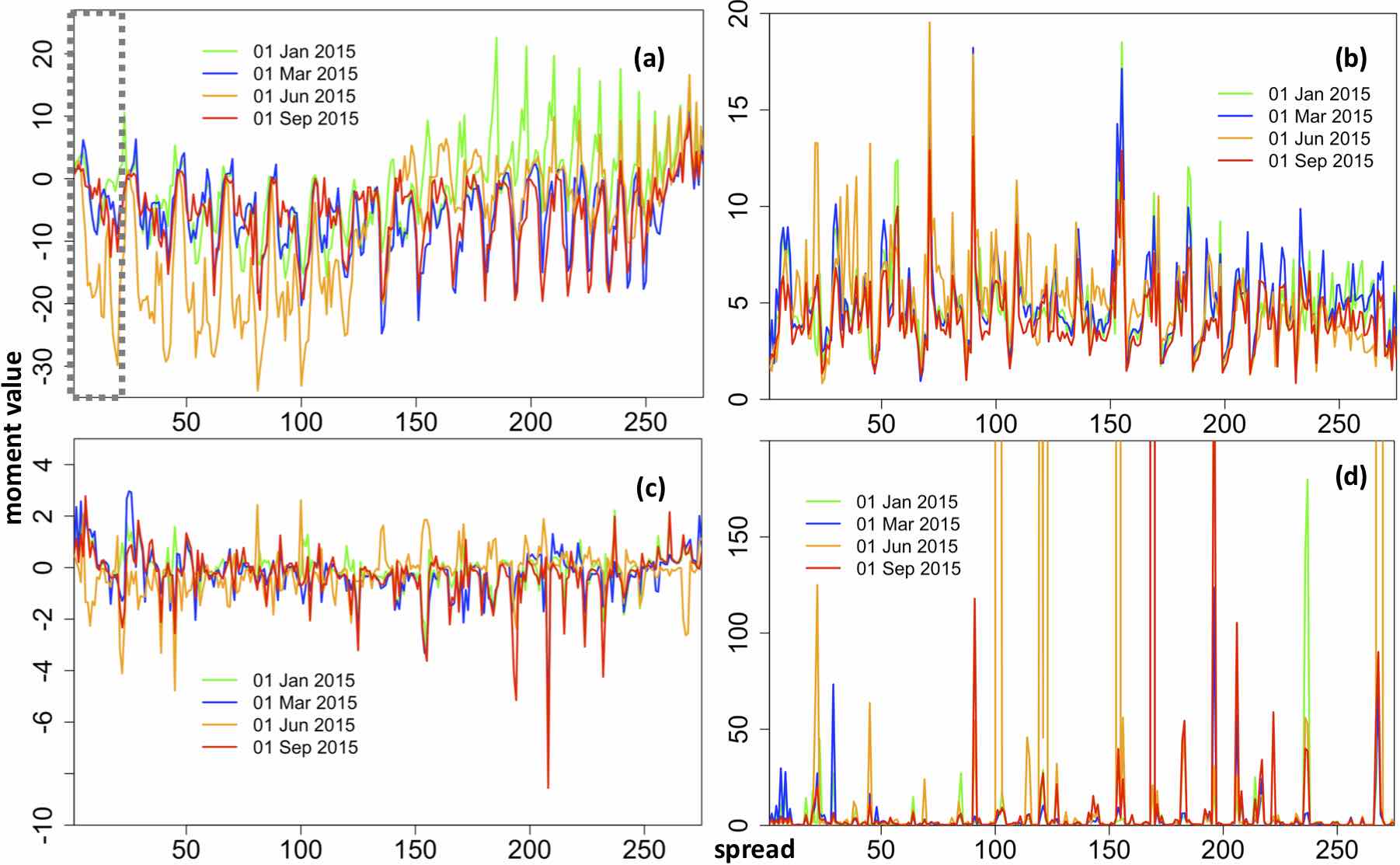}
		\caption{Evolution of fitted central moments throughout the day plotted for 4 days of year 2015 (a) mean (grey box shows midnight spread with all other hours) (b) volatility (c) skewness and (d) kurtosis.}
		\label{fig:4moments_4days2015}
	\end{figure}

	\newpage
	\textbf{Evolution of the 4 latent central \textit{moments} throughout a day} demonstrates how distribution parameters change throughout a day and throughout different times of the year.
	We plot the four parameters for 276 spreads on 01 Jan 2015, 01 Mar 2015, 01 June 2015 and 01 September 2015 i.e. one plot per day of the season.
	The spreads were plotted sequentially starting with 23 spreads for midnight hour with all other hours of the day, $00-01, 00-02,...,00-23$, followed by 22 spreads of hour 01 with all other hours, continuing on until the last spread between hours 22 and 23.
	A total of \textit{276} spreads are plotted for each of the four selected days of year 2015.
	The results show periodic spikes due to points where the spread moves between hour 00 with all other, hour 01 with all other etc.
	We show the changing dynamics of the distribution parameters throughout seasons of the year (see Figure \ref{fig:4moments_4days2015}).

\begin{figure}[h!]
		\centering
		\includegraphics[width=0.77\textwidth]{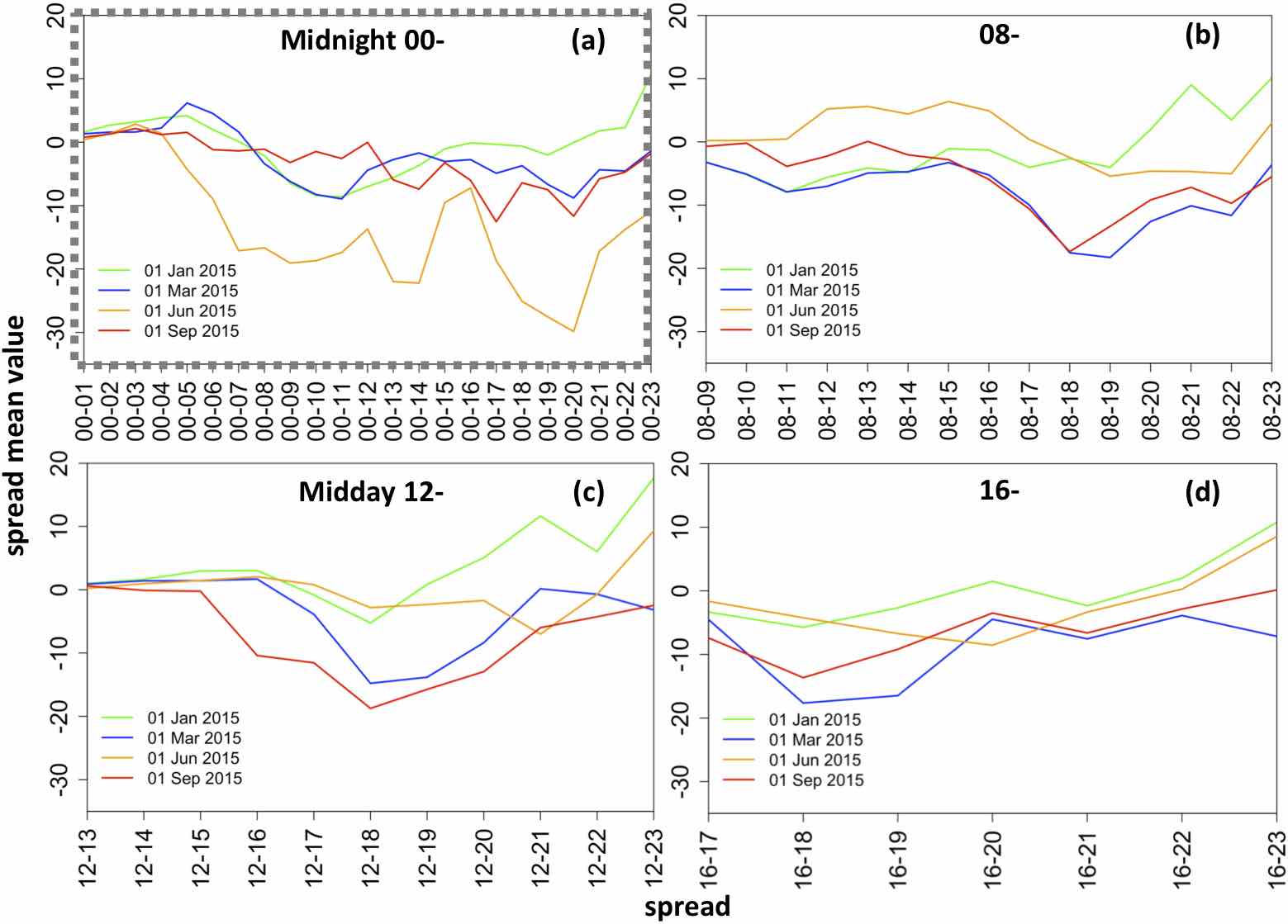}
		\caption{Evolution of fitted $\hat{\mu}$ throughout the day plotted for 4 chosen hrs, year 2015. Plots show hours spread with all other hours of the day that follow it (a) midnight (b) 08 (c) 12 and (d) 16.}
		\label{fig:mu4Days4hrs2015}
	\end{figure}

	Plots detailing the same intra-day information but only for spreads of 4 chosen example hours of the day with all other hours (00-; 08-; 12-; 16-) are given below for clarity.
	Effectively, for example, the Figure showing the mean for different days of the year 2015 (see Figure \ref{fig:mu4Days4hrs2015}) has sub-plots (\textbf{a}-\textbf{d}) which demonstrate zoomed in sections of Figure \ref{fig:4moments_4days2015} depicted by the grey dotted box.
	Note: the sub-plots $x$ values range reduces when going from \textbf{(a)} to \textbf{(d)} since midnight hour 00 spreads with all hours of the day, but 16 hour only spreads with 7 other hours.
	\begin{figure}[h!]
		\centering
		\includegraphics[width=0.77\textwidth]{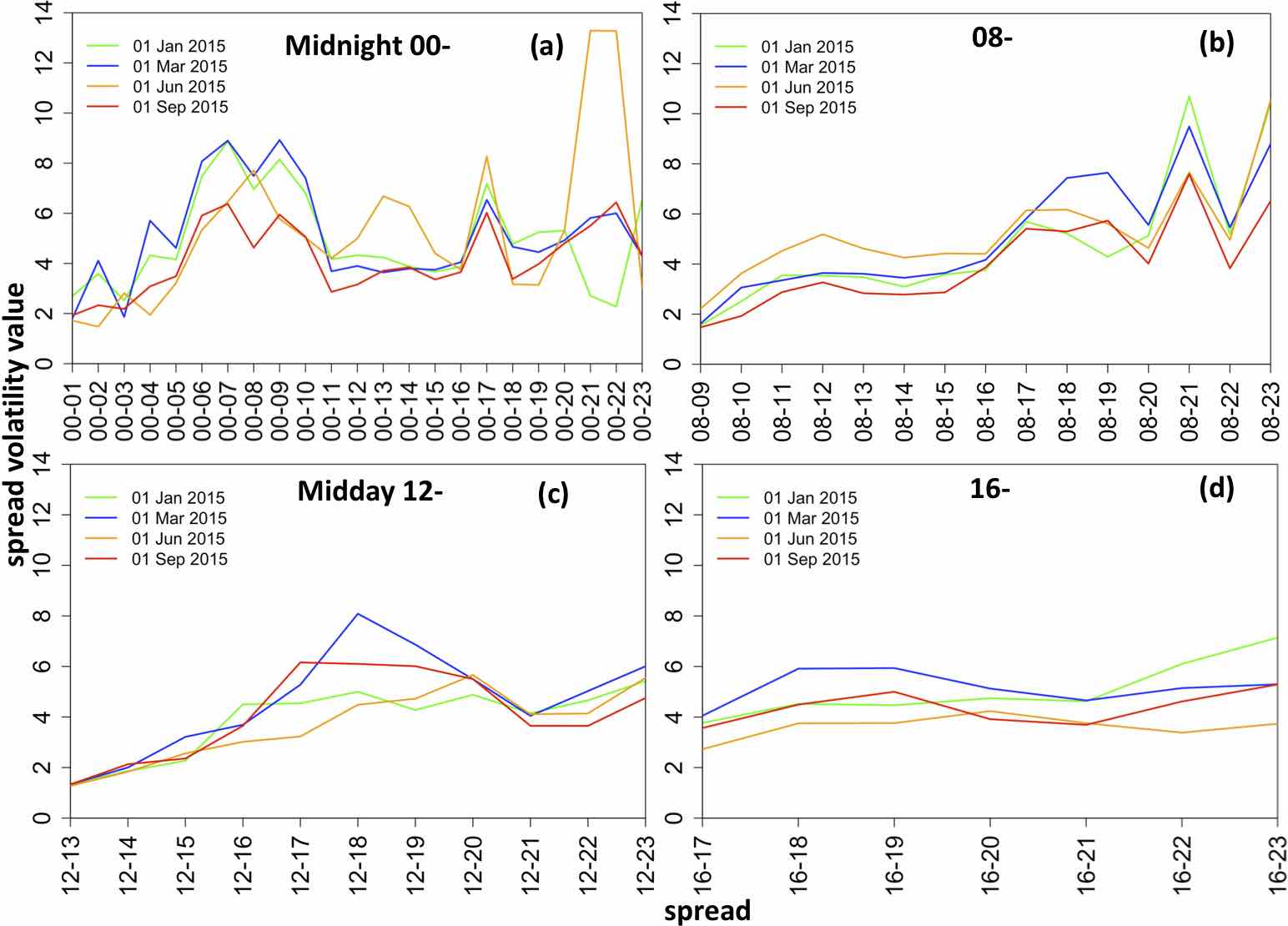}
		\caption{Evolution of fitted $\hat{\sigma}$ throughout the day plotted for 4 chosen hrs, year 2015. Plots show hours spread with all other hours of the day that follow it (a) midnight (b) 08 (c) 12 and (d) 16.}
		\label{fig:sigma4Days4hrs2015}
	\end{figure}
	\begin{figure}[h!]
		\centering
		\includegraphics[width=0.77\textwidth]{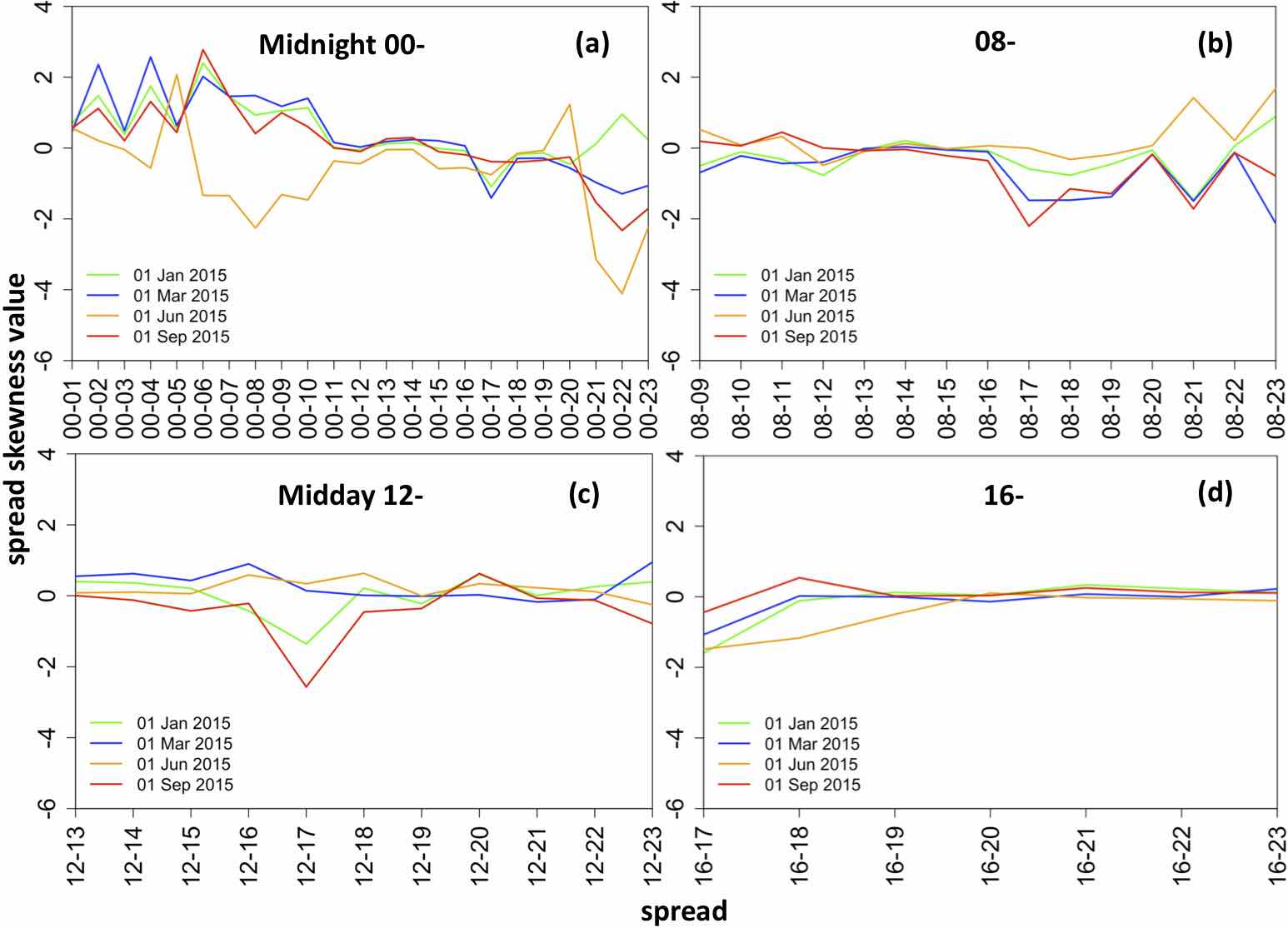}
		\caption{Evolution of fitted $\hat{\nu}$ throughout the day plotted for 4 chosen hrs, year 2015. Plots show hours spread with all other hours of the day that follow it (a) midnight (b) 08 (c) 12 and (d) 16.}
		\label{fig:nu4Days4hrs2015}
	\end{figure}
	\begin{figure}[h!]
		\centering
		\includegraphics[width=0.77\textwidth]{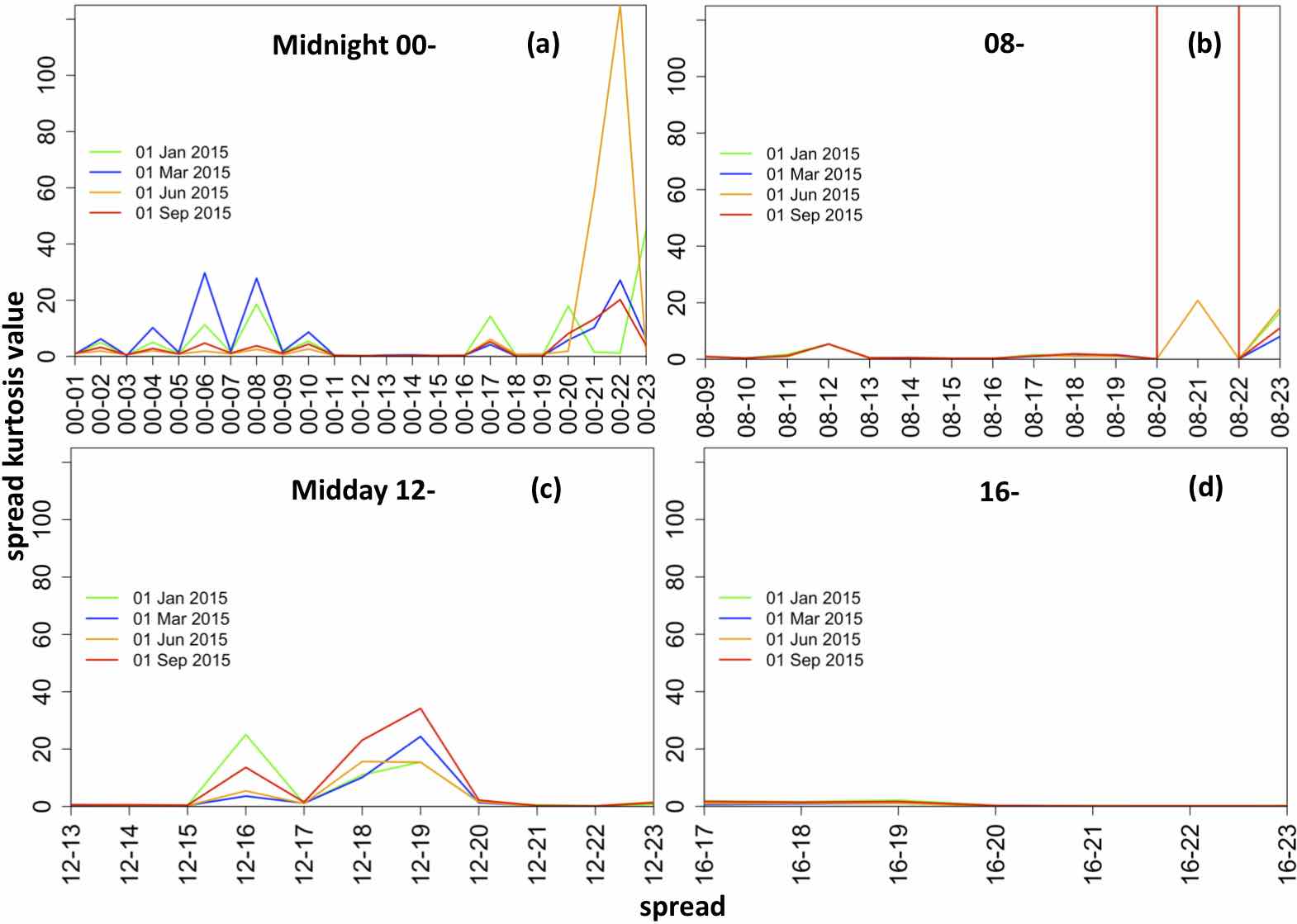}
		\caption{Evolution of fitted $\hat{\tau}$ throughout the day plotted for 4 chosen hrs, year 2015. Plots show hours spread with all other hours of the day that follow it (a) midnight (b) 08 (c) 12 and (d) 16.}
		\label{fig:tau4Days4hrs2015}
	\end{figure}

	\clearpage
	\textbf{Variations in the size and sign of \textit{explanatory variable coefficients} for different spreads of the same day} illustrate the varying impact of drivers (see Tables \ref{tab:coeff_Illustration1} and \ref{tab:coeff_Illustration2} which show the estimated coefficients for selected 4 spreads: 00-08; 08-12; 12-16; 16-20).
	The coefficients were extracted from the model estimated for the first rolling-window frame i.e. $t=1,..., 1534$.
	The displayed values for the coefficients are all significant at 5\% thus forming the equation for that moment.
	Missing values indicate that the independent variable was not significant and thus was omitted from the equation specification for that moment (table column).
	 
	The coefficients correspond to their associated independent variables:
	$\beta_0$ coeff is the intercept; 
	$\beta_1$ coeff for spread of the lagged day-ahead electricity price;
	$\beta_2$ coeff for gas Gaspool forward daily price; 
	$\beta_3$ coeff for coal ARA forward daily price; 
	$\beta_4$ coeff for spread of wind day-ahead forecast;
	$\beta_5$ coeff for spread of solar day-ahead forecast; 
	$\beta_6$ coeff for dummy variable taking value of 1 for weekends/holidays; 
	$\beta_7$ coeff for spread of the day-ahead total load forecast; 
	$\beta_8$ coeff for an interaction load variable.
	
	\begin{table}[h!]
		\centering 
		\begin{tabular}{c|c|c|c|c||c|c|c|c|}
			\cline{2-9}
			& \multicolumn{4}{ c||}{\textbf{Spread: 00-08}}  & \multicolumn{4}{ c| }{\textbf{Spread: 08-12}} \\ \cline{2-9}
			& ${\mu}$  & ${log(\sigma)}$ & ${\nu}$ & ${log(\tau)}$ 
			& ${\mu}$  & ${log(\sigma)}$ & ${\nu}$ & ${log(\tau)}$  \\ \cline{1-9}
			%
			\multicolumn{1}{ |c| }{${\beta_0}$} & -1.485e+1 & 1.725 & 4.561e-1 & 2.154e-1  & 3.112 & 1.256e-1 & -1.049 & 1.682  \\ \hline
			%
			\multicolumn{1}{ |c| }{$s_{t-1}$, ${\beta_1}$ } & 1.220e-1 & -9.593e-3 &  & 5.967e-3 & 2.219e-1 &  & -3.243e-2 &  \\ \hline
			%
			\multicolumn{1}{ |c| }{gas, ${\beta_2}$} & -3.905e-1 & 4.229e-2 &  & 3.171e-2 & 1.578e-1 &  &  &  \\ \hline
			%
			\multicolumn{1}{ |c| }{coal, ${\beta_3}$} & 3.544e-2 & -4.560e-3 &  & -1.173e-2 & -1.353e-1 & 7.057e-3 & 2.194e-2 &  \\ \hline
			%
			\multicolumn{1}{ |c| }{$s_{wind}$, ${\beta_4}$} &  -6.717e-4 &  & -8.320e-5 & -8.139e-5 & -9.562e-4 &  & -6.700e-5 & \\ \hline
			%
			\multicolumn{1}{ |c| }{$s_{solar}$, ${\beta_5}$ } & -7.292e-4 & 2.452e-5 & -4.243e-5 & -5.218e-5 & -6.842e-4 & -2.781e-5 & -1.367e-4 & \\ \hline
			%
			\multicolumn{1}{ |c| }{dummy, ${\beta_6}$} & 1.303e+1 &  & 1.696 & 2.908e-1 & -4.640 & -3.858e-1 & -1.104 &  \\ \hline
			%
			\multicolumn{1}{ |c| }{$s_{load}$, ${\beta_7}$} &  & 3.584e-5 & -2.862e-5 & -1.698e-5 & -1.886e-3 &  & 6.114e-4 & \\ \hline
			%
			\multicolumn{1}{ |c| }{$s_{iLoad}$, ${\beta_8}$} &  &  &  &  & 3.252e-8 &  & -9.848e-9 & \\ \hline
		\end{tabular}
		\caption{Illustration of coefficients for 4 moments for spreads 00-08 and 08-12.} 
		\label{tab:coeff_Illustration1}
	\end{table}

	\begin{table}[h!]
		\centering 
		\begin{tabular}{c|c|c|c|c||c|c|c|c|}
			\cline{2-9}
			& \multicolumn{4}{ c|| }{\textbf{Spread: 12-16}} & \multicolumn{4}{ c| }{\textbf{Spread: 16-20}} \\ \cline{2-9}
			& ${\mu}$  & ${log(\sigma)}$ & ${\nu}$ & ${log(\tau)}$ 
			& ${\mu}$  & ${log(\sigma)}$ & ${\nu}$ & ${log(\tau)}$ \\ \cline{1-9}
			%
			\multicolumn{1}{ |c| }{${\beta_0}$} & 3.444 & 8.324e-1 & -2.023e-1 & 5.037 & -4.541  & 5.176e-1 & 4.167e-1 & -1.661 \\ \hline
			%
			\multicolumn{1}{ |c| }{$s_{t-1}$, ${\beta_1}$ } & 0.08 & -1.543e-2 &  &  & 3.827e-1 &  & -5.269e-3 & 1.688e-2 \\ \hline
			%
			\multicolumn{1}{ |c| }{gas, ${\beta_2}$} & -2.461e-1 & 1.989e-2 & 7.291e-2 & -9.713e-2 &  &4.445e-2 & -5.630e-3 &  \\ \hline
			%
			\multicolumn{1}{ |c| }{coal, ${\beta_3}$} & 9.227e-2 & 4.327e-3 & -2.479e-2 &  & 5.267e-2 &  & -3.148e-3 &  \\ \hline
			%
			\multicolumn{1}{ |c| }{$s_{wind}$, ${\beta_4}$} & -1.237e-3 &  & -6.023e-5 & -1.705e-4 & -6.593e-4 &  & -1.837e-5 &  \\ \hline
			%
			\multicolumn{1}{ |c| }{$s_{solar}$, ${\beta_5}$ } & -1.188e-3 &  & -5.714e-5 &  & -2.863e-4 &  & -1.883e-5 &  \\ \hline
			%
			\multicolumn{1}{ |c| }{dummy, ${\beta_6}$} &  &  &  &  & -3.039 & 1.282e-1 & -1.478e-1 & 3.460e-1 \\ \hline
			%
			\multicolumn{1}{ |c| }{$s_{load}$, ${\beta_7}$} & 5.920e-4 & -6.492e-5 &  & -2.251e-4 & -4.080e-3 & 1.254e-4 & 1.646e-4 & -5.030e-4 \\ \hline
			%
			\multicolumn{1}{ |c| }{$s_{iLoad}$, ${\beta_8}$} &  &  & 3.440e-9 &  & 6.724e-8 & -2.355e-9 & -2.263e-9 & 7.951e-9 \\ \hline
		\end{tabular}
		\caption{Illustration of coefficients for 4 moments for spreads 12-16 and 16-20.} 
		\label{tab:coeff_Illustration2}
	\end{table}
	
	Overall the signs and significances of the coefficients are intuitive. In particular, wind and solar production spreads have negative effects on the mean and skewness of spreads for the morning and afternoon spread pairs shown. Recall that the spread is defined as the former minus the later hours and so a higher wind and solar production spread will generally reduce the the average spreads and also the skewness. This is consistent with the effects of wind and solar production on price levels reported in \citep{gianfreda2017stochastic} and elsewhere.

	\subsubsection{Out-Of-Sample Performance}
	The average Root Mean Squared Errors for the expected values of spreads forecasted with models using the best and Normal distributions over the rolling-window forecasting horizon are given in Table \ref{tab:RMSE_mu_fitted_forecasted}. 
	Each score is calculated by averaging RMSE values across all forecasted spreads (note: if at least 1 time step forecast was missing for a given spread, the forecasted time series was omitted from RMSE calculation).
	A detailed breakdown of the RMSE values for the best and Normal distribution models is given in Appendix Figures \ref{fig:app1_Forecast_EY_RMSE_Skew} and \ref{fig:app1_Forecast_EY_RMSE_NO} respectively.
	The RMSE values are in line with, but slightly higher overall, than the ones found for the validation data set (see Figure \ref{fig:app1_muVal_RMSEbasedBestDistn-ActualRMSE}).
	The RMSE based forecasting power evaluation concludes that the best distribution and Normal distribution models produce compatible results, without significant differences. 
	\begin{table}[h!]
		\centering 
		\begin{tabular}{ |l|l|l| }  
			\hline
			& \textbf{Best Dist. Model} & \textbf{Normal Model} \\ \hline
			\textbf{RMSE Expected Value} & $7.01$ & $6.95$ \\ \hline
			\textbf{Std Error} & $0.134$ & $0.119$ \\ \hline
		\end{tabular}
    	\caption{\textbf{Forecast RMSE} - $E(\widehat{\mathbf{Y}})$ Obtained With Best vs Normal Distribution Models.}
		\label{tab:RMSE_mu_fitted_forecasted}
	\end{table}

	\section{Optimisation of Trading Schedule}
	The battery operation trading schedule is optimised by maximising the trading profit using forecasted spread densities for the day-ahead hourly electricity prices, while meeting a risk criterion of making a profit per MWh traded of more than $c$ per daily cycle with 95\% confidence. 
	We consider $c$ to be the round trip transaction cost for a storage facility in charging and discharging. 
	This covers the technical efficiency loss of the battery between the charging and discharging of energy, as well as the transmission, distribution, trading, balancing, levies and other use of system costs for a facility seeking to operate in the wholesale market. 
	Estimates of these costs vary widely in practice and so for comparison we use a sensitivity analysis approach with $c = 5, 10, 15$ Euro/MWh.
	We impose an assumption of a finite horizon daily operation in the day-ahead forecasting analysis, which constrains the opening and closing battery charge levels to be equal each day.
	We investigate the optimal opening/closing level of charge level by performing calculations for a number of initial battery levels, $b = 0, 0.1 ,0.2,...,0.9$, and selecting the one with the highest profit.
	Therefore, the schedule is optimised with respect to initial battery levels, $b$, of the total capacity of a nominal 1 MWh battery, which could trade in the wholesale market.
	The battery is assumed to be fully (dis)chargeable within 1 hour of trade execution.
	
	The realised Profit and Loss (P\&L) for a day's trade, $PNL_t$, executed using spread hour $s$, is calculated as per Equation \ref{eq:pnl_1Trade} and is based on realised spread value and total round-trip cost.
	\begin{eqnarray} \label{eq:pnl_1Trade}
	PNL^{(s)}_t = \left\{
	\begin{array}{ll}
	\Big(E\big(Y_t^{(s)}\big) - c\Big) * b         ~ \quad  \quad  \quad  \quad \text{if}~~ E\big(Y_t^{(s)}\big) > 0 \\
	\bigg(\Big|E\big(Y_t^{(s)}\big)\Big| - c\bigg) * (1-b) \quad \text{if}~~ E\big(Y_t^{(s)}\big) < 0
	\end{array}
	\right.
	\end{eqnarray}
	
	We backtest the trading schedule over approximately 1 year of data (383 days, time steps $t=1535,...,1917$) and report for each starting battery level: 
	P\&L over the backtest period, $PNL$ (Eq. \ref{eq:av_total_pnl_1Trade}),
	average P\&L over the backtest period, $\overline{PNL}$, (Eq. \ref{eq:av_pnl_1Trade}), 
	standard error of the P\&L average, $s^{\overline{PNL}}$ (Eq. \ref{eq:stdErr_of_av_pnl_1Trade}), 
	number of trades which resulted in a loss after all costs are taken into account, $n_l$, (Eq. \ref{eq:number_Loss_Days_1Trade}), 
	total monetary value resulting from loss days, $l$, (Eq. \ref{eq:tot_Loss_1Trade}), and
	average loss, $\overline{l}$, (Eq. \ref{eq:av_Loss_1Trade}).
	\begin{align}
	PNL                              &= \sum_{t=1}^{383} PNL_t                                                              \label{eq:av_total_pnl_1Trade}  \\
	\overline{PNL}               &= \frac{1}{383}  \sum_{t=1}^{383} PNL_t                                         \label{eq:av_pnl_1Trade} \\
	s^{\overline{PNL}}         &= \frac{1}{\sqrt{383}} \overline{PNL}                                               \label{eq:stdErr_of_av_pnl_1Trade} \\
	n_l                               &= \sum_{i=1}^{k} 1  \quad \quad \quad \quad \text{if}~~ PNL_t < 0                                 \label{eq:number_Loss_Days_1Trade} \\
	l                                   &= \sum_{i=1}^{n_l} PNL_i \quad \quad \text{if}~~ PNL_t < 0                   \label{eq:tot_Loss_1Trade} \\
	\overline{l}                   &= \frac{1}{k} \sum_{i=1}^{k} PNL_i \quad \text{if}~~ PNL_t < 0          \label{eq:av_Loss_1Trade}
	\end{align}
	
	\subsection{Single Trade Per Day}
	\begin{figure}[h!]
		\centering
		\includegraphics[width=0.8\textwidth]{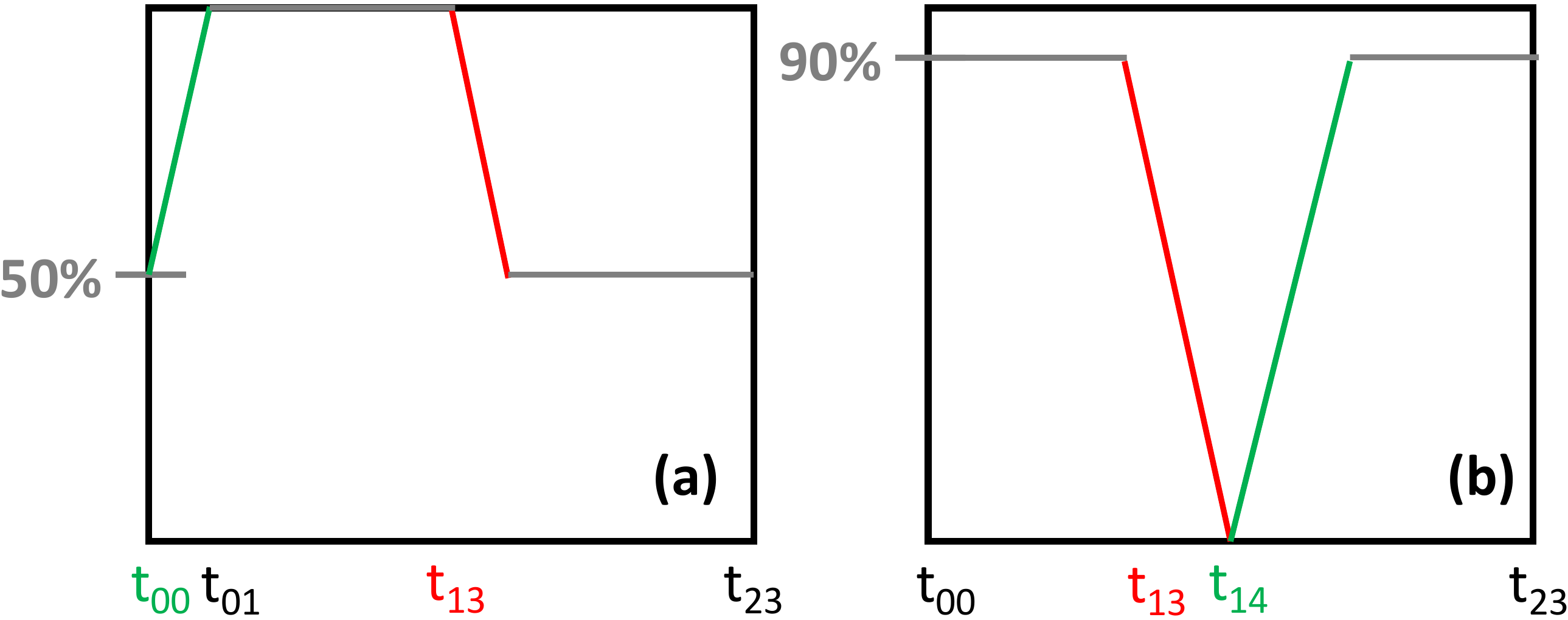}
		\caption{Single trade example for initial battery charge level of (a) 50\% and (b) 90\%.}
		\label{fig:singleTrade}
	\end{figure}
	Figure \ref{fig:singleTrade} \textbf{(a)} depicts a hypothetical trade where the battery is charged to 50\% level at the start of the day. 
	The example trade consists of a charge at 00 midnight from 50\% to full capacity of 100\% (green line - 1 hour charge), followed by a holding period (grey) until discharge at hour 13 (red line) down to 50\% charge level, thus completing a single trade for the day (consisting of two legs of the spread), returning the battery to the initial 50\% charge level.
	An alternative trade is displayed in sub-plot \textbf{(b)} where the initial battery level is 90\%. 
	This charge level is held until discharge at hour 13 down to 0\% charge level, followed by a re-charge at hour 14 back to 90\% charge, thus completing a single trade for the day.
	
	\subsection{Optimal Trade Selection} 
	Initially, the optimal trade to be performed on each day is established using an exhaustive state space search over all available trade executions, performed using Algorithm \ref{algo:singleTrade} (note, the same procedure is repeated for Normal two parameter $(\mu_t,\sigma_t)$ distribution estimated models to benchmark the results).
	At each test point time step, $t=1,...,383$, i.e. 1 day of trading, we consider $s=1,...,276$ potential trades based on forecasted spreads, where for each trade we examine with 95\% confidence the possibility of making a profit of at least $c$ Euro/MWh.
	
	First, we calculate the forecasted expected value of the spreads at each time step $t$ using Equations \ref{eq:SEP1_E(Y)} - \ref{eq:ST5_E(Y)} (note: $E(\widehat{Y}^{(s)}_t) = \hat{\mu}^{(s)}_t$ for distributions JSU and ST1, see Appendix \ref{app:allE(Y)} for calculating expected values of the other distributions used).
	Occasionally the forecasted kurtosis was very large and when this happened it was capped at the value of 100 in order to successfully use the equations for expected value calculation. 
	Next we establish whether the expected value $E(\widehat{Y}^{(s)}_t)$ of a spread number $s$, is positive (later price is lower, therefore discharge then charge) or negative (later price is higher, therefore a profitable trade would require charging first, followed by discharging).
	If the expected value of the spread was missing, the calculations for that spread were skipped (note this happened a total of 1795 for models based on the best chosen distributions and 4159 for models based on Normal benchmark distribution, out of $276\times383=105,708$ possible times).
	For each case, we calculate the critical value corresponding to 95\% confidence interval.
	If forecasted expected value of the spread $s$ on day $t$ is positive, $E(\widehat{Y}^{(s)}_t) > 0$, we access the 5$^{th}$ quantile, $q_5$, since the body of the distribution is to the right of this critical value (later price is lower, hence discharge then charge).
	If $E(\widehat{Y}^{(s)}_t) < 0$ we access the 95$^{th}$ quantile so that body of the distribution is to the left of the critical value (i.e. later price is higher, hence charge then discharge).
	
	Next, we calculate the forecasted profit for each spread at time step $t$, only if the critical value obtained from the quantile estimation exceeds the cost of $c$ Euro/MWh.
	The forecasted profit is found as the weighted difference of the forecasted expected value and the total roundtrip cost. 
	The trade corresponding to the maximum P\&L value at time $t$ is selected as the optimal trade.
	The realised profit is calculated according to the realised outcome for that spread (see Equation \ref{eq:pnl_1Trade}) and the profit \& loss analysis is performed as per Equations \ref{eq:av_pnl_1Trade} -  \ref{eq:av_total_pnl_1Trade} for each trade and with respect to the benchmark Normal type distribution.
	\textit{Note 1}: we do not trade on any day, for which the forecasted profit is below the roundtrip cost $c$ Euro/MWh.
	\textit{Note 2}: if a quantile fails to be estimated for a given model, the critical value $\hat{q}_{xx}$ gets set to 0 (for the best chosen distributions models, this happened at forecast time steps, $t = 251, 264, 304, 349$).
	
	\begin{algorithm} 
		\caption{Single Trade Per Day Model}
		\label{algo:singleTrade}
		\begin{algorithmic}[1]
			\State Initialise \textit{forecasted} profit and loss matrix, $ \widehat{PNL} \gets \mathbf{0} \in R^{276 \times 383}$ 
			\State Initialise \textit{realised} profit and loss vector, $PNL \gets \mathbf{0} \in R^{383}$
			\For{each forecast time step $t = 1,...,383$}
				\For{each spread number $s=1,...,276$}
					\State Using estimated model $\widehat{M}^{(s,t )}$ extract forecasted moments $\boldsymbol{\hat{\theta}}^{(s)}_t = [\hat{\mu}^{(s)}_t , \hat{\sigma}_t^{(s)} , \hat{\nu}_t^{(s)} , \hat{\tau}_t^{(s)}]^T$
					\If {dist. parameters $\boldsymbol{\hat{\theta}}^{(s)}_t$ were successfully forecasted} 
						\State Calculate $E(\widehat{Y}^{(s)}_t)$ using Eqs. \ref{eq:SEP1_E(Y)}-\ref{eq:ST5_E(Y)} (note, $E(\widehat{Y}^{(s)}_t) = \hat{\mu}_t^{(s)} $ for JSU and ST1).
						\If {$E(\widehat{Y}^{(s)}_t) > 0$}
							\State Extract 5$^{th}$ quantile, $\hat{q}_{05}$ (i.e. 95\% of distribution is on the right) 
						\Else
							\State Extract 95$^{th}$ quantile, $\hat{q}_{95}$ (i.e. 95\% of distribution is on the left) 
						\EndIf
						\If {$|\hat{q}_{xx}| > c$} 
							\If {$E(\widehat{Y}^{(s)}_t) > 0$}
								\State $\widehat{PNL}^{(s)}_t  \gets (E(\widehat{Y}^{(s)}_t) -c)*b$ forecasted profit for spread $s$
							\Else
								\State $\widehat{PNL}^{(s)}_t  \gets (|E(\widehat{Y}^{(s)}_t) -c)*(1-b)$ forecasted profit for spread $s$
							\EndIf
						\EndIf
					\EndIf
				\EndFor
				\State Execute trade corresponding to spread $s_{max}$, where $s_{max} \gets \text{argmax}_s \widehat{PNL}_t$
				\If {${\mu}^{(s_{max})}_t > 0$}
					\State $PNL_t \gets \Big({\mu}^{(s_{max})}_t - c\Big)*b$
				\Else
					\State $PNL_t \gets  \Big(|{\mu}^{(s_{max})}_t| - c\Big)*(1-b)$
				\EndIf
			\EndFor
			\State Obtain simple summary statistics using Eqs. \ref{eq:av_total_pnl_1Trade} - \ref{eq:av_Loss_1Trade}
		\end{algorithmic}
	\end{algorithm}

	\subsubsection{Roundtrip cost - 5 Euro/MWh}
	The roundtrip cost of 5 Euro/MWh is considered first, the results of which are reported in Table \ref{tab:res1trade_5Euro} (note: models based on the best chosen distribution did not get used for trading on 2 days, while those using the Normal distribution did not trade on 0 days).
	\begin{table}[h!]
		\centering 
		\begin{tabular}{ |l|l|l|l|l|l|l|l|ll|l|l| }  
			\hline
			\textbf{$b$} & \textbf{0} & \textbf{0.1} & \textbf{0.2} & \textbf{0.3} & \textbf{0.4} & \textbf{0.5} & \textbf{0.6} & \textbf{0.7} & & \textbf{0.8} & \textbf{0.9} \\ 
			\hline
			\hline
			\textbf{$PNL$, BD}    & $6537$  & $5887$ & $5197.8$ & $4556.7$ & $3926$ & $3347.5$ & $3129$ & $3361.3$ & & $3742.2$ & $4172$ \\
			\hline
			\textbf{$\overline{PNL}$}         & $17.24$  & $15.45$ & $13.64$ & $11.95$ & $10.30$ & $8.79$ & $8.22$ & $8.82$ & & $9.82$ & $10.95$ \\
			\hline
			\textbf{$s^{\overline{PNL}}$} &$0.76$ & $0.68$ & $0.60$   & $0.52$ & $0.45$ & $0.37$ & $0.39$ & $0.49$ & & $0.57$ & $0.65$ \\
			\hline
			\textbf{$n_l$} &$1$ & $1$ & $1$ & $1$ & $1$ & $1$ & $1$ & $2$ & & $3$ & $3$ \\ 
			\hline
			\textbf{$l$}    &  $-2$  & $-1.8$ & $-1.6$ & $-1.4$ & $-1.2$ & $-1$ & $-0.8$ & $-2.7$ & & $-3.6$ & $-3.8$ \\ 
			\hline
			\textbf{$\overline{l}$}        & $-2$  & $-1.8$ & $-1.6$ & $-1.4$ & $-1.2$ & $-1$ & $-0.8$ & $-1.35$ & & $-1.2$ & $-1.27$ \\ 
			\hline
			\hline
			\textbf{$PNL$, NO}    & $6639$  & $5975.1$ & $5294.6$ & $4637.1$ & $3981.4$ & $3398$ & $3215.8$ & $3449.4$ & & $3731.6$ & $4062$ \\
			\hline
			\textbf{$\overline{PNL}$}      & $17.33$  & $15.60$ & $13.82$ & $12.10$ & $10.40$ & $8.87$ & $8.40$ & $9.00$ & & $9.74$ & $10.61$ \\
			\hline
			\textbf{$s^{\overline{PNL}}$} &$0.80$ & $0.72$ & $0.64$ & $0.56$ & $0.48$ & $0.41$ & $0.44$ & $0.52$ & & $0.61$ & $0.70$ \\
			\hline
			\textbf{$n_l$} &$6$ & $6$ & $5$ & $5$ & $5$ & $4$ & $4$ & $5$ & & $11$ & $20$ \\
			\hline
			\textbf{$l$} &$-18$ & $-16.2$ & $-11.2$ & $-10$ & $-9$ & $-6.5$ & $-6$ & $-6.7$ & & $-18.8$ & $-41.4$ \\
			\hline
			\textbf{$\overline{l}$}     &  $-3$ & $-2.7$ & $-2.24$ & $-2$ & $-1.8$ & $-1.63$ & $-1.5$ & $-1.34$ & & $-1.71$ & $-2.07$ \\
			\hline
		\end{tabular}
		\caption{\textbf{Cost 5 Euro/MWh - Single Trade}. Results for Best Dist (BD) vs Normal (NO).} \label{tab:res1trade_5Euro}
	\end{table}
	
	The results of 5 Euro/MWh roundtrip cost trading indicate that our method and the benchmark approach both have the same predictive power, with average P\&L results for each initial battery level not having significantly different results at 95\% level. 
	The best initial battery level transpires to be 0\% charge, which is explained by the fact that typically electricity would be cheaper during the night due to lack of demand (which makes it cheaper to charge the battery), and thus making the most profit by discharging at some point during peak demand in the day.
	The total P\&L value across all battery levels of Normal type models is better by approximately $\frac{44384-43856.5}{43856.5}*100= 1.2$\% over the backtest period, however this does not reflect the number of encountered loss days and their total monetary value.
	The ratio of total number of loss days is $\frac{71}{15}=4.7$ times more for the Normal type, with the monetary value of these losses $\frac{143.8}{19.9}=7.2$ times higher over models based on the best chosen distributions. 
	
	Overall, the 5 Euro/MWh cost basis is not large enough to reveal the advantage of using our method when making trading decisions, however this becomes obvious when the cost basis is increased to more realistic scenarios of 10 and 15 Euro/MWh (see below).

	\subsubsection{Roundtrip cost - 10 Euro/MWh}
	The P\&L results for 10 Euro/MWh roundtrip cost trading scenario are reported in Table \ref{tab:res1trade_10Euro} (note: models based on the best chosen  distribution did not get used for trading on 57 days for b=0 and 49 days for all other starting battery levels, while Normal did not trade on 62 days).
	
	\begin{table}[h!]
		\centering 
		\begin{tabular}{ |l|l|l|l|l|l|l|l|l|l|l| }  
			\hline
			\textbf{$b$} & \textbf{0} & \textbf{0.1} & \textbf{0.2} & \textbf{0.3} & \textbf{0.4} & \textbf{0.5} & \textbf{0.6} & \textbf{0.7} & \textbf{0.8} & \textbf{0.9} \\ 
			\hline
			\hline
			\textbf{$PNL$, BD}    & $4542$  & $4092.3$ & $3606.4$ & $3175.7$ & $2736.4$ & $2341.5$ & $2164.6$ & $2180$ & $2258.8$ & $2371.3$ \\
			\hline
			\textbf{$\overline{PNL}$}  & $13.93$ & $12.25$ & $10.80$ & $9.51$ & $8.19$ & $7.01$ & $6.48$ & $6.53$ & $6.76$ & $7.10$ \\
			\hline
			\textbf{$s^{\overline{PNL}}$} & $0.85$ & $0.75$ & $0.66$ & $0.58$ & $0.50$ & $0.41$ & $0.41$ & $0.52$ & $0.60$ & $0.70$ \\
			\hline
			\textbf{$n_l$} & $8$ & $8$ & $8$ & $7$ & $7$ & $6$ & $6$ & $11$ & $11$ & $11$ \\
			\hline
			\textbf{$l$}   &  $-17$  & $-15.3$ & $-13.6$ & $-11.2$ & $-9.6$ & $-7$ & $-5.8$ & $-10.2$ & $-9.8$ & $-9.4$ \\
			\hline
			\textbf{$\overline{l}$}       & $-2.13$  & $-1.91$ & $-1.7$ & $-1.6$ & $-1.37$ & $-1.17$ & $-0.97$ & $-0.93$ & $-0.89$ & $-0.85$ \\
			\hline
			\hline
			\textbf{$PNL$, NO}   &  $3972$  & $3578.5$ & $3185$ & $2793.1$ & $2408.6$ & $2077.5$ & $1895.4$ & $1793.6$ & $1723.4$ & $1672.1$ \\
			\hline			
			\textbf{$\overline{PNL}$}         &  $12.57$  & $11.15$ & $9.92$ & $8.70$ & $7.50$ & $6.47$ & $5.90$ & $5.59$ & $5.37$ & $5.21$ \\
			\hline
			\textbf{$s^{\overline{PNL}}$} & $0.93$ & $0.83$ & $0.74$ & $0.64$ & $0.55$ & $0.47$ & $0.44$ & $0.48$ & $0.54$ & $0.61$ \\
			\hline
			\textbf{$n_l$} & $25$ & $26$ & $26$ & $26$ & $26$ & $26$ & $26$ & $27$ & $27$ & $27$ \\
			\hline
			\textbf{$l$}   &  $-85$   & $-76.6$ & $-68.2$ & $-59.8$ & $-51.4$ & $-42.5$ & $-34.4$ & $-29.1$ & $-21.4$ & $-13.7$ \\
			\hline
			\textbf{$\overline{l}$}        &  $-3.4$  & $-2.95$ & $-2.62$ & $-2.3$ & $-1.97$ & $-1.63$ & $-1.32$ & $-1.08$ & $-0.79$ & $-0.51$ \\
			\hline
		\end{tabular}
		\caption{\textbf{Cost 10 Euro/MWh - Single Trade}. Results for Best Dist (BD) vs Normal (NO).} \label{tab:res1trade_10Euro}
	\end{table}
	
	The higher roundtrip cost of 10 Euro/MWh, reveals the advantage of using Skew type and similar distribution based models: the average P\&L is significantly better than Normal model's at 95\% confidence, and the best initial battery charge level is again 0\% charge level (i.e. fully discharged).
	The P\&L over the tested period across all battery levels is $\frac{29469-25099.2}{25099.2}*100 = 17.4$\% higher using our approach over the Normal. 
	The ratio of total number of loss days for Normal to Skew models is $\frac{262}{80}=3.3$,  
	with the monetary value of losses being $\frac{482.1}{108.9} = 4.4$ times higher for the Normal types.

	\subsubsection{Roundtrip cost - 15 Euro/MWh} 
	The P\&L results for 15 Euro/MWh roundtrip cost trading scenario are reported in Table \ref{tab:res1trade_15Euro} (note: models based on the best chosen  distribution did not get used for trading on 223 for b=0 and 210 days for all other starting battery levels, while Normal did not trade on 255 days (out of the total tested period of 383 days).
	
	\begin{table}[h!]
		\centering 
		\begin{tabular}{ |l|l|l|l|l|l|l|l|l|l|l| }  
			\hline
			\textbf{b} & \textbf{0} & \textbf{0.1} & \textbf{0.2} & \textbf{0.3} & \textbf{0.4} & \textbf{0.5} & \textbf{0.6} & \textbf{0.7} & \textbf{0.8} & \textbf{0.9} \\ 
			\hline
			\hline
			\textbf{$PNL$, BD}    & $2409$ & $2176.7$ & $1911.2$ & $1689.3$ & $1484.4$ & $1267.5$ & $1155.8$ & $1074.8$ & $992.8$ & $913.9$ \\
			\hline
			\textbf{$\overline{PNL}$}          & $15.06$ & $12.58$ & $11.05$ & $9.76$ & $8.58$ & $7.33$ & $6.68$ & $6.21$ & $5.74$ & $5.28$ \\
			\hline
			\textbf{$s^{\overline{PNL}}$} & $1.46$ & $1.25$ & $1.09$ & $0.95$ & $0.81$ & $0.67$ & $0.63$ & $0.65$ & $0.71$ & $0.79$ \\
			\hline
			\textbf{$n_l$} & $12$ & $13$ & $13$ & $13$ & $12$ & $12$ & $12$ & $12$ & $12$ & $12$ \\
			\hline
			\textbf{$l$}  & $-27$ & $-24.4$ & $-21.8$ & $-19.2$ & $-15.4$ & $-13$ & $-10.6$ & $-8.2$ & $-5.8$ & $-3.4$ \\
			\hline
			\textbf{$\overline{l}$}     &  $-2.25$ & $-1.88$ & $-1.68$ & $-1.48$ & $-1.28$ & $-1.08$ & $-0.88$ & $-0.68$ & $-0.48$ & $-0.28$ \\ 
			\hline
			\hline
			\textbf{$PNL$, NO}   & $1463$ & $1326.3$ & $1189.6$ & $1052.9$ & $937.8$ & $807.5$ & $698.6$ & $610.3$ & $531.2$ & $445.6$ \\
			\hline
			\textbf{$\overline{PNL}$}         & $12.72$ & $10.36$ & $9.29$ & $8.23$ & $7.33$ & $6.31$ & $5.46$ & $4.77$ & $4.15$ & $3.48$ \\
			\hline
			\textbf{$s^{\overline{PNL}}$} & $1.68$ & $1.39$ & $1.23$ & $1.07$ & $0.92$ & $0.76$ & $0.64$ & $0.58$ & $0.59$ & $0.63$ \\
			\hline
			\textbf{$n_l$} & $22$ & $23$ & $23$ & $23$ & $23$ & $23$ & $23$ & $23$ & $23$ & $23$ \\
			\hline
			\textbf{$l$}  & $-74$ & $-66.7$ & $-59.4$ & $ -52.1$ & $-44.8$ & $-37.5$ & $-30.2$ & $-22.9$ & $-15.6$ & $-8.3$ \\
			\hline
			\textbf{$\overline{l}$}        & $-3.36$ & $-2.9$ & $-2.58$ & $-2.27$ & $-1.95$ & $-1.63$ & $-1.31$ & $-1$ & $-0.68$ & $-0.36$ \\
			\hline
		\end{tabular}
		\caption{\textbf{Cost 15 Euro/MWh - Single Trade}. Results for Best Dist (BD) vs Normal (NO).} \label{tab:res1trade_15Euro}
	\end{table}
	
	The average P\&L is significantly better than Normal model's at 95\% confidence, and the best initial battery charge level is again 0\% charged.
	The P\&L over the tested period across all battery levels is $\frac{15075.4 - 9062.8}{9062.8}*100 = 66.3$\% higher for the Skew and similar distribution types selected as best, over the Normal. 
	The ratio of total number of loss days for Normal to Skew models is $\frac{229}{123}=1.9$,  
	with the monetary value of losses being $\frac{411.5}{140.8} = 2.9$ times higher for the Normal types.

	\section{Conclusion}
	We have demonstrated the value of detailed, computationally-intensive modelling of intra-day power price spread densities using a flexible four parameter distributional form, generally the skew-t. 
	This allows the dynamic conditional parameter estimates to follow stochastic evolutions driven by exogenous factors, most importantly the day ahead demand, wind and solar forecasts. 
	These forecasts fit well in backtesting and support the optimal daily scheduling of a storage facility, operating on a single cycle. 
	The model outperforms baseline comparisons to a normal density model.
	The model specification and validation process is computationally intensive, and whilst modelling simplifications could be introduced, accuracy is important. 
	Overall, this formulation and application shows the merits of a computationally intensive approach to accurate specifications and these are likely to be more attractive in practice than methods based upon analytical simplifications of the stochastic price processes.
	The optimal choice of spreads to trade do vary daily and the need to utilise forecasts in well-specified models is evident, as is the delicate balance between expected profits and risk. 
	Furthermore, the algorithmic nature of the modelling presented would lead naturally to the potential for algorithmic trading by battery asset owners, which may be more economical to small enterprises than outsourcing their trading to larger service providers.
	In terms of optimisation, we have indicated the value of the methodology in supporting optimal storage on a one cycle per day basis. 
	Further extensions to two or more cycles per day is clearly possible and would most likely further endorse the value of accurate spread density specifications.

	\clearpage
	\bibliography{paper.bib}

	\clearpage
	\appendix
	\appendixpage
	\addappheadtotoc
	\section{Data Collection and Pre-processing} \label{app:dataCollection}
	\subsection{German Day-Ahead Electricity Price}
	Day-ahead prices for each hour of the day ahead, downloaded from \url{http://www.energinet.dk/en/el/engrosmarked/udtraek-af-markedsdata/Sider/default.aspx}. 
	
	\subsection{Total Wind Day Ahead Forecast}
	Sum of wind forecasts for the four control areas, each downloaded from \url{https://data.open-power-system-data.org}.
	Missing data for two control areas of the following dates was interpolated:
	\textbf{(a)} 50 Hertz (MW): 
	$01.01.2017-31.03.2017$ hourly wind forecast data replaced with average of four 15 min intervals starting on the hour: e.g. for data of 23.00 hour is the average of $23:00 - 23:15, 23:15 - 23:30, 23:30 - 23:45, 23:45 - 00:00$ obtained from \url{http://www.50hertz.com/en/Grid-Data/Wind-power/Archive-Wind-power}. 
	\textbf{(b)} TransnetBW (MW): 
	the data for the following dates was missing (and also missing from \url{https://www.transnetbw.com/en/transparency/market-data/key-figures}) $23.03.2012; 12.10.2012;  21.02.2013; 31.03.2013; 30.03.2014; 07.09.2014; 29.03.2017$ and thus were replaced with an average of the same hour from the past 7 days.
	Note the March dates experience a change in the clock, therefore for consistency of data the average (for each hour for end of March missing data) was shifted by one hour: i.e. for hour 06, an average of 7 previous days of hour 05 was taken instead of 06. 
	
	\subsection{Total Solar Day Ahead Forecast}
	Each control area's forecast is obtained by finding the average of 15 min forecasts starting from the hour (e.g. for 23.00 data point this is the average of $23:00 - 23:15, 23:15 - 23:30, 23:30 - 23:45, 23:45 - 00:00$), with data downloaded from \url{https://data.open-power-system-data.org}. 
	The total solar day-ahead forecast is found by summing the average for each of the four control regions.
	A number of days contained non-zero entries for forecasted sunshine for hours 22.00 and 23.00 in the night (e.g. $30.06.2014$ contained entries224 and 15 respectively).
	In order aid smooth regression estimation, any non-zero entries for hour 22.00/23.00 have been overwritten with 0 in order to be consistent with the majority of the data for those hours (approx. 10 entries in total). 
	Note: the processed data set therefore did not contain any forecasted sunshine for hours: $22.00, 23.00, 24.00, 00.00, 01.00, 02.00, 03.00$.
	Some data was missing and was therefore replaced for the following control areas:
	\textbf{(a)} 50 Hertz (MW): 
	$01.01.2017-31.03.2017$ hourly data replaced with solar forecast, average of four 15 min intervals starting on the hour: e.g. for 23.00 data is the average of $23:00 - 23:15, 23:15 - 23:30, 23:30 - 23:45, 23:45 - 00:00$ (data downloaded from \url{http://www.50hertz.com/en/Grid-Data/Photovoltaics/Archive-Photovoltaics})
	$13.05.2014$ (original data from the same website suggests a forecast of full sunshine from 00 midnight to 23hr of 2000+ MW; it was therefore replaced with an av. of last 7 days of forecasts for each hour).
	\textbf{(b)} Amprion (MW):
	Date $28.04.2014$ was missing thus hourly data replaced with solar forecast, av. of four 15 min intervals starting on the hour: e.g. for 23.00 data is the av. of $23:00 - 23:15, 23:15 - 23:30, 23:30 - 23:45, 23:45 - 00:00$ (data downloaded from \url{http://www.amprion.net/en/photovoltaic-infeed}). 
	\textbf{(c)} TransnetBW (MW): 
	The days $31.03.2013; 30.03.2014; 07.09.2014; 29.03.2015$ were missing from \url{https://www.transnetbw.com/en/transparency/market-data/key-figures} and thus were replaced with an av. of the hour from the past 7 days.
	Note March dates experience a change in the clock, therefore for consistency of data, the average (for each hour for end of March missing data) was shifted by one hour: i.e. for hour 06, an av. of 7 previous days of hour 05 was taken instead of 06. 
	Incorrect data for $13.12.2014$ shows sunshine at night and for 24 hours of that day (200+ MW), hence this was replaced with an av. of last 7 days of forecasts for each hour.
	
	\subsection{Day Ahead Total Load}
	Example hour 23.00 data point: is the sum of the average of segments $23:00 - 23:15, 23:15 - 23:30, 23:30 - 23:45, 23:45 - 00:00$ for each control region. 
	The load data used was \texttt{'load\char`_old'} from $01.01.2012 - 31.12.2015$ and then data \texttt{'load\char`_new'} was used from $01.01.2016 - 31.03.2017$.
	The "forecasted load" was calculated as: in order to submit a bid for tomorrow a trader would need to submit all data by 12.00noon. 
	Therefore they would use the realised total actual load from today's hours starting from hour 10.00am and work backwards to yesterday 11.00am (since 11am is calculated as the average actual load for 4 x 15 min segments starting from 11.00 am and finishing at 12.00.
	It is most likely that 12.00 would not be met, thus an 11.00 cut off is suggested.
	Original Source: \url{https://transparency.entsoe.eu/load-domain/r2/totalLoadR2/show}. 
	
	\subsection{Steam Coal ARA 1 Month Forward}
	Benchmark index for steam coal delivered to the Amsterdam Rotterdam Antwerp region of the Northwest Europe (ARA) - one price per day. 
	Original Source: Bloomberg ticker API21MON.
	
	\subsection{Gas Day-Ahead Forward}
	Germany Gaspool (GPL) natural gas day-ahead forward (delivered next working day) - one price per day. 
	Original Source: Bloomberg ticker BHDAHD.
	
	\subsection{Dummy}
	The weekly seasonality and holidays are included into a single dummy variable: it takes on value 1 for Saturday/Sunday and the following German state holidays: New Year's Day, Good Friday, Easter Monday, Labour Day, Ascension Day, Whit Monday, German unit day, Christmas Day, Boxing day and New Years Eve. 
	The following holidays have been omitted because of the focus on German TSO market: Austrian public holidays / regional holidays, such as Pentecost, Corpus Christi Assumption of Mary, Reformation day, All Saints day and Repentance Day.
	Data created by hand.

	%
	\section{Expected Value Calculation - Four Parameter Distributions} \label{app:allE(Y)}
	Large fitted/forecasted kurtosis values are capped at 100 for algorithmic convergence. 
	In the case of Johnson's $S_u$ and skew t type 1 distributions, the expected value does equal the distribution mean, $E(Y_t) = \mu_t$.
	\subsection{SEP1 Distribution} \label{app:SEP1}
	The probability density function of skew exponential power type 1 distribution \citep{azzalini1985class} is given by
	\begin{equation*}
	{f}_Y(y_t|\mu_t,\sigma_t,\nu_t,\tau_t) = \frac{2}{\sigma_t} f_{Z_1}(z_t) F_{Z_1}(\nu_t z_t) 
	\end{equation*}
	where
	$Z_1 \sim PE2(0,\tau_t^{1/\tau_t},\tau_t)$ and 
	$PE2$ is the exponential type 2 distribution.
	
	The expected value of the random variable $Y_t$ is given by 
	\begin{equation} \label{eq:SEP1_E(Y)}
	E(Z_t) = \text{sign}(\nu_t) \tau_t^{1/\tau_t} \frac{\Gamma(\frac{2}{\tau_t})}{\Gamma(\frac{1}{\tau_t})} pBEo\Big( \frac{\nu_t^{\tau_t}}{1+\nu_t^{\tau_t}}, \frac{1}{\tau_t}, \frac{2}{\tau_t} \Big)
	\end{equation}
	where $\Gamma(\cdot)$ is the Gamma function, $pBEo(q,a,b)$ is the cumulative distribution function of the beta distribution $BEo(a,b)$, evaluated at q, with shape parameters a and b.
	Certain values of $\nu_t$ and $\tau_t$, resulted in a numerical error when calculating exponent, $\nu_t^{\tau_t}$, when this happened the value of this quantity was set to 1. 
	
	\subsection{SEP2 Distribution} \label{app:SEP2}
	The probability density function of skew exponential power type 2 distribution \citep{azzalini1985class, diciccio2004inferential} is given by
	\begin{equation*}
	{f}_Y(y_t|\mu_t,\sigma_t,\nu_t,\tau_t) = \frac{2}{\sigma} f_{Z_1}(z_t) \Phi(\omega_t)
	\end{equation*}
	where
	$\omega_t=\text{sign}(z_t) |z_t|^{\tau_t/2} \nu_t \sqrt{\frac{2}{\tau_t}}$, 
	$Z_1 \sim PE2(0,{\tau_t}^{1/{\tau_t}},\tau_t)$,
	$\Phi(\omega_t)$ is the cdf of the standard normal random variable evaluated at $\omega_t$.
	
	The expected value of the random variable $Y_t$ is given by 
	\begin{equation} \label{eq:SEP2_E(Y)}
	E(Z_t) = \frac{ 2 {\tau_t}^{{1}/{\tau_t}} \nu_t }{\sqrt{\pi} \Gamma(\frac{1}{\tau_t}) (1+\nu_t^2)^{2/\tau_t + 0.5}  }  
	\sum^{\infty}_{n=0} \frac{\Gamma(\frac{2}{\tau_t} + n + 0.5)}{(2n+1)!!} \Big( \frac{2\nu_t^2}{1+\nu_t^2} \Big)^n
	\end{equation}
	where
	$(2n+1)!! = 1.3.5...(2n-1)$.
	Note, when calculating the summation term, we set a cap of $n=10$ iterations is set, which was found to provide sufficient convergence.
	
	
	\subsection{ST2 Distribution} \label{app:ST2}
	The probability density function of skew t distribution type 2 distribution \citep{azzalini2003distributions} is
	\begin{equation*}
	{f}_Y(y_t|\mu_t,\sigma_t,\nu_t,\tau_t) = \frac{2}{\sigma} f_{Z_1}(z_t) F_{Z_2}(w_t) 
	\end{equation*}
	where
	$w_t = \nu_t \lambda_t^{0.5} z_t$,
	$\lambda_t = \frac{\tau_t+1}{\tau_t + z_t^2}$,
	$Z_1 \sim TF(0,1,\tau_t)$,
	$Z_2 \sim TF(0,1,\tau_t+1)$,
	$TF(\cdot)$ is a t distribution with $\tau_t > 0$ degrees of freedom.
	
	The expected value of the random variable $Y_t$ is given by 
	\begin{equation} \label{eq:ST2_E(Y)}
	E(Z_t) = \frac{\nu_t \tau_t^{0.5} \Gamma(\frac{\tau_t - 1}{2}) } {\pi^{0.5}(1+\nu_t^2)^{0.5}\Gamma(\frac{\tau_t}{2})}
	\end{equation}
	where kurtosis $\tau_t > 1$.
	Note, the expected value $E(Z_t)$ was set to 0 any time that $(\tau_t-1) < \text{tol}$, where $\text{tol}$ was set to $0.05$, in order to limit large outputs from $\Gamma(n)$ when $n \rightarrow 0$. 
	
	\subsection{ST5 Distribution} \label{app:ST5}
	The probability density function of skew t distribution type 5 distribution \citep{jones2003skew} is
	\begin{equation*}
	{f}_Y(y_t|\mu_t,\sigma_t,\nu_t,\tau_t) = \frac{c_t}{\sigma_t} \Bigg( 1 + \frac{z_t}{(a_t+b_t+z_t^2)^{0.5}} \Bigg)^{a_t+0.5} \Bigg( 1 - \frac{z_t}{(a_t+b_t+z_t^2)^{0.5}} \Bigg)^{b_t+0.5} 
	\end{equation*}
	where
	$c=\big( 2^{a_t+b_t-1}(a_t+b_t)^{0.5} B(a_t,b_t) \big)^{-1}$, 
	$\nu_t = \frac{(a_t-b_t)}{\big(a_t b_t (a_t+b)_t\big)^{0.5}}$,
	$\tau_t=\frac{2}{(a_t+b_t)}$,
	$B(a_t,b_t)$.
	
	The expected value of the random variable $Y_t$ is given by 
	\begin{equation} \label{eq:ST5_E(Y)}
	E(Z_t) =\frac {(a_t-b_t)(a_t+b_t)^{0.5} \Gamma(a_t-0.5)\Gamma(b_t-0.5)}{2 \Gamma(a_t)\Gamma(b_t)}
	\end{equation}
	Note: this corrects a typo in the specification of $E(Z_t)$ of the Manual \citep{rigby2009flexible,stasinopoulos2014instructions}, which specifies $\Gamma(a_t-0.5)\Gamma(a_t-0.5)$.
	
	Once the parameters $\nu_t$ and $\tau_t$ are estimated, re-arranging the system of two equations (see Appendix \ref{App:ST5_syst_eq} for derivation) results in the following expressions for $a_t$ and $b_t$
	\begin{align}
	a_t    &= \frac{2}{\tau_t} - b_t \\
	\nu_t &= \frac{ 2(1-b_t \tau_t) } { \big(2b_t(2-b_t \tau_t)\big)^{0.5} }
	\end{align}
	where $b_t$ is found first using Newton-Raphson root searching algorithm with the R function $\mathtt{uniroot}$ (for certain values of $\nu_t,\tau_t$ the root of this function does not exist, in this case we set $b_t=0$).
	Note: in order to limit large outputs from $\Gamma(n)$ when $n \rightarrow 0$, a restriction was imposed on values of $a_t,b_t$ to be positive and for $(a_t-0.5) < \text{tol}$, $(b_t-0.5) < \text{tol}$, where $\text{tol} = 0.05$, and if this condition was not met we set $E(Z_t) = 0$.
	
	\subsubsection{ST5 System of Equations} \label{App:ST5_syst_eq}
	The parameters $a$ and $b$ are derived by rearranging the system of two equations, where $\nu$ and $\tau$ are known.
	\begin{align}
	\nu  &= \frac{(a-b)}{\big(ab(a+b)\big)^{0.5}}  \label{eq:ST5_nu}  \\
	\tau &= \frac{2}{(a+b)} \label{eq:ST5_tau}
	\end{align}
	
	Hence, by rearranging Equation \ref{eq:ST5_tau}, $a$ equals:
	\begin{equation}
	a = \frac{2}{\tau} - b
	\end{equation}
	
	Substituting $a$ into Equation \ref{eq:ST5_nu} we obtain
	\begin{align}
	\nu  &= \frac{\frac{2}{\tau} - b -b} { \big[  b(\frac{2}{\tau} - b)(\frac{2}{\tau} - b + b)  \big]^{0.5} } \nonumber \\
	\nu  &= \frac{\frac{2}{\tau} - 2b} { \big[  b(\frac{2}{\tau} - b)(\frac{2}{\tau})  \big]^{0.5} } \nonumber \\
	\nu  &= \frac{ \frac{2(1- b\tau)}{\tau} } { \big[  b(\frac{2-b\tau}{\tau}) \frac{2}{\tau}  \big]^{0.5} } \nonumber \\
	\nu  &= \frac{ \frac{2(1- b\tau)}{\bcancel{\tau}} } { \big[  2b(\frac{2-b\tau}{ \bcancel{\tau} }) \frac{1}{\bcancel{\tau}}  \big]^{0.5} } \nonumber 
	\end{align}
	\begin{align}
	\nu  &= \frac{2(1- b\tau)} {\big[ 2b(2-b\tau)\big]^{0.5}} \label{eq:ST5_nu2}
	\end{align}
	
	Equation \ref{eq:ST5_nu2} is rearranged to have the form $f(x) = 0$ (see Equation \ref{eq:ST5_nu3}) and solved for $b$ using Newton-Raphson root searching algorithm. 
	\begin{equation} \label{eq:ST5_nu3}
	0  = \frac{2(1- b\tau)} {\big[ 2b(2-b\tau)\big]^{0.5}} - \nu
	\end{equation}

	\section{Electricity Prices - Preliminary Data Analysis}
	\subsection{Prices Histograms, Yearly Distributions}
	Day ahead electricity price data is plotted for a selection of hours, showing the evolution of fitted distribution over years 2012-2017..
	The distribution of choice 'ST5' was used to fit the histogram data with GAMLSS library. 
	Note that the price of 0 is highlighted with a vertical black line. 
	\begin{figure}[h!]
		\centering
		\includegraphics[width=0.7\textwidth]{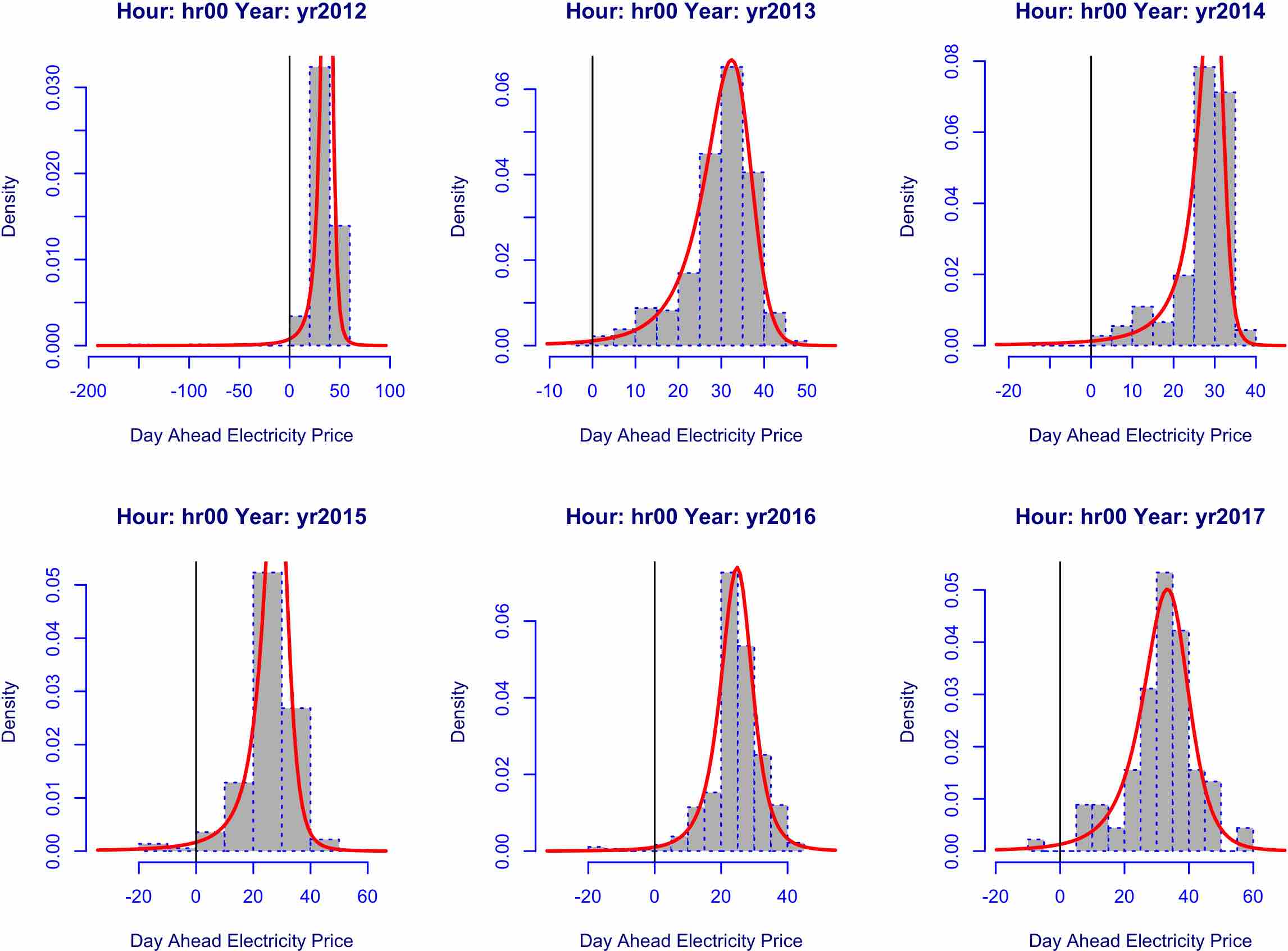}
		\caption{Electricity price for hr 00; years 2012-2017.}
		\label{fig:App1_histHr00_spotPrice}
	\end{figure}
	\begin{figure}[h!]
		\centering
		\includegraphics[width=0.7\textwidth]{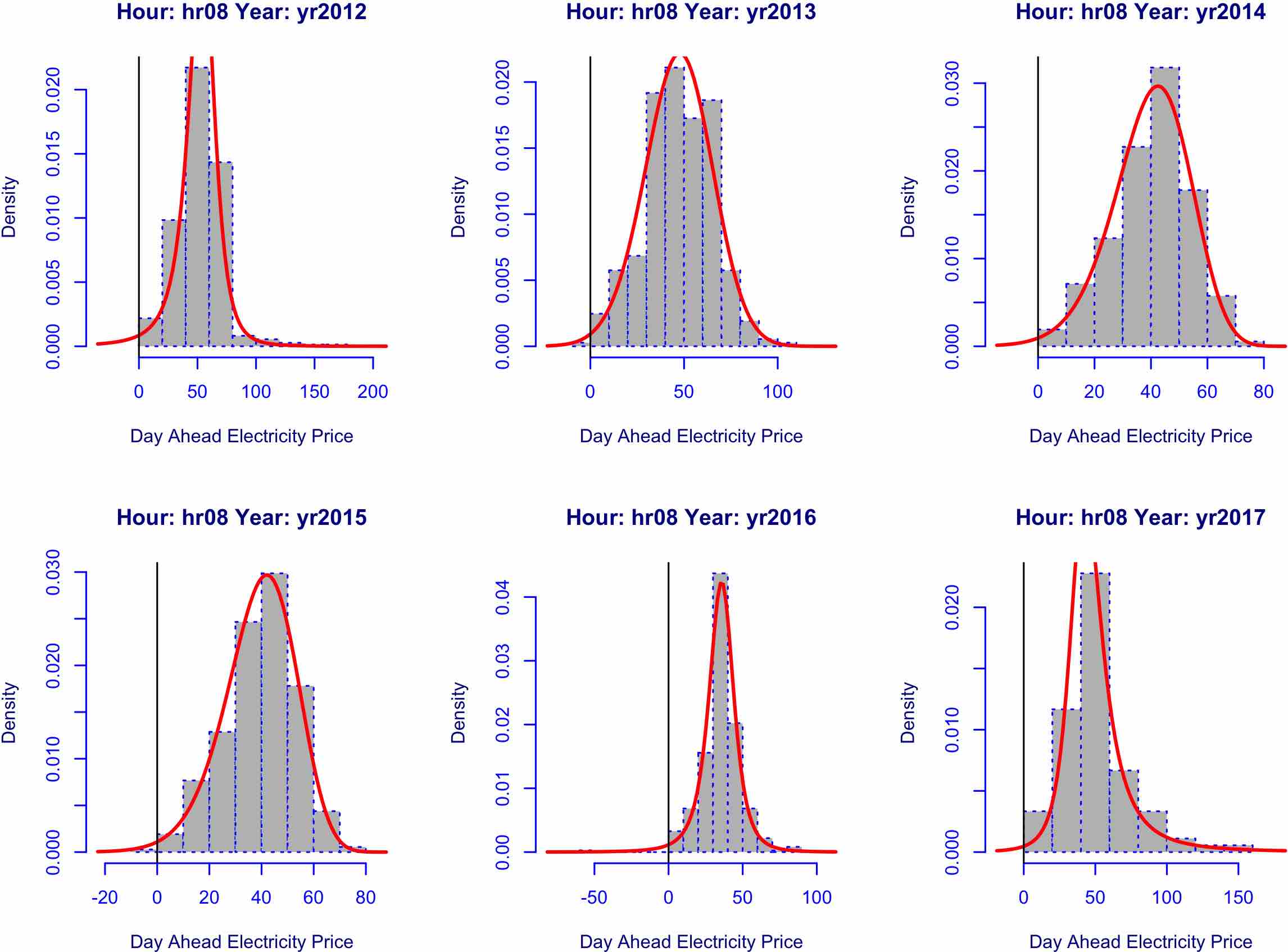}
		\caption{Electricity price for hr 08; years 2012-2017.}
		\label{fig:App1_histHr08_spotPrice}
	\end{figure}
	\begin{figure}[h!]
		\centering
		\includegraphics[width=0.7\textwidth]{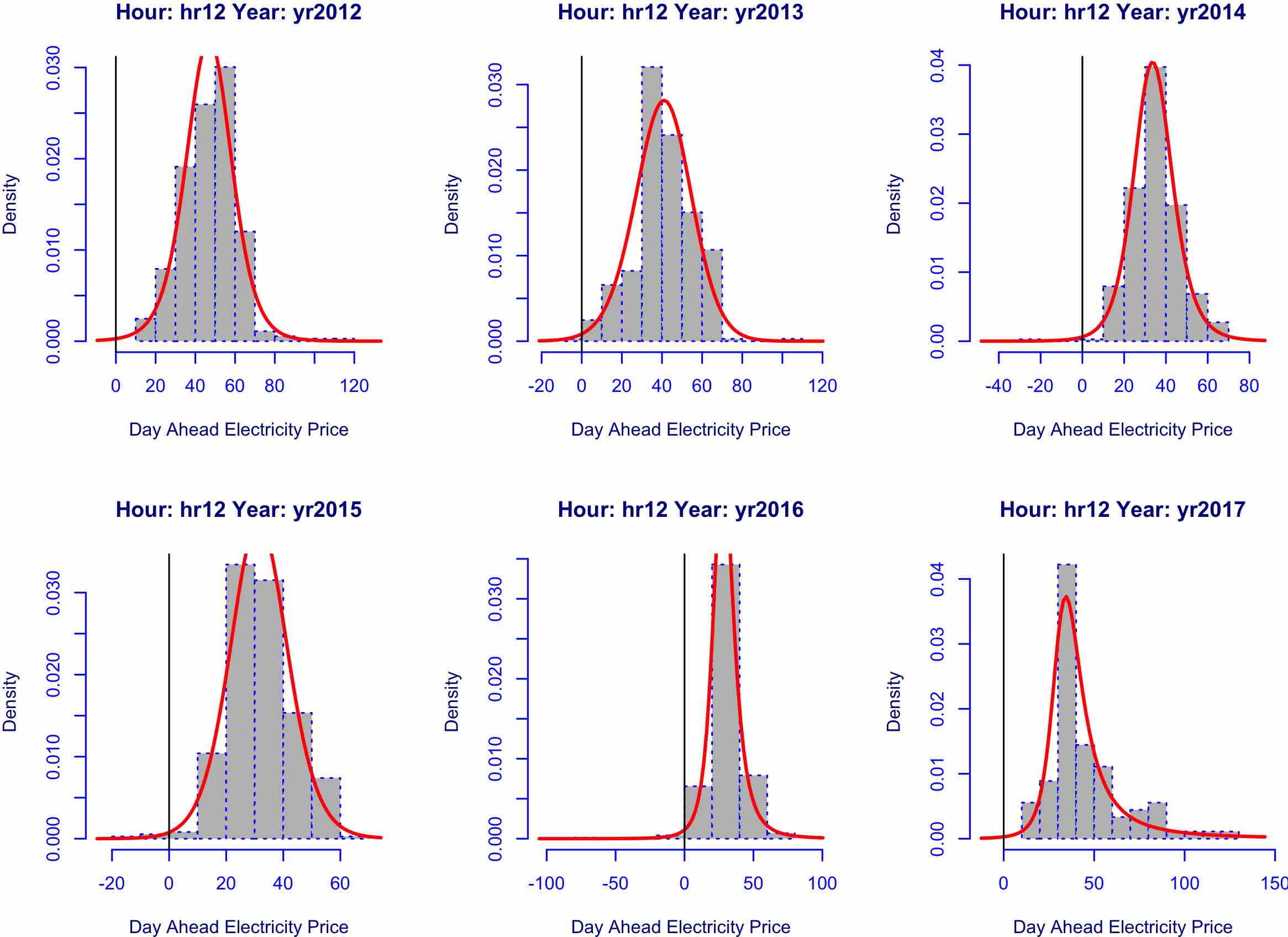}
		\caption{Electricity price for hr 12; years 2012-2017.}
		\label{fig:App1_histHr12_spotPrice}
	\end{figure}
	\begin{figure}[h!]
		\centering
		\includegraphics[width=0.7\textwidth]{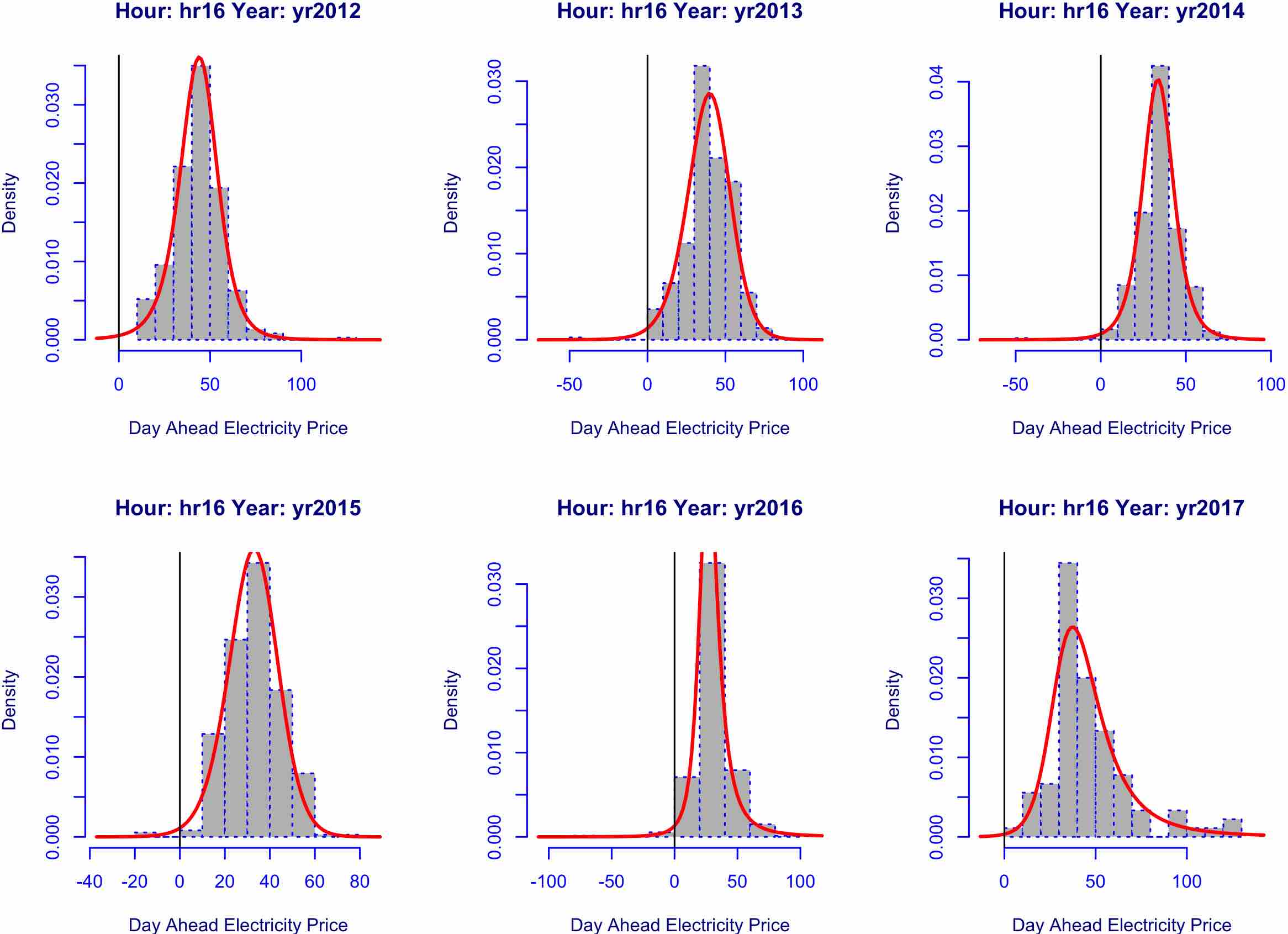}
		\caption{Electricity price for hr 16; years 2012-2017.}
		\label{fig:App1_histHr16_spotPrice}
	\end{figure}
	\begin{figure}[h!]
		\centering
		\includegraphics[width=0.7\textwidth]{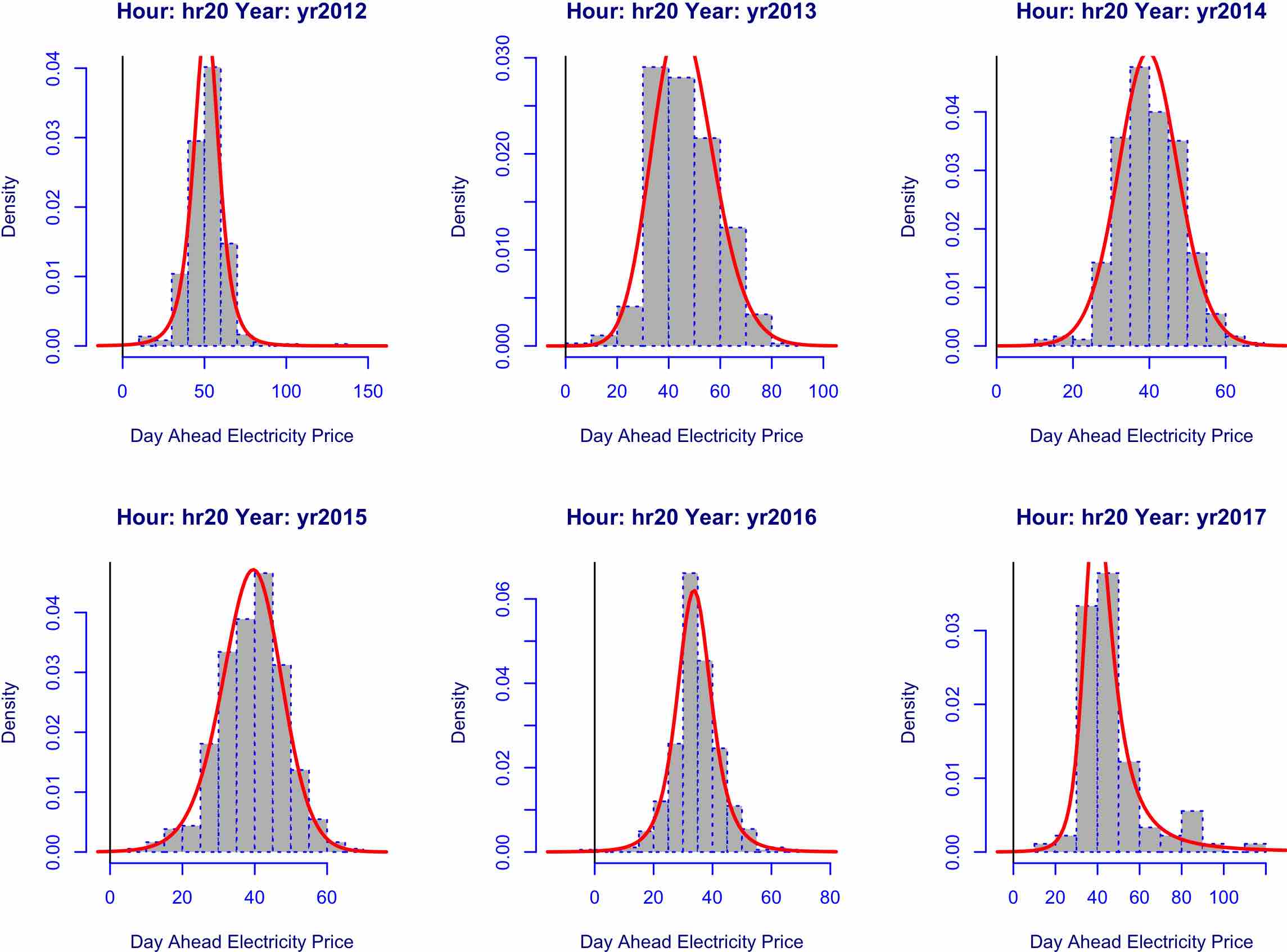}
		\caption{Electricity price for hr 20; years 2012-2017.}
		\label{fig:App1_histHr20_spotPrice}
	\end{figure}
	
	\subsection{Prices Joint Distribution, Variance-Covariance Matrix}
\begin{figure}[h!]
		\centering
		\includegraphics[width=1.0\textwidth]{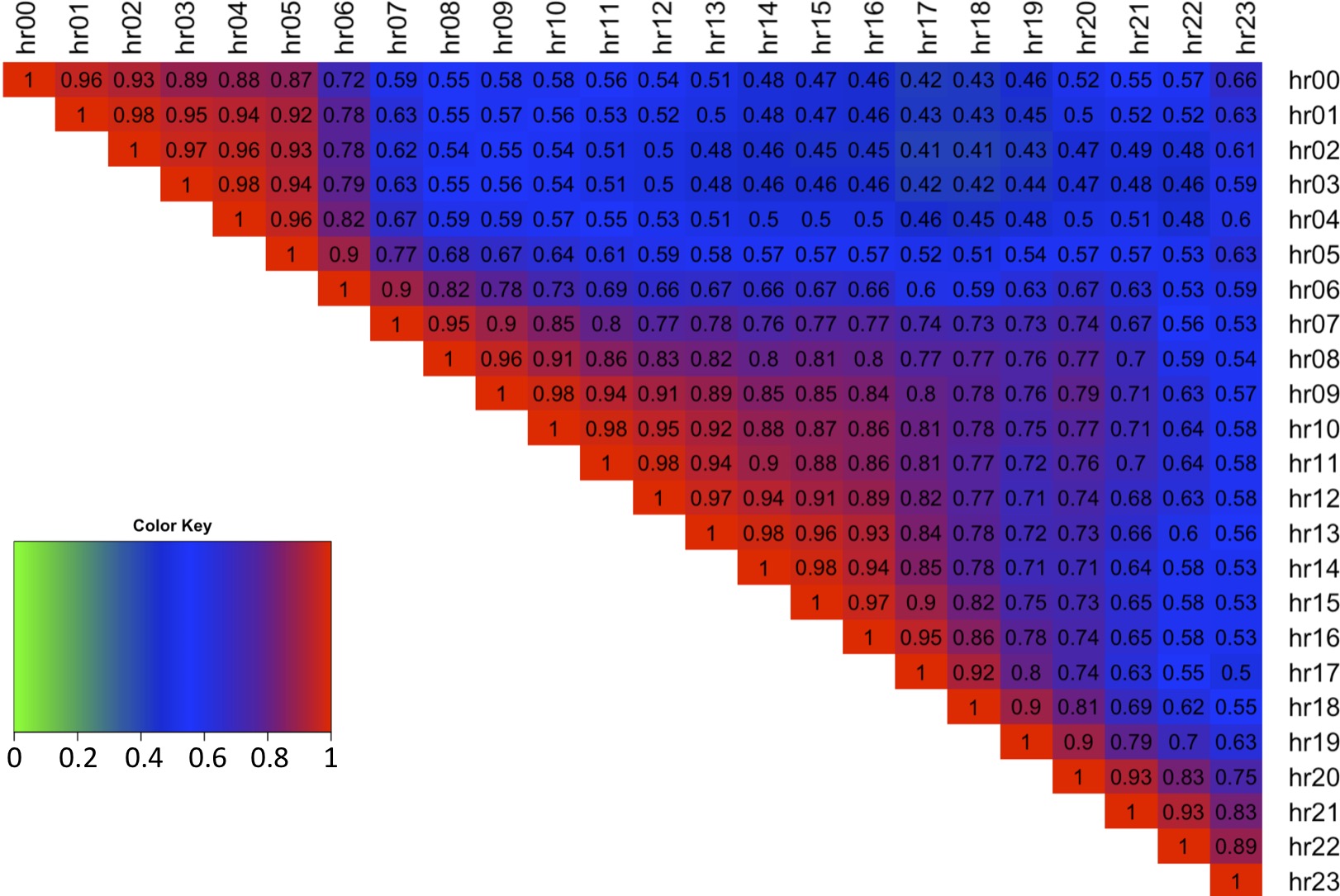}
		\caption{Variance-covariance matrix of intraday electricity prices.}
		\label{fig:App1_prices_VarCov}
	\end{figure}

	In order to be able to take differences between forecasted prices the assumption of independence for joint hourly price distribution needs to be satisfied; i.e. the joint distribution variance-covariance matrix of hourly price data should have 0 off-diagonal covariances.
	The variance-covariance matrix for jointly distributed intraday electricity prices (all data) is displayed in Figure \ref{fig:App1_prices_VarCov}.
	Strong positive covariances between prices at different hour of the day can be seen, with the smallest value of 0.41.
	Therefore intraday hourly prices are therefore not independent and the approach of modelling individual hours and then taking differences to establish the most profitable spreads would not be correct. 

	\clearpage
	\section{Electricity Spreads - Preliminary Data Analysis}
	\subsection{Spreads Stationarity Test} \label{App:subsec_ADF_spreads}
	The spreads data was tested using the Augmented Dickey Fuller (\texttt{adf.test()}) test with 10 lags and a linear trend.
	All of the spreads are stationary at 1\% significance level. 
	
	\subsection{Spreads Histograms, Yearly Distributions} 
	\begin{figure}[h!]
		\centering
		\includegraphics[width=0.8\textwidth]{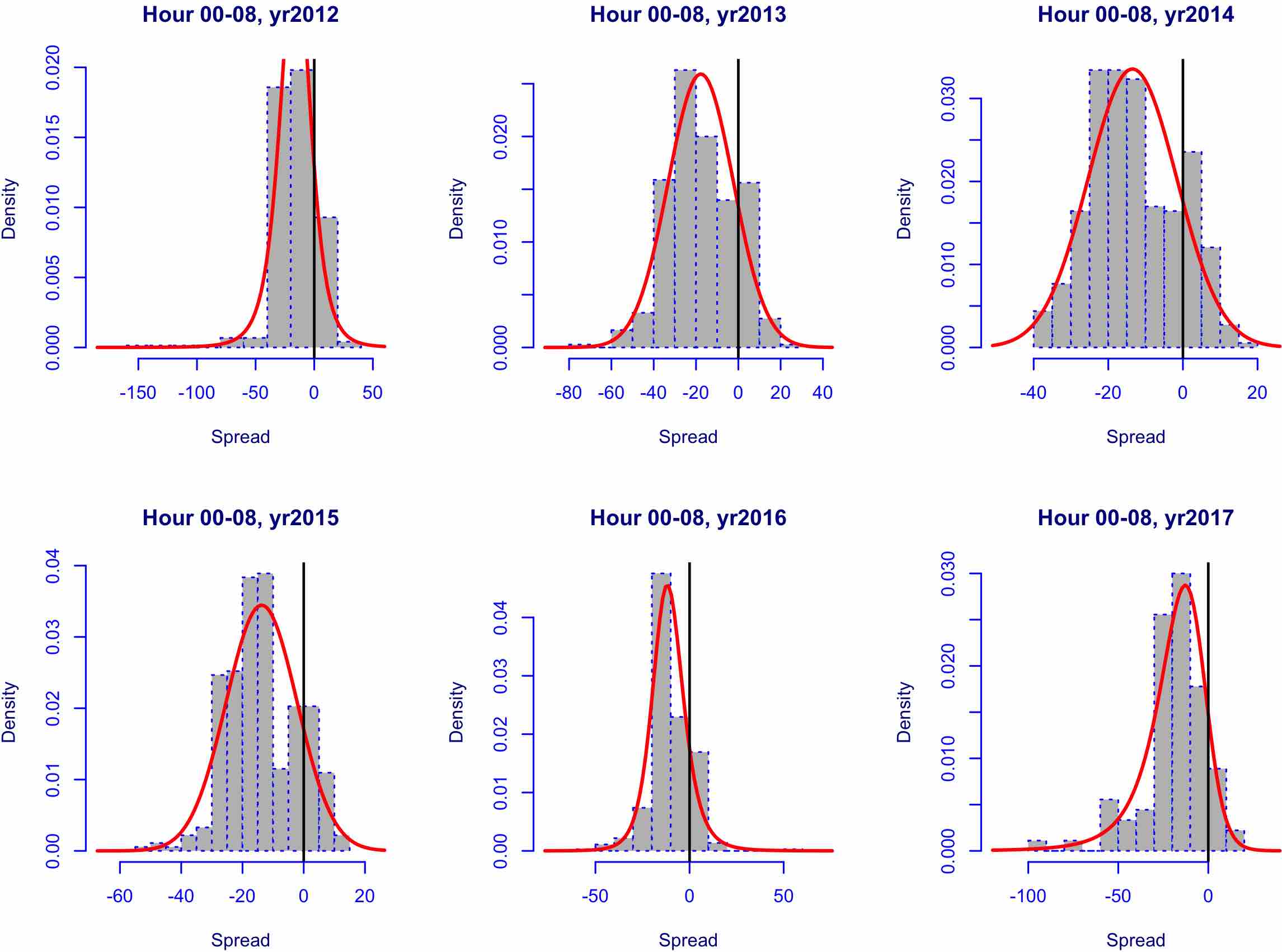}
		\caption{Spread for hr 00-08 of electricity prices; years 2012-2017.}
		\label{fig:App1_histHr00-08}
	\end{figure}
	\begin{figure}[h!]
		\centering
		\includegraphics[width=0.8\textwidth]{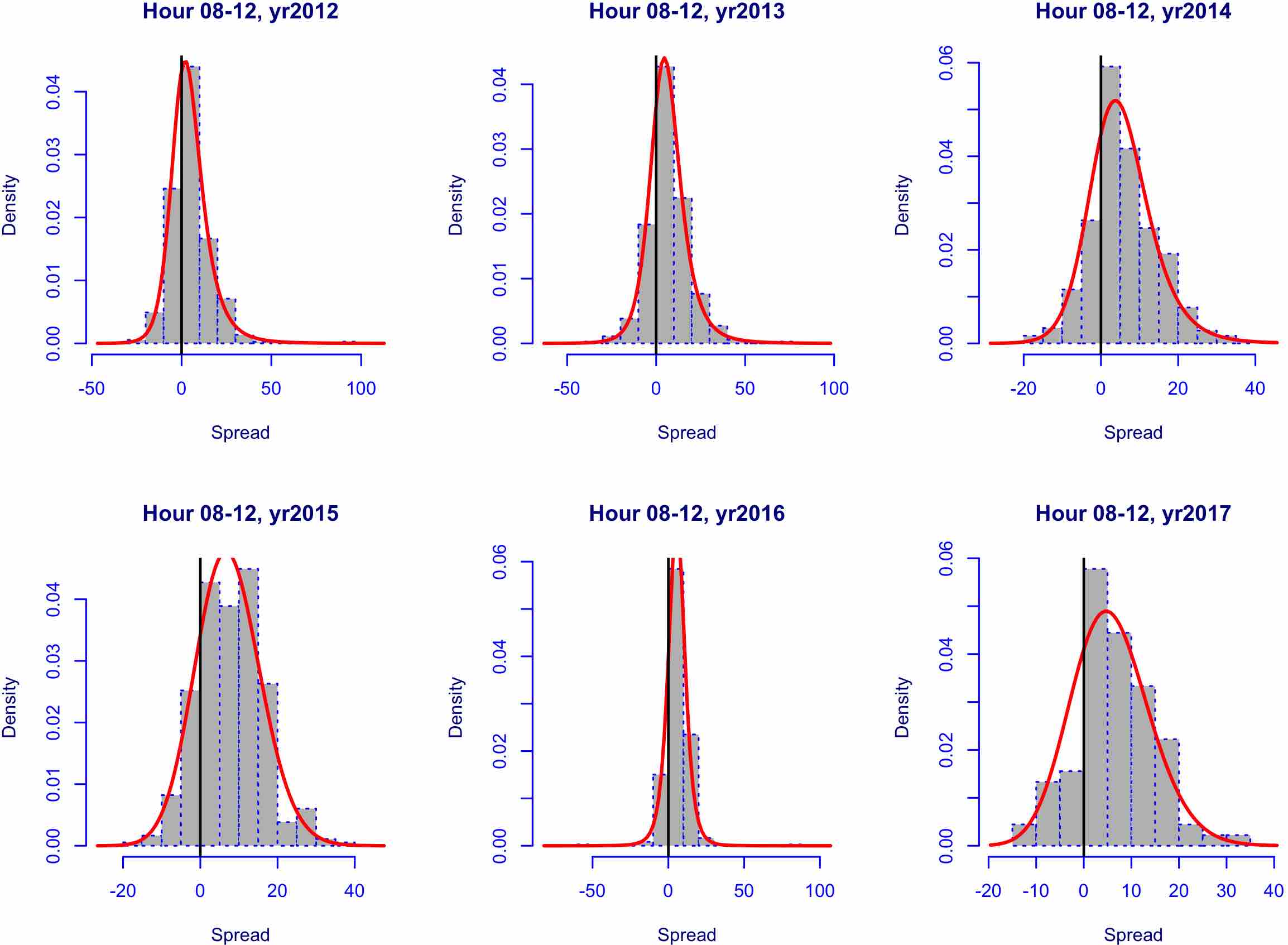}
		\caption{Spread for hr 08-12 of electricity prices; years 2012-2017.}
		\label{fig:App1_histHr08-12}
	\end{figure}
	\begin{figure}[h!]
		\centering
		\includegraphics[width=0.8\textwidth]{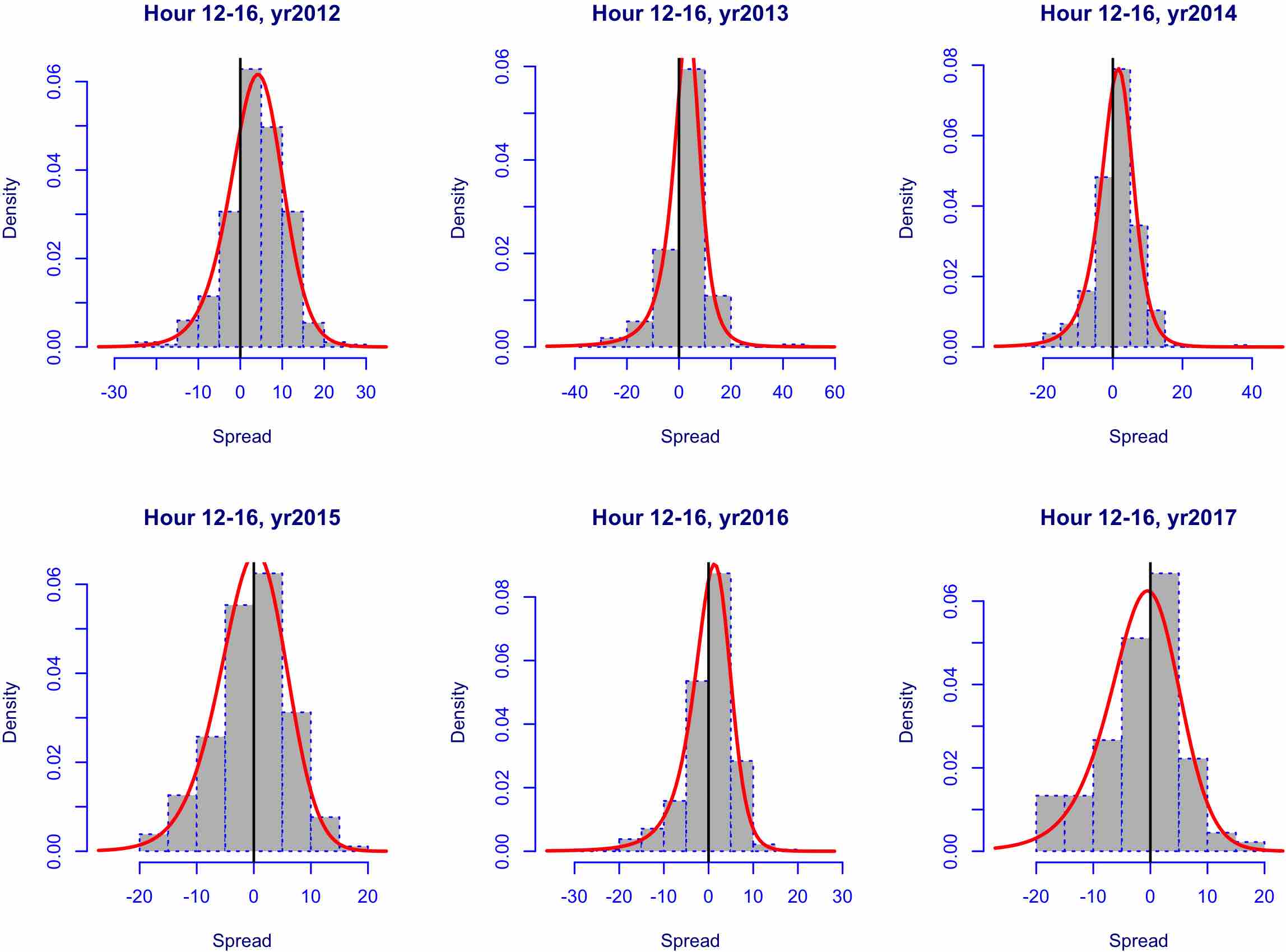}
		\caption{Spread for hr 12-16 of electricity prices; years 2012-2017.}
		\label{fig:App1_histHr12-16}
	\end{figure}
	\begin{figure}[h!]
		\centering
		\includegraphics[width=0.8\textwidth]{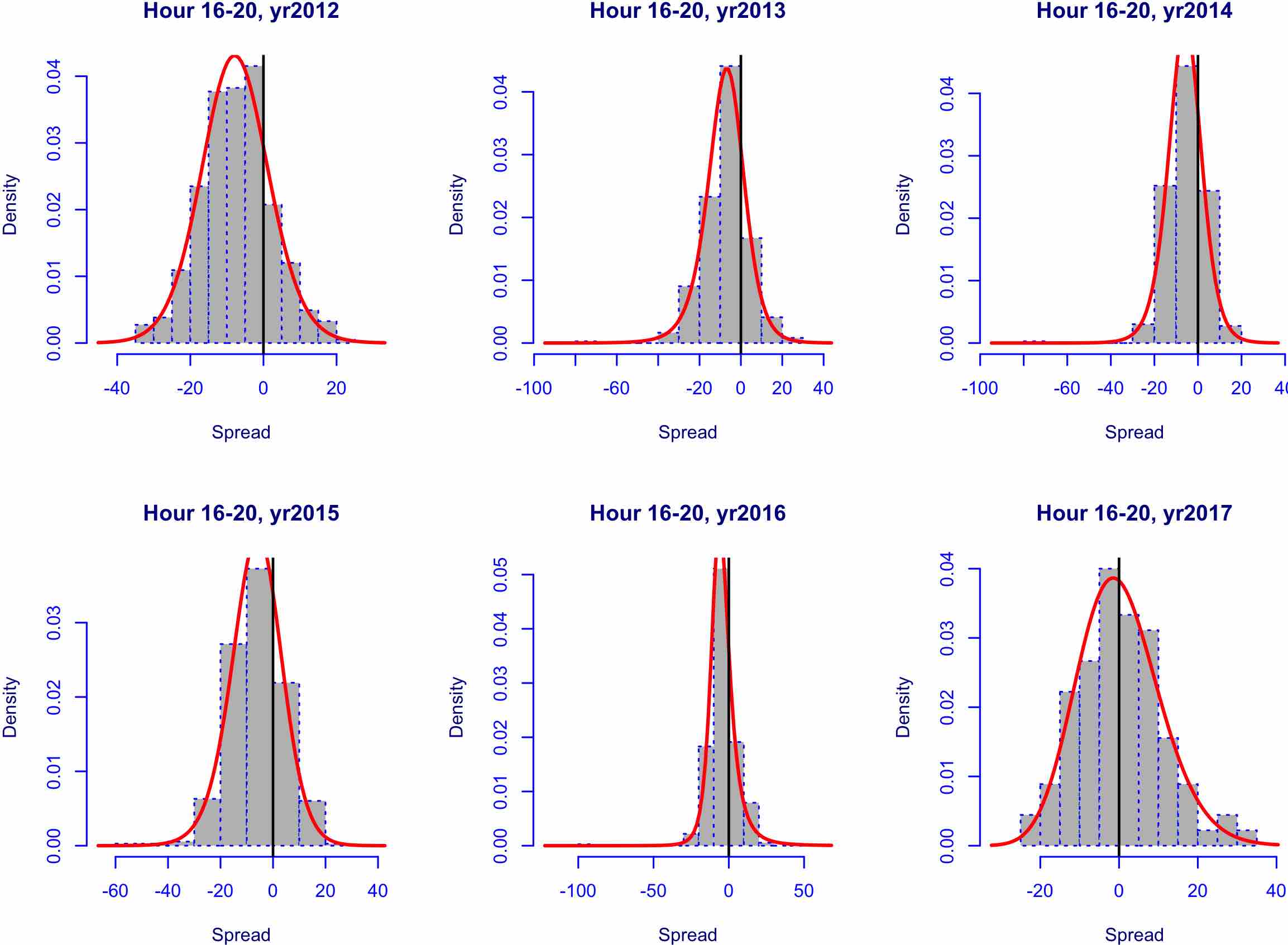}
		\caption{Spread for hr 16-20 of electricity prices; years 2012-2017.}
		\label{fig:App1_histHr16-20}
	\end{figure}
	
	The histograms for 5 different intraday electricity price spreads (00am-08am, 08am-12noon, 12noon-16, 16-20) are plotted and summarised by year 2012 - 2017.
	The figures depict ongoing changes in distribution of electricity spot price over the last 5 years, and highlight the presence of strong skewness and show the variation between distribution shapes for different hours of the day.
	Next the skewness of the spread data is analysed.

	\clearpage
	\subsection{Spreads, Higher Moments Analysis} 
\begin{figure}[h!]
		\centering
		\includegraphics[width=1.0\textwidth]{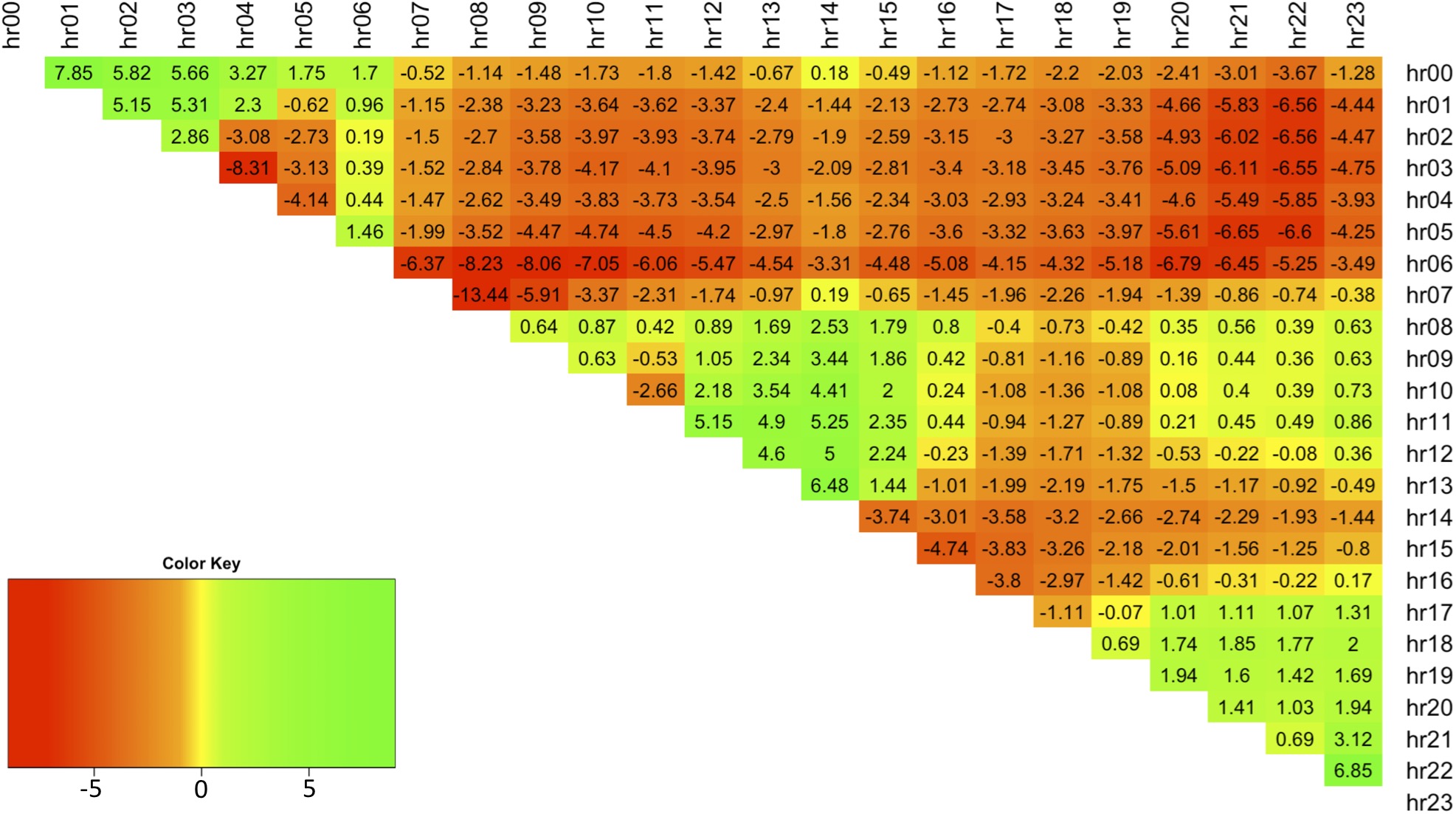}
		\caption{Skewness for each electricity hour spread.}
		\label{fig:App1_spreads_skewness}
	\end{figure}

	The spread prices for all hours have skewness in the range [-8.31, 7.85] (see Figure \ref{fig:App1_spreads_skewness} - note data is symmetrical about the diagonal thus only upper diagonal is plotted).
	The plot shows that typically the spreads are negatively skewed (note spreads are obtained by $hr_{earlier} - hr_{later}$).
	
\begin{figure}[h!]
		\centering
		\includegraphics[width=1.0\textwidth]{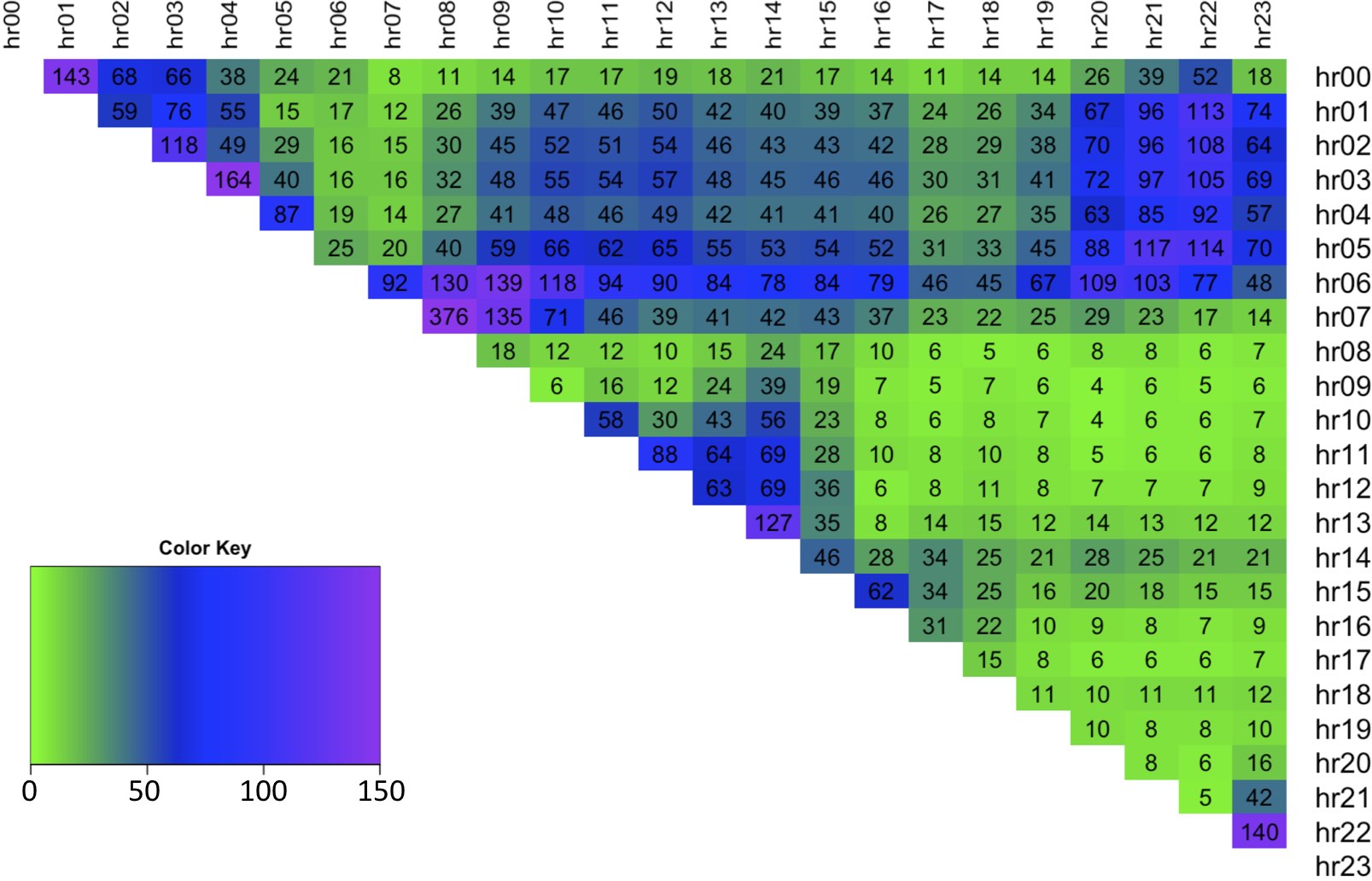}
		\caption{Excess kurtosis for each electricity hour spread.}
		\label{fig:App1_spreads_kurtosis}
	\end{figure}

	Figure \ref{fig:App1_spreads_kurtosis} display kurtosis for all spreads.
	The spreads between all hours possess positive excess kurtosis and thus indicate that spreads have Leptokurtic distributions. 
	The minimum excess kurtosis is 3.8, while the maximum is 140.

	\clearpage
	\section{Spreads, Best Distribution Fit (Simple)}
\begin{figure}[h!]
		\centering
		\includegraphics[width=1.0\textwidth]{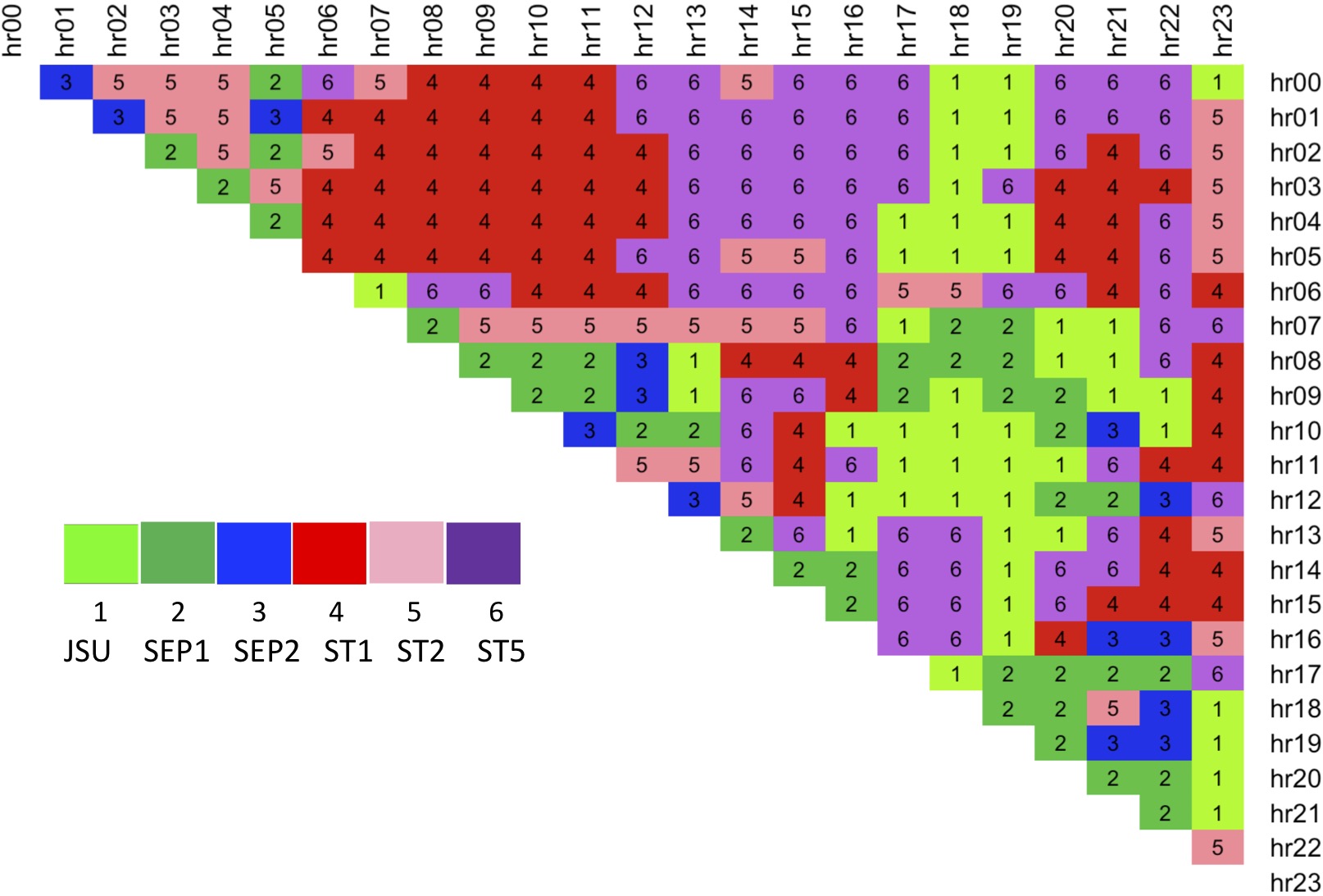}
		\caption{Continuous four parameter distribution chosen as 'best fit' to training spread data using \texttt{gamlss <- $\mathbf{y} \sim 1$} and the AIC criterion.}
		\label{fig:App1_spreads_bestDistSimple}
	\end{figure}

	The parametric distribution fit is performed using GAMLSS function \texttt{gamlss <- $\mathbf{y} \sim 1$} fitted to spread price data.
	Note: upon occasion GAMLSS failed to fit the distribution to the spreads data, in which case that distribution  was omitted for that spread. 
	This problem was encountered for spread hours and distributions: 
	00-01 SEP1; 
	03-04 SEP2;
	12-13 SEP1;
	The best distribution is selected for each spread and plotted (see Figure \ref{fig:App1_spreads_bestDistSimple}), where the following colour coding is applied:
	light green (1) - JSU;
	green (2) - SEP1;
	blue (3) - SEP2;
	red (4) - ST1;
	pink (5) - ST2;
	purple (6) - ST5;

	\clearpage
	\section{Spreads - Best Distribution Fit (Factor-based)} \label{app:DistnFit_SpreadsAdv}
	The following spreads hours were not learnt for distributions:\\
	\textbf{JSU} 21-22\\
	\textbf{SEP1} 00-02; 00-20; 01-02; 01-09; 01-17; 01-21; 01-22; 02-03; 02-04; 02-05; 02-09; 02-10; 02-17; 02-20; 02-21; 02-23; 02-07; 03-08; 03-21; 04-07; 04-10; 04-21; 05-07; 05-09; 05-21; 05-22; 06-07; 06-08; 06-09; 06-10;  06-17; 06-20; 06-22; 07-08; 07-09; 07-10; 12-14; 13-14; 13-21; 21-23.\\
	\textbf{SEP2} 01-02; 01-07; 01-21; 02-03; 02-04; 02-05; 02-18; 02-21; 02-22; 03-04; 03-07; 03-21; 04-07; 04-21; 05-07; 05-09; 05-21; 06-09; 06-10; 06-22; 07-08; 07-10; 07-21; 07-22; 12-13; 13-21; 21-22.\\
	\textbf{ST1} 01-02; 02-03; 02-06; 05-21; 06-20; 13-21. \\
	\textbf{ST2} 02-03. \\
	\textbf{ST5} 01-02; 02-03.

	\section{Training Spread Data Fit}
\begin{figure}[h!]
		\centering
		\includegraphics[width=1.0\textwidth]{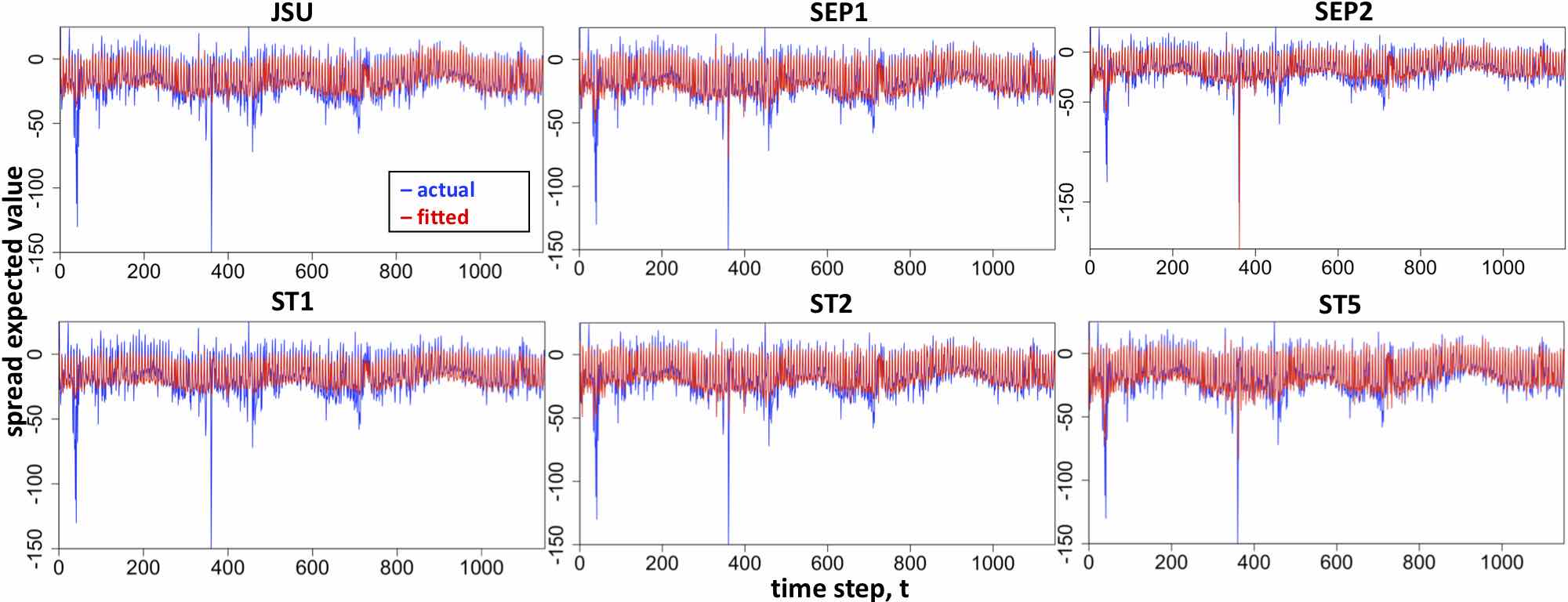}
		\caption{\textbf{Spread Hour 00-08 Exp. Value Plot} - True $E(\mathbf{Y}^{(s)})$ vs fitted spread, $E(\mathbf{\widehat{Y}}^{(s)})$.}
		\label{fig:app1_muTrain_00_08}
	\end{figure}

\begin{figure}[h!]
		\centering
		\includegraphics[width=1.0\textwidth]{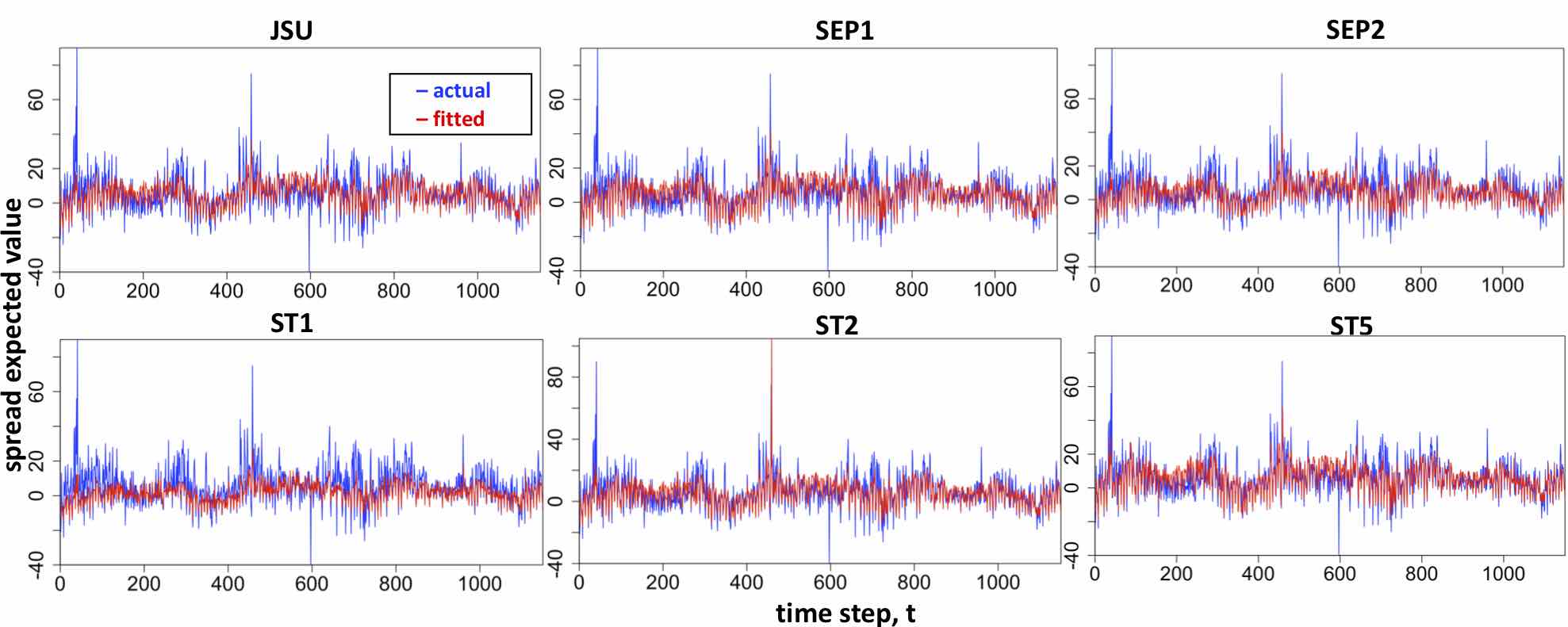}
		\caption{\textbf{Spread Hour 08-12 Exp. Value Plot} - True $E(\mathbf{Y}^{(s)})$ vs fitted spread, $E(\mathbf{\widehat{Y}}^{(s)})$.}
		\label{fig:app1_muTrain_08_12}
	\end{figure}

	The expected value fit over the training data $t=1,...,1150$ is plotted for 00-08 and 08-12 spread hours using models built with the six continuous distributions.
	The plots show a good fit to the data with slight underestimation (see Figures \ref{fig:app1_muTrain_00_08} and \ref{fig:app1_muTrain_08_12}).

	\section{Validation Spread Data Fit} \label{app:valDataFit}
\begin{figure}[h!]
		\centering
		\includegraphics[width=1.0\textwidth]{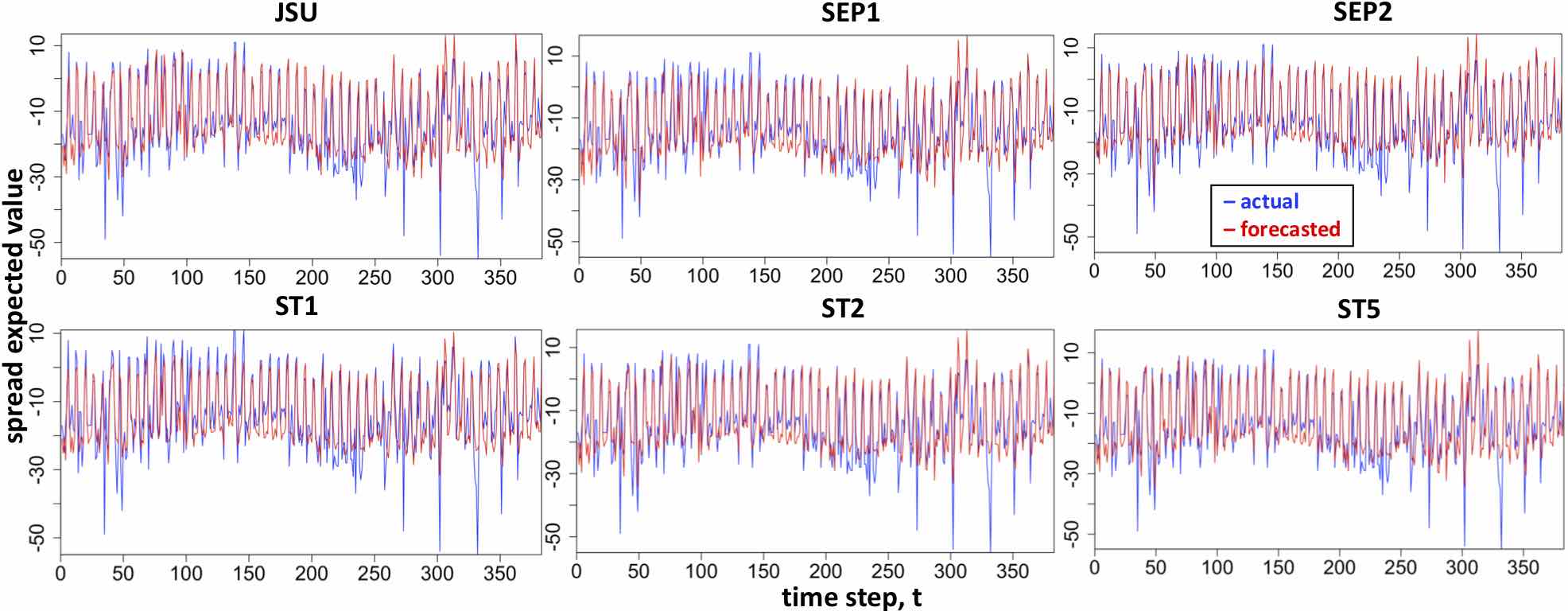}
		\caption{\textbf{Spread Hour 00-08 Exp. Value Plot} - True $E(\mathbf{Y}^{(s)})$ vs forecasted spread, $E(\mathbf{\widehat{Y}}^{(s)})$.}
		\label{fig:app1_muVal_00_08}
	\end{figure}

\begin{figure}[h!]
		\centering
		\includegraphics[width=1.0\textwidth]{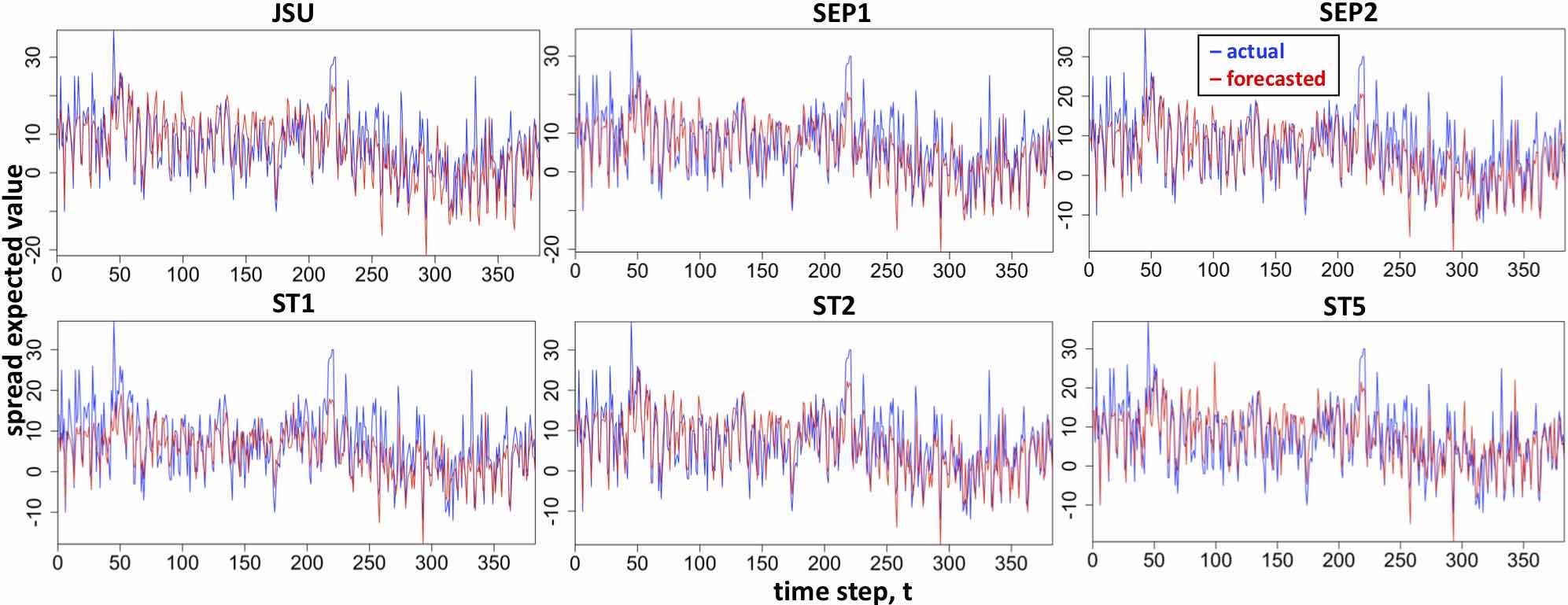}
		\caption{\textbf{Spread Hour 08-12 Exp. Value Plot} - True $E(\mathbf{Y}^{(s)})$ vs forecasted $E(\mathbf{\widehat{Y}}^{(s)})$ spread.}
		\label{fig:app1_muVal_08_12}
	\end{figure}

	The expected value fit over the validation data $t=1151,...,1534$ is plotted for 00-08 and 08-12 spread hours using models built with the six continuous distributions (see Figures \ref{fig:app1_muVal_00_08} and \ref{fig:app1_muVal_08_12}).
	
	The results show that the predicted spread prices (using the unseen validation data set) is of a closer fit than during the training phase. 
	The difference in the predictive power of different distributions is evident, since the predicted values differ.

	\section{Spreads - Best Distribution Fit (Factor-based)} \label{app:rmse_bestDist}
\begin{figure}[h!]
		\centering
		\includegraphics[width=0.8\textwidth]{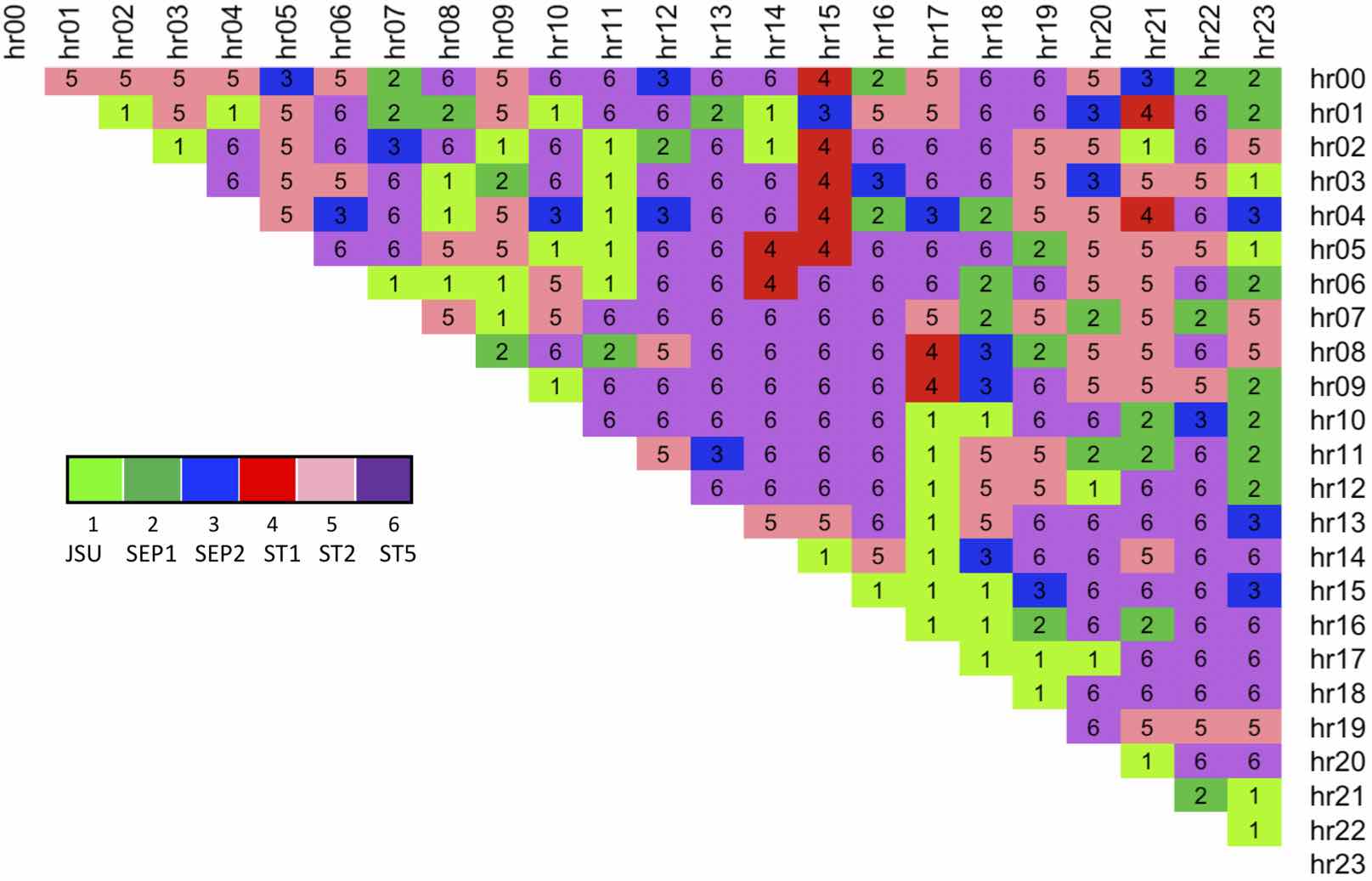}
		\caption{Best dist. for forecasting exp. value of validation spread, $E(Y_t)$, based on lowest RMSE.}
		\label{fig:app1_muVal_RMSEbasedBestDistn}
	\end{figure}

\begin{figure}[h!]
		\centering
		\includegraphics[width=0.9\textwidth]{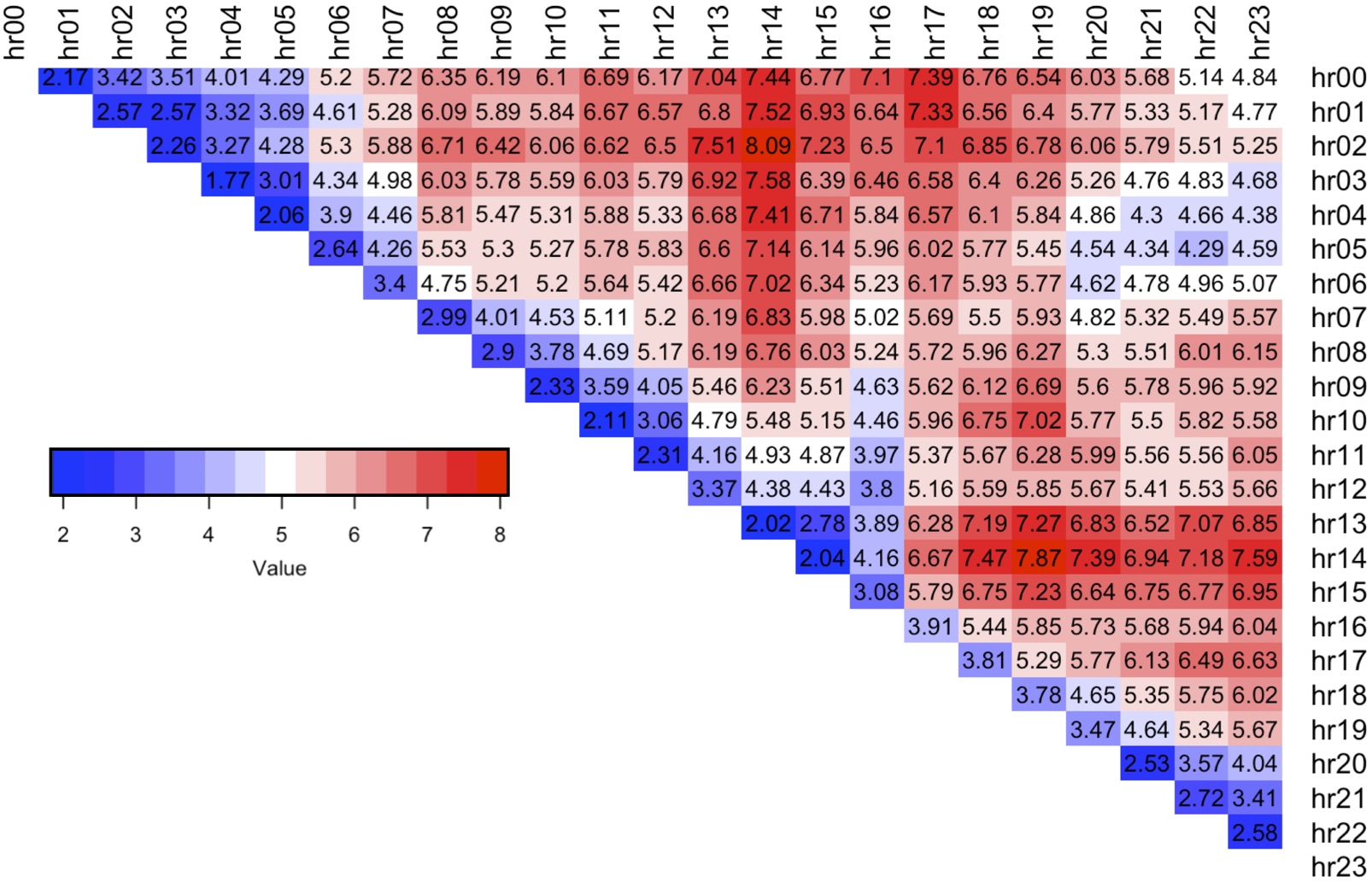}
		\caption{RMSE of selected best dist. for forecasting expected value, $E(Y_t)$, validation data.}
		\label{fig:app1_muVal_RMSEbasedBestDistn-ActualRMSE}
	\end{figure}

	An RMSE based calculation was performed for selecting the best distribution for each spread using the forecasting performance over the validation data set.
	The distribution resulting in the lowest RMSE was selected as the best and the results for each spread are plotted in Figure \ref{fig:app1_muVal_RMSEbasedBestDistn}. 
	The actual RMSE values corresponding to the best distributions are shown in Figure \ref{fig:app1_muVal_RMSEbasedBestDistn-ActualRMSE}, with blue indicating the lowest error.

	\section{Validation Stage - Quantile Estimation}
	The number of times quantiles were not obtained for the $D^{i}, i=1,..,6$ distributions during the validation stage is given in Table \ref{app_tab:failedQuantilesValidationData}.
	The table shows that ST1 and ST5 distrubitons struggled to extract quantiles the most.
	\begin{table}[h!]
		\centering 
		\begin{tabular}{ |l|l|l|l|l|l| }  
			\hline
			\textbf{JSU} & \textbf{SEP1} & \textbf{SEP2} & \textbf{ST1} & \textbf{ST2} & \textbf{ST5} \\ 
			\hline
			$0$ & $19$ & $4$ & $176$ & $147$ & $0$ \\
			\hline
		\end{tabular}
		\caption{Distributions and number of times for which quantile estimates at $t$ were not obtained.} \label{app_tab:failedQuantilesValidationData}
	\end{table}
	
	The detailed information on the distributions for which the R function providing quantile measures did not converge to a solution are given below:\\
	\textbf{Hrs 00-05}: distributions ST1 (1 time step) and ST2 (1 time step) \\
	\textbf{Hrs 00-11}: distribution ST1 (19 time steps) \\
	\textbf{Hrs 00-15}: distribution ST2 (2 time steps) \\
	\textbf{Hrs 00-23}: distributions ST1 (4 time steps) and ST2 (6 time steps)\\
	\textbf{Hrs 01-04}: distributions SEP1 (2 time steps), ST1 (3 time steps) and ST2 (4 time steps)\\
	\textbf{Hrs 01-07}: distribution ST2 (1 time step) \\
	\textbf{Hrs 01-12}: distributions ST1 (4 time steps) and ST2 (5 time steps)\\
	\textbf{Hrs 01-15}: distribution ST1 (2 time steps) \\
	\textbf{Hrs 01-16}: distribution ST1 (6 time steps) \\
	\textbf{Hrs 01-12}: distributions ST1 (4 time steps) and ST2 (5 time steps)\\
	\textbf{Hrs 01-23}: distributions ST1 (3 time steps) and ST2 (1 time step)\\
	\textbf{Hrs 02-04}: distributions ST1 (2 time steps) and ST2 (2 time steps)\\
	\textbf{Hrs 02-12}: distributions ST1 (3 time steps) and ST2 (3 time steps)\\
	\textbf{Hrs 02-23}: distributions ST1 (2 time steps) and ST2 (1 time step)\\
	\textbf{Hrs 03-04}: distribution ST1 (1 time step) \\
	\textbf{Hrs 03-12}: distributions ST1 (11 time steps) and ST2 (5 time steps)\\
	\textbf{Hrs 03-13}: distributions SEP1 (1 time step), ST1 (7 time steps) and ST2 (12 time steps)\\
	\textbf{Hrs 03-16}: distributions ST1 (9 time steps) and ST2 (12 time steps)\\
	\textbf{Hrs 04-12}: distributions ST1 (12 time steps) and ST2 (8 time steps)\\
	\textbf{Hrs 04-13}: distributions ST1 (21 time steps) and ST2 (20 time steps)\\
	\textbf{Hrs 04-14}: distribution ST1 (3 time steps)\\
	\textbf{Hrs 04-15}: distributions ST1 (6 time steps) and ST2 (4 time steps)\\
	\textbf{Hrs 05-07}: distributions ST1 (2 time steps) and ST2 (1 time step)\\
	\textbf{Hrs 05-12}: distributions SEP1 (2 time steps), ST1 (10 time steps) and ST2 (6 time steps)\\
	\textbf{Hrs 05-13}: distribution ST1 (19 time steps)\\
	\textbf{Hrs 05-14}: distribution SEP1 (1 time step)\\
	\textbf{Hrs 06-07}: distributions ST1 (2 time steps) and ST2 (2 time steps)\\
	\textbf{Hrs 06-11}: distribution ST1 (2 time steps)\\
	\textbf{Hrs 06-12}: distributions ST1 (2 time steps) and ST2 (1 time step)\\
	\textbf{Hrs 06-13}: distributions ST1 (2 time steps) and ST2 (2 time steps)\\
	\textbf{Hrs 06-15}: distribution ST1 (1 time step) \\
	\textbf{Hrs 08-19}: distribution ST1 (1 time step) \\
	\textbf{Hrs 09-10}: distributions ST1 (2 time steps) and ST2 (3 time steps)\\
	\textbf{Hrs 10-13}: distribution ST2 (1 time step)\\
	\textbf{Hrs 10-14}: distributions ST1 (1 time step) and ST2 (1 time step)\\
	\textbf{Hrs 11-13}: distribution SEP1 (2 time steps)\\
	\textbf{Hrs 11-14}: distribution SEP1 (1 time step)\\
	\textbf{Hrs 11-20}: distribution ST2 (1 time step)\\
	\textbf{Hrs 11-23}: distribution ST2 (1 time step)\\
	\textbf{Hrs 12-13}: distribution SEP1 (1 time step)\\
	\textbf{Hrs 12-15}: distribution SEP2 (1 time step)\\
	\textbf{Hrs 13-22}: distributions SEP1/SEP2/ST1/ST2 (1 time step each)\\
	\textbf{Hrs 14-17}: distributions SEP1 (1 time step)\\
	\textbf{Hrs 14-19}: distributions SEP1 (5 time steps)\\
	\textbf{Hrs 14-21}: distributions SEP1 (1 time step)\\
	\textbf{Hrs 14-22}: distributions SEP1/SEP2/ST1/ST2 (1 time step each)\\
	\textbf{Hrs 15-19}: distributions ST2 (12 time steps)\\
	\textbf{Hrs 15-20}: distributions ST2 (1 time step)\\
	\textbf{Hrs 15-21}: distributions SEP2 (1 time step)\\
	\textbf{Hrs 15-22}: distributions ST2 (1 time step)\\
	\textbf{Hrs 16-19}: distributions ST1/ST2 (1 time step each)\\
	\textbf{Hrs 16-20}: distributions ST1 (7 time steps) and ST2 (8 time steps)\\
	\textbf{Hrs 16-21}: distributions ST1 (1 time step) and ST2 (6 time steps)\\
	\textbf{Hrs 17-19}: distributions ST1/ST2 (1 time step each)\\
	\textbf{Hrs 17-20}: distributions ST2 (1 time step)

	\newpage
	\section{Rolling-Window RMSE - Skew Based}
\begin{figure}[h!]
		\centering
		\includegraphics[width=1.0\textwidth]{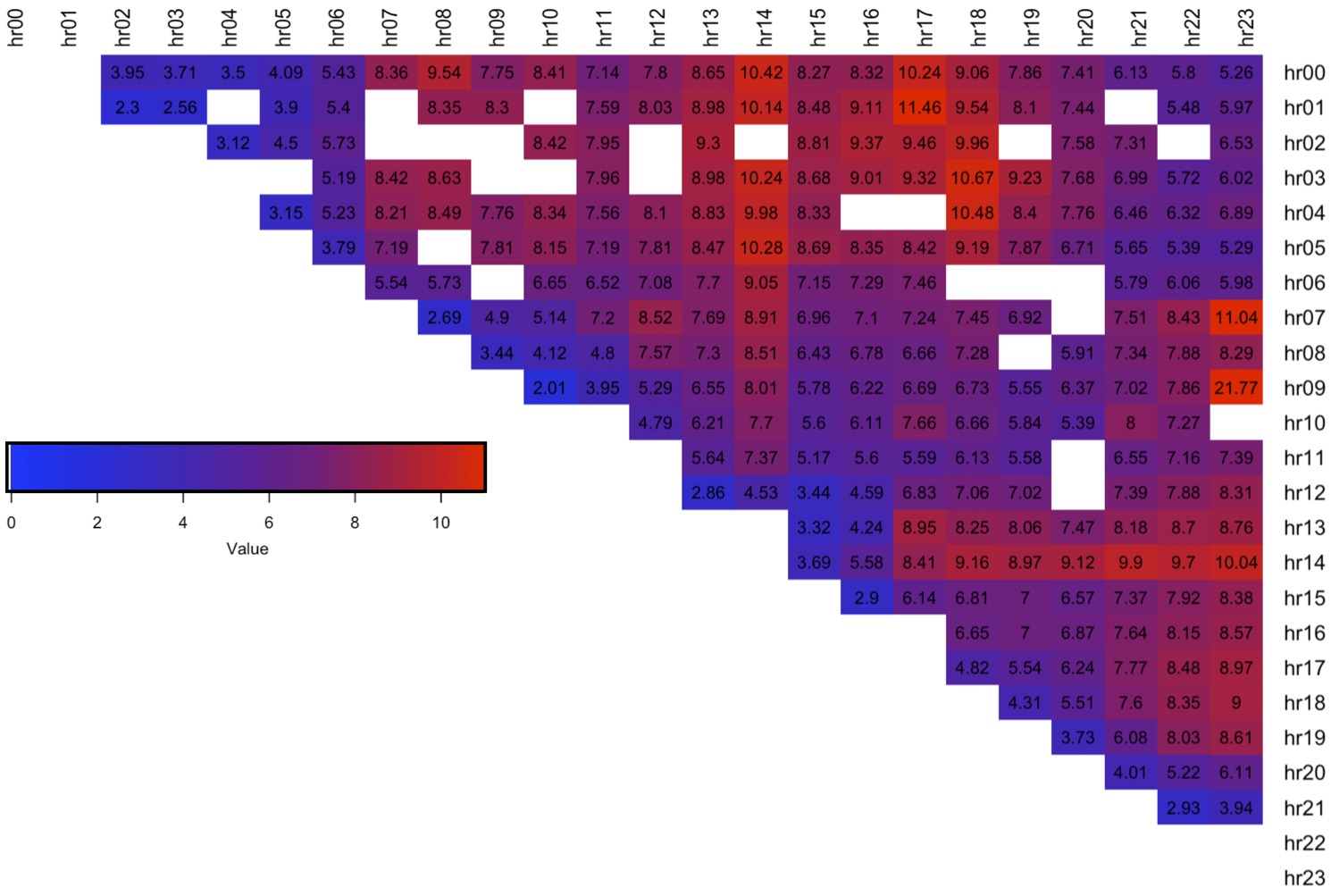}
		\caption{RMSE of skew type distributions for forecasting expected value, $E(Y)$, over rolling-window test data.}
		\label{fig:app1_Forecast_EY_RMSE_Skew}
	\end{figure}

	The RMSE values of \textit{skew type} distribution models for the forecasted expected value of the rolling-window test data is given in Figure \ref{fig:app1_Forecast_EY_RMSE_Skew}.
	The results show that nighttime spreads with hour 14.00, 17.00 and 18.00 had the biggest error when forecasting the level.
	
	\section{Rolling-Window RMSE - Normal Based}
\begin{figure}[h!]
		\centering
		\includegraphics[width=1.0\textwidth]{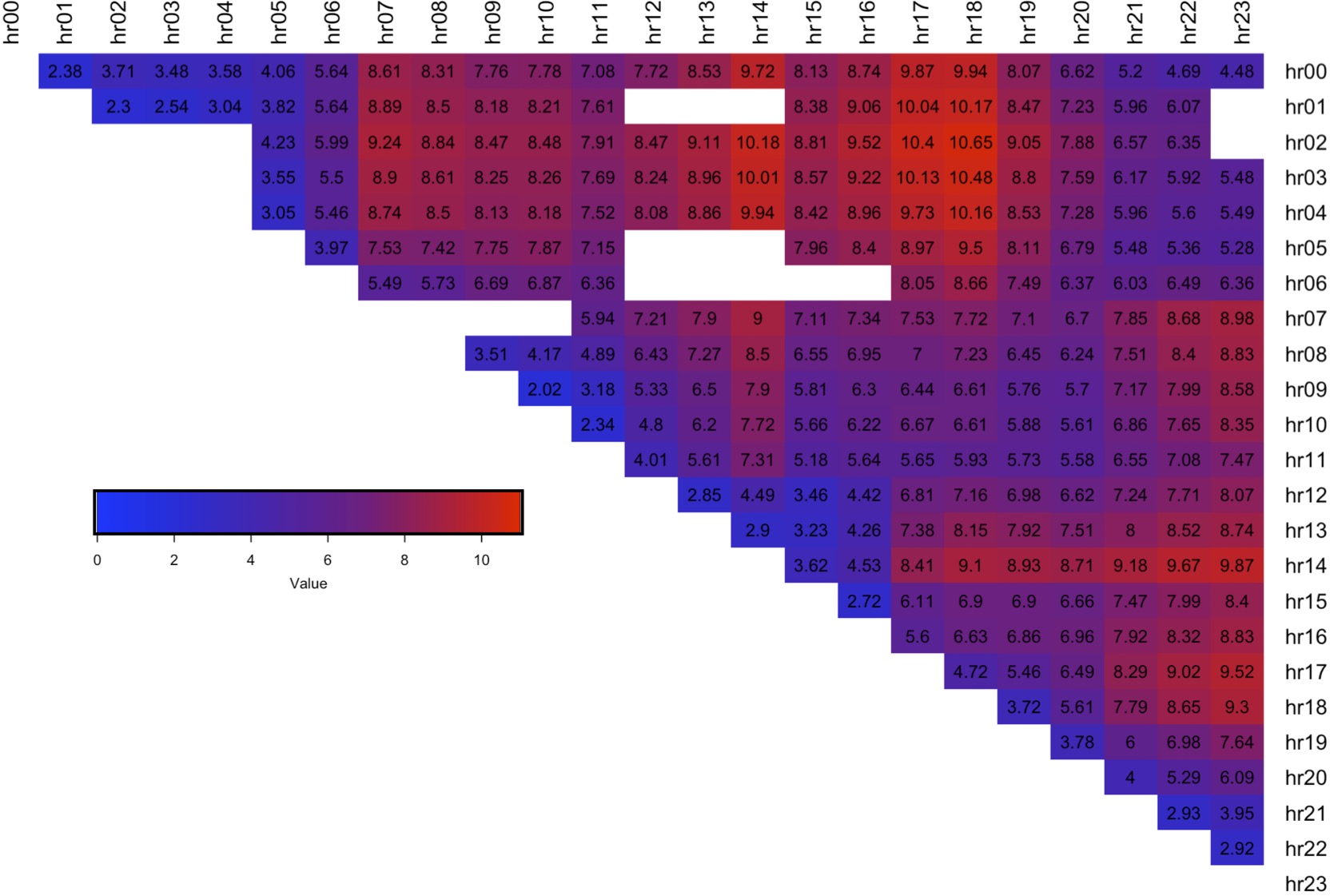}
		\caption{RMSE of normal type distributions for forecasting expected value, $E(Y)$, over rolling-window test data.}
		\label{fig:app1_Forecast_EY_RMSE_NO}
	\end{figure}

	The RMSE values of the benchmark \textit{Normal type} models for the forecasted expected value of the rolling-window test data is given in Figure \ref{fig:app1_Forecast_EY_RMSE_NO}.
	The results show that nighttime spreads with hour 14.00, 17.00 and 18.00 had the biggest error when forecasting the level.

\end{document}